\documentclass[12pt,a4paper]{article}

\usepackage{epstopdf}
\usepackage{amsmath}
\usepackage{amssymb}
\usepackage{latexsym}
\usepackage{graphicx}

\title{Critical points and symmetries of a free energy function for 
biaxial nematic liquid crystals} 
\author{David Chillingworth\\
                            \\
         Mathematical Studies AU\\
        University of Southampton\\
         Southampton SO17 1BJ\quad UK\\
          {\tt drjc@soton.ac.uk}
}

\begin{document}
\maketitle

\def\C{{\mathbf C}}
\def\R{{\mathbf R}}
\def\Z{{\mathbf Z}}
\def\D{{\mathbf D}}
\def\K{{\mathbf K}}
\def\u{{\mathbf u}}
\def\v{{\mathbf v}}
\def\w{{\mathbf w}}
\def\U{{\mathbf U}}
\def\V{{\mathbf V}}
\def\SS{{\mathbf S}}
\def\QQ{{\mathbf Q}}

\def\A{{\mathcal A}}
\def\CC{{\mathcal C}}
\def\DD{{\mathcal D}}
\def\FF{{\mathcal F}}
\def\P{{\mathcal P}}
\def\RR{{\mathcal R}}
\def\KK{{\mathcal K}}
\def\LL{{\mathcal L}}
\def\WW{{\mathcal W}}

\def\etp3{e^{\frac{2\pi i}3}}

\def\dbd#1#2{\frac{\partial#1}{\partial#2}}

\def\bz{\bar z}
\def\bw{\bar w}
\def\bq{\bar Q}
\def\oz{\overline Z}
\def\ow{\overline W}
\def\tq{\tilde q}
\def\bl{{\boldsymbol \lambda}}
\def\bn{{\mathbf n}}
\def\bu{{\mathbf u}}

\def\tu{\tilde u}
\def\tw{\tilde w}
\def\tq{\tilde q}

\def\tr{\mathop{ \mathrm{tr} }}
\def\til{\tilde}
\def\wt{\widetilde}

\def\del{\delta}
\def\eps{\varepsilon}
\def\beps{{\mathbf \varepsilon}}
\def\vf{\varphi}
\def\lam{\lambda}
\def\om{\omega}
\def\Om{\Omega}

\newcommand{\threej}[2]{%
 \ensuremath{\begin{pmatrix} 
 #1 \\ 
 #2 
 \end{pmatrix}}}

\def\nhd{neighbourhood\ }

\def\bsk{\bigskip}
\def\msk{\medskip}
\def\ssk{\smallskip}

\def\proof{{\em Proof}.\enspace}
\def\endproof{{\hfill $\Box$}}
\def\rem{\noindent{\em Remark.}\ }

\newtheorem{theo}{Theorem}
\newtheorem{prop}{Proposition}
\newtheorem{lemma}{Lemma}
\newtheorem{cor}{Corollary}
\newtheorem{example}{Example}

\begin{abstract}
We describe a general model for the free energy function for a homogeneous medium of mutually interacting molecules, based on the formalism for a biaxial nematic liquid crystal set out by Katriel {\em et~al.} (1986) in an influential paper in {\em Liquid Crystals} {\bf 1} and subsequently called the KKLS formalism. The free energy is expressed as the sum of an entropy term and an interaction (Hamiltonian) term.
Using the language of group representation theory we identify the order parameters as averaged components of a linear transformation, and characterise the full symmetry group of the entropy term in the liquid crystal context as a wreath product $SO(3)\wr Z_2$.  The symmetry-breaking role of the Hamiltonian, pointed out by Katriel {\em et~al.}, is here made explicit in terms of centre manifold reduction at bifurcation from 
isotropy.  We use tools and methods of equivariant singularity theory to reduce the bifurcation study to that of a $D_3\,$-invariant function on $\R^2$, ubiquitous in liquid crystal theory, and to describe the \lq universal' bifurcation geometry in terms of the superposition of a familiar swallowtail controlling uniaxial equilibria and another less familiar surface controlling biaxial equilibria. In principle this provides a template for {\em all} nematic liquid crystal phase transitions close to isotropy, although further work is needed to identify the absolute minima that are the critical points representing stable phases.
\end{abstract}

\section{Introduction}
In its simplest terms a liquid crystal material is viewed as a fluid consisting of a large number of molecules, identical in shape but
with potentially different orientations in 3-dimensional space that affect their mutual interactions. 
A molecule will tend to change its orientation in 
$\R^3$ under the influence of other molecules and/or external forces. The resulting changes in optical properties of a liquid crystal sample are the key to the currently ubiquitous liquid crystal display technology~\cite{DS,FI,SL}.
\msk

Equilibrium states or {\em phases} of a liquid crystal are typically modelled as critical points of a a suitable {\em free energy} function defined on a space of {\em order parameters} that measure the degree and nature of alignment of the molecules. Stable states correspond to absolute minima of the free energy.
In the paper \cite{KK} updated in \cite{LN} the authors set out a molecular field 
theory formalism for a free energy function for spatially homogeneous uniaxial and biaxial nematic liquid crystals. In this formalism the free energy has two components: a molecular interaction term (typically quadratic) and an entropy term.
The main focus is on a small number of order parameters as typically used in the liquid crystal literature, but in an Appendix to~\cite{KK} and Section~III of~\cite{LN} they indicate how to generalise to an arbitrary number of order parameters. 
In~\cite{LN} the formalism is put into practice through the calculation of leading terms in the Taylor series for the free energy function close to isotropy in terms of four standard order parameters widely used in studies of biaxiality (with equally widely varying notation: see ~\cite{RO}).
\msk

In this paper we describe the model (which, following the authors in~\cite{LN}, we call the KKLS model) in a general coordinate-free geometric setting which makes evident the natural symmetry group of the free energy function.  In  particular we note that the symmetry group $G$ of the entropy term is larger than is commonly supposed in the literature.  We then specialise to the more familiar setting in which the space of molecular variables is $5$-dimensional (the space $V$ of $3\times3$ real symmetric traceless matrices or $Q$-{\em tensors}) which gives rise to a $25$-dimensional space of order parameters ({\em supertensors}), namely the space $L(V)$ of linear transformations $V\to V$.
\msk

To study the critical point behaviour of a real-valued function on $L(V)$ that is invariant under the action of the appropriate group $G$ it is natural to turn to the classical theory of invariants and to seek a Hilbert basis (also called an integrity basis) for these functions, that is a finite set of $G$-invariant polynomials in terms of which all other $G$-invariant functions $h$ can be expressed. 
Here this is too ambitious a task, although elsewhere~\cite{CLT} we are able to exhibit the Molien generating function that counts the number of such (independent) polynomials at each degree. 
\msk

Fortunately, there are two reductions of the problem that by different routes lead to
essentially equivalent simplifications from 25 dimensions to precisely two. As we show, the bifurcations of equilibria from the isotropic phase (no mutual alignment) take place in a $5$-dimensional subspace isomorphic to $V$ with the natural symmetry group $SO(3)$ acting by conjugacy.  For unusual group-theoretic reasons the study of $SO(3)$-invariant functions on $V$ is no different from the study of $D_3$-invariant functions on $\R^2$, which therefore suffices for the analysis of uniaxial and biaxial bifurcation from isotropy.  This is the setting for the classical Landau-de~Gennes theory of phase transitions for biaxial nematic liquid crystals~\cite{DG}, naturally much studied in the literature: a thorough and careful mathematical analysis can be found in~\cite{AL}.  
\msk

Alternatively, the common assumption can be made that at equilibrium the ensemble average over $G$ of a matrix $Q\in V$ has the same eigenframe as $Q$, an assumption supported by physical intuition 
but so far lacking a rigorous proof in the present context.  With this assumption the 
$5$-dimensional space $V$ reduces to the $2$-dimensional space $D$ of traceless diagonal $3\times3$ matrices, so then $L(D)$ is $4$-dimensional and is the space of $4$ order parameters familiar in studies of biaxial liquid crystals.  Again, we show that bifurcation from isotropy occurs in a $2$-dimensional subspace with the natural $D_3$ action, thus leading again to the Landau-de~Gennes formulation. 
\msk

The purpose of this paper is first to clarify the nature of order parameters and free energy functions for liquid crystals, and in particular their symmetries, within the wider context of group representation theory, and then to invoke ideas and techniques from equivariant singularity theory to provide rigorous results on robustness of the bifurcation scenario, including precise criteria for the safe neglect of higher order terms. Many of the results in the literature on bifurcation of uniaxial and biaxial equilibria, and in particular those described for the KKLS model in~\cite{LN}, can be interpeted and their simplifying assumptions and approximations tested within this \lq universal' framework. In carrying out the reduction from four order parameters to the two that govern the bifurcations from isotropy, we are able to give explicit expressions for the coefficients of the universal model in terms of those calculated numerically for the specific KKLS free energy function on $\R^4$ in~\cite{LN}.  For an overview of the mathematical and physical assumptions that lie behind the KKLS and related models together with notes on the historical context and many key literature references we defer to~\cite{LN}.
\msk

The structure of the paper is as follows.  In Section~\ref{s:algebra} we introduce some of the basic formalism of group actions (representations) and invariant functions. Section~\ref{s:KKLSmodel} describes the KKLS model and its symmetries. In Section~\ref{s:classprog} the general (if unfeasible) classification programme of equilibria for the KKLS model is formulated, redeemed by a result that allows drastic dimension-reduction when treating bifurcation from isotropy. Section~\ref{s:reducedmodels}
sets out notation for the reduced KKLS model with four order parameters, presenting a Hilbert basis for the functions invariant under the natural symmetry characterised by the action of a $72$-element group.  In Section~\ref{s:expandf} the Taylor series of the free energy is considered up to degree~$8$, and its explicit conversion to the sum of a nondegenerate quadratic from in two variables plus a residual function in two other variables is carried out up to degree~$6$ (for comparison with KKLS data from~\cite{LN})
at a general point of instability of the isotropic state.   
Finally, Section~\ref{s:bifnormal} gives a bifurcation analysis for the normal form of the residual function in terms of the geometry of the familiar swallowtail for uniaxial equilibria and a less familiar configuration for biaxial equilibria.
There are four Appendices giving technical results needed in the main text.
\msk

Among the aims of the Programme on {\em The Mathematics of Liquid Crystals} hosted at the Isaac Newton Institute, Cambridge in 2013 was to encourage mathematicians familiar with techniques of equivariant bifurcation theory and  singularity theory to engage with the wealth of interesting and important problems arising in the study of liquid crystals, while at the same time alerting experts in liquid crystals to the availability of these powerful mathematical tools. It is hoped that this paper may contribute to the cultural exchange.
\bsk

\noindent{\em Remarks on terminology and notation}.
\msk

\noindent In this paper we use notation that characterises a group as an abstract object (up to isomorphism), while in much of the liquid crystal literature a group name represents an abstract group together with a specified {\em action} of that group on euclidean space (Sch\"onflies notation). See~\cite{MZ} for a thorough discussion of this and related issues.  

In the bifurcation theory literature it is important to distinguish between the {\em variables} that are to be measured, and the {\em parameters} which are the coefficients of the governing equations and can in principle be controlled.  This is a potential source of confusion in liquid crystal theory where the variables are the {\em order parameters}. 
\section{Algebraic preliminaries} \label{s:algebra}
We begin by setting out some basic facts about group actions and invariant functions.
\medskip
\subsection{Group actions} \label{s:actions}
Let $G$ denote a finite or, more generally, a compact Lie group (such as the 
rotation group $SO(n)$), and let $\alpha$ denote an {\em action} of $G$ on a linear 
space $E$.  This means that to every element $g\in G$ there is an associated invertible linear map $\alpha(g):E\to E$ such that group multiplication corresponds to composition of linear maps: for all $g_1,g_2\in G$ we have 
\begin{equation}  \label{e:action}
\alpha(g_1g_2)=\alpha(g_1)\alpha(g_2)
\end{equation}
with the usual understanding (not shared by algebraists) that the right-hand map is applied first.  A linear action is also called a {\em representation}. 
The action $\alpha$ is {\em absolutely irreducible} if the only elements of $L(E)$ that commute with $\alpha(g)$ for all $g\in G$ are the scalar multiples of the identity map.
In particular this implies that no proper linear subspace $D\subset E$ is $\alpha$-invariant (i.e. mapped into itself) by all $\alpha(g)$ for $g\in G$.
For any $x\in E$ the {\em orbit} of $x$ under the action $\alpha$ is the set $\{\alpha(g)x:g\in G\}$, that is the set of all points in $E$ to which $x$ is taken by the action $\alpha$ of $G$ (the partners of $x$ under the $G$-symmetry $\alpha$).
\msk

The notion of different actions of the same group is familiar in quantum mechanics, for example, where for a given $g\in G=SO(3)$ the  $(2j+1)\times (2j+1)$ matrix $\DD^{(j)}(g)$ denotes the effect of $g$ as a natural linear transformation in $2j+1$ dimensions: see for example~\cite{MR},\cite{AE}.
Of course any group $G$ can act {\em trivially} on any space $E$ by taking $\alpha(g)$ to be the identity transformation of $E$ for all $g\in G$.
\msk

Given an action $\alpha$ of $G$ on $E$ there are three natural actions of $G$ on the linear space $L(E)$ of linear maps $E\to E$  arising from $\alpha$: the {\em left action} $\lambda_\alpha$ in which $g\in G$ takes $M\in L(E)$ to $\alpha(g)M$, the {\em right action} $\rho_\alpha$  in which $g$ takes $M$ to $M\alpha(g^{-1})=M\alpha(g)^{-1}$ 
(the inversion is to ensure that (\ref{e:action}) holds in the correct order), and the combined {\em conjugacy action} $\til\alpha=\lambda_\alpha\rho_\alpha$ where
\begin{equation} \label{e:conj}
\til\alpha(g):M\to \alpha(g)M\alpha(g)^{-1}.
\end{equation}
For each $g\in G$ the conjugacy transformation (\ref{e:conj}) is a
{\em linear} map from $L(E)$ to $L(E)$; thus $\til\alpha(g)\in L(L(E))$.
When a basis is chosen for $E$, elements of $L(E)$ are represented by matrices (or tensors) $(a_{ij})$, and elements of $L(L(E))$ by {\em supertensors} $\bigl(a_{ij}^{kl}\bigr)$.
\medskip

If $E$ is equipped with  a scalar product (inner product) denoted by ${\cdot}$ 
then associated to every $A\in L(E)$ is the {\em dual} 
linear map $A^*\in L(E)$ defined by 
$Au\cdot v=u\cdot A^*v$ for every $u,v\in E$. In
terms of (real) matrices this is simply the {\em transpose}, and
the map $A\in L(E)$ is {\em symmetric} if $A^*=A$.
We denote the set of symmetric elements of $L(E)$ by $Sym(E)$, and the subset of those with {\em zero trace} by $Sym_0(E)$.
\medskip

The linear map $A\in L(E)$ is {\em orthogonal} if $A^*A$ is the identity map, so that $A$ preserves the scalar product on $E$.   The action $\alpha$ is {\em orthogonal} if $\alpha(g)\in L(E)$ is orthogonal for every $g\in G$, in which case 
\[
\alpha(g)u\cdot\alpha(g)v=u\cdot v
\]
for every $g\in G$ and $u,v\in E$. By a standard process of averaging it is easy to show that an inner product can always be chosen to make a given action $\alpha$ orthogonal: see~\cite[Theorem II(1.7)]{BD} or~\cite[Theorem 4.4.3]{CL} for example.  This greatly simplifies the handling of matrices describing symmetries.  In particular, if $\alpha$ is orthogonal then $Sym(E)$ and $Sym_0(E)$ are preserved by the conjugacy action~$\til\alpha$ on $L(E)$.
\msk

A natural scalar product on the space $L(E)$ arises from that on $E$
and given by
\[
  A\cdot B :=  \tr(A^*B)
\]
where $\tr$ denotes {\em trace}. The following identities are easily verified:
\begin{prop}  \label{p:trace}
\[
 A\cdot B =B\cdot A =A^*\cdot B^* =B^*\cdot A^* 
\]
for all $A,B\in L(E)$.
\endproof
\end{prop}
A further elementary result is important for our purposes:
\begin{prop}  \label{p:inner}
If the action $\alpha$ on $E$ is orthogonal then the action $\til\alpha$ on $L(E)$ is also orthogonal. 
\end{prop}
\proof By definition 
\begin{align*}
  \til\alpha(g)(B)\cdot \til\alpha(g)(C) &=\tr\bigl((\til\alpha(g)(B))^* \til\alpha(g)(C)\bigr)  \\
            &=\tr\bigl( (\alpha(g)B\alpha(g)^{-1})^*\alpha(g)C\alpha(g)^{-1} \bigr)    \\ 
            &=\tr\bigl( \alpha(g)\,B^*C\,\alpha(g)^{-1} \bigr)  \\
            &=\tr(B^*C) = B\cdot C 
\end{align*}
using Proposition \ref{p:trace} and the fact that $\alpha(g)^*=\alpha(g)^{-1}$ by orthogonality.
\endproof
%
\subsection{Invariant functions}
A real-valued function $f:E\to\R$ is said to be {\em invariant} under
the action $\alpha$ of $G$ on $E$ when it is the case that 
\[
f(\alpha(g)x)=f(x)
\]
 for all $g\in G$ and all $x\in E$.  It is a classical 
theorem due to Hilbert that if $G$ is finite or compact then there is a finite set of $G$-invariant polynomials (a {\em Hilbert basis} or {\em integrity basis}) such that every $G$-invariant polynomial function on $E$ can be expressed as a polynomial function of these basic polynomials.  This result was extended by Schwarz \cite{SC} to show that 
every $G$-invariant smooth (i.e. $C^\infty$) function on $E$ can be expressed as a smooth function of the polynomials in a Hilbert basis.  Thus, for example, every smooth function $V=Sym_0(\R^3)\to\R$ invariant under the conjugacy action of $SO(3)$ can be written as a smooth function of the two invariants $\tr Q^2$ and $\tr Q^3$ for $Q\in V$.  This result, well known in liquid crystal theory, follows as a simple application of a powerful and elegant general technique in invariant theory that we now describe for later use. 
\subsection{Molien series}   \label{s:molintro}
The Hilbert basis for a given group action is not         
unique, and finding one in practice can be a considerable challenge~\cite{SF}.  However, a valuable tool exists
enabling us to calculate in advance how many elements a Hilbert
basis will have.  The {\em Molien series}~\cite{CL},\cite{MZ},\cite{SF}
for the given action $\alpha$ of $G$ on $E$ is a formal power series
\[
\P_\alpha(t)=\sum_{d=0}^\infty r_dt^d
\]
where the coefficient $r_d$ is the number of linearly independent
$G$-invariant polynomial functions on $E$ of homogeneous degree $d\,$.
Remarkably, in the case of a finite group $G$ there is an explicit expression for $\P_\alpha(t)$ as
\begin{equation}  \label{e:molien}
\P_\alpha(t)=\frac1{|G|}\sum_{g\in G}\frac1{\det(I-t\alpha(g))}
\end{equation}
where $|G|$ is the number of elements in $G$, with an analogous
expression for an infinite compact group using an integral rather
than a sum. Furthermore, the series $\P_\alpha(t)$ is expressible as a rational function which gives more detailed information about the algebraic relationships between the invariants~\cite{MZ},\cite{SF}.
\msk

\noindent\textbf{Example}\enspace\emph{Action of $D_3$ on $\R^2$}
\ssk

The Molien function~(\ref{e:molien}) for the natural action $\delta$ of $D_3$ on $\R^2$ as the symmetries of an equilateral triangle is easily found to be
\begin{equation}  \label{e:molien_D3}
\P_\delta(t)=\frac1{(1-t^2)(1-t^3)}
\end{equation}
which conveys the information that a Hilbert basis consists of one polynomial of degree~$2$ and one of degree~$3$, there being no algebraic relation between them as the terms $t^2,t^3$ appear only in the denominator.  For suitable coordinates in $\R^2$ the basis elements may be taken to be $x^2+y^2$ and $x^3-3xy^2$. 
\msk

\noindent\textbf{Example}\enspace\emph{Action of $SO(3)$ on $V=Sym_0(\R^3)$}
\ssk

An element $g\in SO(3)$ is a $3\times 3$ orthogonal matrix with eigenvalues 
$\{1,e^{\pm i\theta}\}$ for some angle $\theta$, and for the conjugacy action $\til\alpha$ the eigenvalues of $\til\alpha(g)$ are $\{1,e^{\pm i\theta},e^{\pm 2i\theta}\}$.
To evaluate the Molien series using the integral version of the formula~(\ref{e:molien}) we make use of the Weyl Integral Formula (as described in~\cite{BD} for example) which shows that such an integral can be decomposed into an integral over a {\em maximal torus} in $G$ and over the quotient of the group by this torus.  In the case when the integrand is a class function (invariant under conjugation in the group) then the latter integral becomes trivial, and the whole integral reduces to an integral over the maximal torus only, at the cost of introducing a further term into the integrand that we call the {\em Weyl factor}. The integral over the circle is expressed in terms of angular variables and can be evaluated using residue calculus.
\msk

 {In our case the maximal \lq torus' is a circle given by the matrices}
\[
\begin{pmatrix}
 1 & 0 & 0 \\ 0 & \cos\theta & -\sin\theta \\  0 & \sin\theta & \cos\theta
\end{pmatrix}
\]
for $0\le\theta<2\pi$, and integration is over the unit circle regarded as the circle $|z|=1$ in the complex plane. The Weyl factor can be found to be 
\[
\tfrac12(1-z)(1-\bar z)= (1-\cos\theta)
\]
(cf.~\cite{LS}) and so the Molien integral~(\ref{e:molien}) for $\til\alpha$ becomes
\[
\P_{\til\alpha}(t)=\frac12\,\frac i{2\pi}\,\int_{|z|=1}
    \frac{(1-z)^2}{\prod_{j=-2}^2(1-tz^j)}\frac{dz}{z^2}.
\] 
Since the aim is to find $\P_{\til\alpha}(t)$ as an infinite series in $t$ we regard $t$ as a small real variable and in any case with $|t|<1$, this restriction to be exploited when evaluating residues. In our case the relevant residues are at $z=0$ and at $z=t,\pm\sqrt{t}$ and easy calculations give
\begin{equation}
\P_{\til\alpha}(t)=\frac 1{(1-t^2)(1-t^3)}
\end{equation}
which shows that the ring of $\til\alpha$-invariant functions on $V$ has two generators of homogeneous degree~$2,3$ respectively. These may be taken to be
\[
\tr Q^2,\quad \tr Q^3
\]
although the latter could be replaced by $\det Q$ since the Cayley-Hamilton equation shows that $\tr Q^3=3\det Q$ when $trace\,Q=0$.
\msk

It is no coincidence that the Molien function $\P_{\til\alpha}(t)$ coincides with that of the natural representation $\delta$ of $D_3$ on $\R^2$.  This is because of special properties of $\til\alpha$: every $\til\alpha$-invariant function on $V$ is uniquely determined by its values on the $2$-dimensional subspace $D$ of $V$ consisting of diagonal matrices, and the $\del$-orbit of a point in $D$ is precisely the intersection of its $\til\alpha$ orbit with~$D$.
%
\section{The KKLS model} \label{s:KKLSmodel} 

We now turn to the main focus of this paper which is the KKLS model~\cite{KK,LN} for the construction of a free energy function on an appropriate  space of order parameters.
Following~\cite{KK,LN} we begin with the key assumption that properties of a molecule relevant to its interactions with its neighbours are characterised by an $n$-tuple of real numbers (orientational order or \lq shape' parameters) $\QQ=(Q_1,\ldots,Q_n)\in\R^n$, and that each element $\Om$ of the rotation group $SO(3)$ when applied to a given molecule induces a linear transformation $\QQ(\Om)$ of $\R^n$ that is a measure of the interaction between a molecule and its rotated counterpart. This places us immediately in the mathematical setting of an action $\alpha$ of $G=SO(3)$ on $E=\R^n$ as discussed in Section~\ref{s:actions}, although given $\Om\in SO(3)$ we write $\alpha(\Om)\QQ$ rather than $\QQ(\Om)$. Here and throughout we assume homogeneity of the medium, so that local properties of the liquid crystal are the same everywhere and spatial derivatives of $\QQ$ do not come into play.
\msk

A well-established technique to finesse the intractable task of calculating mutual interactions of huge numbers of molecules is to apply instead a {\em mean field theory} where each particular molecule is seen as reacting to the {\em average} influence of all the other molecules. Evaluating this average corresponds to integrating $\alpha(\Om)$  with respect to an appropriate probability distribution on the group $SO(3)$, this being determined by the relevant physical theory. The components of the average $\langle\alpha\rangle\in L(\R^n)$ with respect to a suitably chosen basis for $\R^n$ are the  {\em order parameters} for the liquid crystal.
\msk

In an equilibrium state the alignment of a particular molecule must necessarily be consistent with the creation of the mean field in the first place, thus posing a significant mathematical problem.  We first formulate this problem in a general coordinate-free setting, while following the paradigm of~\cite{KK,LN}, and we then apply it in the context of molecules with effective $D_{2h}$ symmetry (Sch\"onflies notation) or $Z_2^3$ symmetry (algebraic notation) where each $Z_2$ factor acts by rotation through angle $\pi$ about one of the coordinate axes. 
As far as interactions are concerned the molecule can be represented by an ellipsoid whose departure from sphericity is given by a real symmetric traceless $3\times3$ {\em susceptibility matrix} $\QQ\in V=Sym_0(\R^3)\cong\R^5$, therefore $\alpha(\Om)\in L(V)\cong\R^{25}$.  As we shall see, however, further assumptions can often be made that enable reduction to much lower dimensions.
\msk

In the KKLS and other  models for equilibrium states (phases) of liquid crystals, an equilibrium  corresponds to a critical point of a suitable {\em free energy} function
on the space of order parameters, with stable phases corresponding to global or absolute minima.  The KKLS model is characterised by the fact that the free energy has two components: a molecular interaction or {\em Hamiltonian} term and an {\em entropy} term.  We focus first on the latter and return afterwards to the former.
\subsection{The entropy component} \label{s:entropydefs}
We follow the general formulation of~\cite{KK,LN}, expressed here in the more abstract context of Section~\ref{s:actions}.

Consider an arbitrary compact group $G$ with orthogonal action $\alpha$ on a linear space $E$, together with a probability distribution $\phi$ on $G$ expressed in terms of the action $\alpha$: thus $\phi$  has the form $\phi(g)=\psi(\alpha(g))$ for some suitable $\psi:L(E)\to\R$ which we assume smooth. 
\msk

There are two important physical quantities to be derived from $\phi$.
The first is the {\em ensemble average}
\begin{equation}  \label{e:defW}
W_\phi:= \langle\alpha\rangle_\phi=\int_G \alpha(g)\,\phi(g)  \in  L(E)
\end{equation}
which represents the average $\alpha$-action of $g\in G$ with respect to the particular 
probability distribution $\phi$, the integral being taken over the natural normalised invariant (Haar) measure on $G$.  This is a quantity that in principle might be measured in the laboratory, whereas the probability distribution itself could not.  If $\dim E=n$ then the $n^2$ entries of $W_\phi$ (with respect to a chosen basis) are the {\em order parameters} describing the state of this system. 
\msk

The second quantity derived from $\phi$ is the {\em entropy} (more accurately, the {\em entropy difference} between the isotropic state characterised by the uniform probability distribution on $G$ and the state given by $\phi$), defined by
\begin{equation}  \label{e:defS}
S_\phi:= -k \int_G \phi(g)\log \phi(g)
\end{equation}
where $k$ is Boltzmann's constant. 
\msk

Following the KKLS methodology~\cite{KK} the probability distribution $\phi$ is now chosen in such as way as to maximise the entropy, given the (observed) value of $W$.  A standard use of calculus of variations shows that such a probability distribution takes the form
$\phi_\eta$ where
\begin{equation}
\phi_\eta(g)=\phi(g,\eta):= \frac1{Z(\eta)}e^{\eta\cdot \alpha(g) }
\end{equation}
for some $\eta\in L(E)$ (effectively a Lagrange multiplier), the {\em partition function} $Z(\eta)$ being chosen so that $\phi(\cdot,\eta)$ is indeed a probability distribution and so satisfies $\int_G \phi(g,\eta)=1\,$, that is
\begin{equation}  \label{e:defZ}
  Z(\eta)=\int_G e^{\eta\cdot\alpha(g) }.
\end{equation}
We then see that $W_{\phi_\eta}=W(\eta)$ can be equivalently expressed as 
\begin{equation}  \label{e:logZ}
W(\eta)= \nabla \log Z(\eta)
\end{equation}
and it follows also from (\ref{e:defZ}) and (\ref{e:defW}) that the entropy $S_{\phi_\eta}=S(\eta)$ may be expressed as
\begin{equation}  \label{e:altS}
         S(\eta) = k\bigl(\log Z(\eta)-\eta\cdot W(\eta)\bigr).
\end{equation}
Next we discuss the all-important symmetry properties of this entropy function.
\subsection{Symmetries of the entropy}  \label{s:sym}
The ensemble average $W$ and entropy function $S$ exhibit much symmetry arising from the natural symmetries of the partition function $Z$ which we discuss first.
\begin{prop}  \label{p:symZ}
The partition function $Z:L(E)\to\R$ is invariant under
\begin{enumerate}
\item[\emph{(i)}] the left action $\lambda_\alpha$ of\/ $G$ on $L(E)\,$;
\item[\emph{(ii)}] transposition (i.e. duality) in $L(E)$.
\end{enumerate}
\end{prop}
\proof  If $h\in G$ and $\eta\in L(E)$ then
\begin{equation}
   Z(\alpha(h)\eta)=  \int_G e^{\alpha(h)\eta\cdot\alpha(g)}  
                =  \int_G e^{\eta\cdot\alpha(h)^*\alpha(g)} 
                =  \int_G e^{\eta\cdot\alpha(h^{-1}g)} 
                = Z(\eta)
\end{equation}
using the facts that $\alpha(h)^*=\alpha(h^{-1})$ by orthogonality and that $h^{-1}g$ runs through the whole of $G$ (for fixed $h$) when $g$ does.
This establishes (i).

For (ii) we note
\begin{equation}
   Z(\eta^*) =  \int_G  e^{\eta^*\cdot\alpha(g)}  
             =  \int_G  e^{\eta\cdot\alpha(g)^*} 
             =  \int_G  e^{\eta\cdot\alpha(g^{-1})}     = Z(\eta)
\end{equation}
again by orthogonality and the fact that $g^{-1}$ runs through $G$ as $g$ does. \endproof
\medskip  

Since from (\ref{e:logZ}) the map 
\[
W:L(E)\to L(E):\eta\mapsto W(\eta)
\] 
is the gradient of the function $Z:L(E)\to\R$ we 
have the following automatic consequence:
\begin{cor}
The map $W$ is {\em equivariant} with respect to {\em (i)} and {\em (ii)}, that is
\begin{enumerate}
\item[\emph{(i}$\,'$\emph{)}]  $W(\alpha(h)\eta) = \alpha(h) W(\eta) \qquad\text{for all\ } h\in G\,$;
\item[\emph{(ii}$\,'$\emph{)}]  $W(\eta^*) = \bigl[ W(\eta) \bigr]^*$
\end{enumerate}
for all $\eta\in L(V)$. \endproof
\end{cor}

Finally, since the action $\alpha$ is orthogonal and the duality operator ${}^*$ preserves the inner product on $L(E)$ (see Proposition~\ref{p:trace}) we have an automatic consequence of the formula (\ref{e:altS}):
\begin{cor}  \label{e:entsym}
The entropy function $S:L(E)\to\R$ is invariant with respect to the symmetries
{\em (i)} and {\em (ii)} of Proposition~\ref{p:symZ}.  \endproof
\end{cor}

\noindent {\em Remark}. The fact that $S$ is invariant under (i) is quite
natural, as it simply reflects coordinate-independence.  Invariance
under (ii) is more subtle, and arises from the fact that averaging over a
group of orthogonal matrices is the same as averaging over their transposes.
This symmetry plays a significant role in formulation of the general theory and construction of Hilbert bases~\cite{CLT} but, notwithstanding some observations in~\cite{LN}, appears to be little exploited in the liquid crystal literature.  In Section~\ref{s:invs} we discuss this further and describe explicitly the contribution that it makes to the KKLS model for the free energy. 
\subsubsection{Structure of the extended symmetry group} 
Here we clarify the algebraic structure of the group $\wt G$ of
symmetries of the partition function $Z:L(E)\to\R$ and (Corollary~\ref{e:entsym}) of the entropy function $S:L(E)\to\R$. The action of this group on $L(E)$ is given by~(i) and~(ii) in Proposition~\ref{p:symZ}. Since the left action $\lambda_\alpha$ of $G$ on $L(E)$ is an orthogonal action, as is the transposition operator $\tau:A\to A^*$, it follows that $\wt G$ can be viewed as a subgroup of the group of orthogonal transformations of $L(E)$.
\msk

The left and right actions $\lambda=\lambda_\alpha\,,\rho=\rho_\alpha$ of $G$ on $L(E)$ satisfy
\[
\rho(g)\lambda(g)=\lambda(g)\rho(g), \qquad   \tau\rho(g)=\lambda(g)\tau
\]
for every $g\in G$, so the action of any element
of $\wt G$ can be expressed uniquely in one or other of the forms
\[
\rho(g)\lambda(h)\quad\text{or}\quad \tau\,\rho(g)\lambda(h)
\]
for $g,h\in G$. 
The elements without $\tau$ form a subgroup $G_0$ of $\wt G$ of index $2$
isomorphic to $G\times G$. 
In group-theoretical terms the group $\wt G$ has the form of a semidirect product
\[
\wt G \cong (G\times G) \rtimes Z_2
\]
or equivalently a {\em  wreath product} $G\wr Z_2$. To summarise:

\begin{prop} \label{p:wreath} Given an action $\alpha$ of $G$ on $E$, the partition function $Z$ and hence also the entropy function $S$ on $L(E)$ are invariant under the action of the wreath product $G\wr Z_2$ on $L(E)$ generated by left multiplication and transposition.\endproof
\end{prop}
In the important case where $G=SO(3)$ and $E=V=Sym_0(\R^3)$ the group $\wt G$ is a subgroup of the orthogonal group $O(25)$. Since the transposition operator $\tau:L(V)\to L(V)$ has eigenvalues
$1,-1$ with multiplicities $15,10$ respectively it follows that in fact
$\wt G<SO(25)$.
\subsection{The Hamiltonian and the free energy} \label{s:Ham}
The second key component of the KKLS model is the {\em Hamiltonian}, a real-valued function of $W$ that in physical terms measures the thermodynamic internal energy of the system.  The simplest plausible form for $H$ is
$H(W)=-\frac12u\,W\cdot W$ where $u$ is a constant; this corresponds to long-range particle interactions with no inter-particle correlations (see \cite{KK}).  A slightly more general expression~\cite{LU},\cite{LN} is a quadratic function of the form
\begin{equation}   \label{e:defH}
   H(W)= -\frac12 BW{\cdot}W
\end{equation}
where $B$ is a symmetric linear map $L(E)\to L(E)$.  We assume that $B$ is equivariant with respect to the left action $\lambda$ of $G$ on $L(E)$ (intrinsic frame-independence when $G=SO(3)$), so that $H$ is $\lambda$-invariant.
\medskip

Finally, the (Helmholtz) {\em free energy} function $\FF:L(E)\to\R$ is defined
by
\begin{equation}   \label{e:defF}
   \FF:= H(W) - T\,S
\end{equation}
where here $T$ denotes absolute temperature. 
Since $W,S$ are functions of $\eta$ as in (\ref{e:defW}),(\ref{e:altS}) we could regard $\FF$ as a function of $\eta$.  However, the physical motivation for this construction \cite{LN}  is that equilibrium states of the liquid crystal
correspond to values of the order parameter $W$ that are critical points of the free energy viewed as a function of $W$.  To find equilibrium states we thus need first to convert $S(\eta)$ into a function of $W$, that is to invert the map $\eta\mapsto W(\eta)$. By the Inverse Function Theorem (IFT) this can always be done locally in a \nhd of any given $\eta\in L(E)$ in view of the following result.  
\msk

First some terminology: we say that the action $\alpha$ {\em spans} $L(E)$ if the set $\alpha(G):=\{\alpha(g):g\in G\}$ does not lie in any proper affine subspace (hyperplane) of $L(E)$.  We prove in Appendix~D that this holds for the conjugacy action of $G=SO(3)$ on $V$, and leave as an easy exercise that it holds for the natural action of $D_3$ on $\R^2$. For a finite group $G$ an obvious necessary condition is that $|G|\ge\dim L(E)$.   
\begin{prop}  \label{p:invert}
If the action $\alpha$ spans $L(E)$ then the derivative $DW(\eta):L(E)\to L(E)$ is invertible for every $\eta\in L(E)$.
\end{prop}
\proof  We generalise the argument given in \cite{KK}.  Differentiating
\[
Z(\eta)W(\eta)=\int_G\alpha(g)e^{\eta\cdot\alpha(g)}
\]
with respect to $\eta$ and applying this to $\om\in L(E)$ gives
\[
(\nabla Z(\eta)\cdot\om)W(\eta)+Z(\eta)\,DW(\eta)\om 
                        = \int_G(\om\cdot\alpha(g))\,\alpha(g)e^{\eta\cdot\alpha(g)}
\]
which on dividing by $Z(\eta)$ and using~(\ref{e:logZ}) becomes 
\begin{equation} \label{e:divZ}
(W(\eta)\cdot\om) W(\eta) + DW(\eta)\om = \langle(\om\cdot\alpha)\alpha\rangle.
\end{equation}
Recalling $W(\eta)=\langle\alpha\rangle$ we take the inner product of~\eqref{e:divZ} with $\om$ to obtain
\begin{align}
\om\cdot DW(\eta)\om &=\langle(\alpha\cdot\om)^2\rangle 
                       - (\langle\alpha\rangle\cdot\om)^2 \\
                    &=\langle(\alpha\cdot\om-\langle\alpha\rangle\cdot\om)^2\rangle.
\end{align}
The right hand side cannot vanish for $\om\ne0\in L(E)$ unless $\alpha(g)-\langle\alpha\rangle$ is orthogonal to $\om$ for all $g\in G$, but in this case $\alpha$ fails to span $L(E)$.
\endproof
\msk

\rem In~\cite{KK} the proof is concluded by invoking the condition that  at high temperature $\langle\alpha\rangle=0$ on physical grounds (isotropy is the only stable equilibrium) and that the orientational order parameters are independent so that no nontrivial linear combination of them can give zero.
The spanning assumption here is more general.
\msk

Proposition~\ref{p:invert} implies that the map $W:L(E)\to L(E)$ is smoothly invertible on some \nhd of any given $\eta\in L(E)$, given the assumption about the group action. (Note that the IFT does not by itself imply that $W$ is globally invertible.) Therefore, expressing $\eta$ as a smooth function $\eta=\eta(W)$ for $W$ in an appropriate open set in $L(E)$ we may write the entropy $S(\eta)$ as $S(\eta(W))$ and the free energy as
\begin{equation}
   F= F(W) = H(W) - T\,S(\eta(W)). 
\end{equation}
The problem we then have to address is the following:
\medskip

\noindent\textbf{Problem}: find the critical points $W\in L(E)$ of $F:L(E)\to\R$.
\medskip

The global minima of $F$ are of primary importance, as they correspond to stable equilibrium states.  Local minima correspond to metastable states, while the critical points that are not minima correspond to unstable states not normally physically observable.  Nevertheless, the configuration of unstable states plays a key role in organising the interactions and basins of attraction of the stable states, and so it is important to understand the overall configuration of critical points and how it responds to changes in temperature $T$ as well as other parameters.
\msk

It is important to note that while the Hamiltonian $H:L(E)\to\R$ is invariant under the left action $\lambda$ of $G$, there is no reason why it should be $\tau$-invariant also and typically it is not. This symmetry-breaking role of the Hamiltonian, pointed out in~\cite{KK},\cite{LN}, is crucial to our analysis of bifurcation from isotropy in Section~\ref{s:expandf} below.
\subsubsection{Fixed-point formulation}
With the entropy
\begin{equation}
    S=S(\eta,W)=k \bigl( \log Z(\eta) - \eta\cdot W \bigr)
\end{equation}
regarded as a function of the two independent variables $\eta$ and $W$, equation~(\ref{e:logZ}) describes the locus where $\nabla_\eta S(\eta,W)=0$. The full gradient $\nabla S$ on this manifold is therefore just $\nabla_W S(\eta,W)=-k\eta$.  Hence  finding critical points of the free energy is equivalent to solving the problem  
\begin{equation} \label{e:newprob}
                            \nabla H(W) + kT\eta=0.
\end{equation}
Thus (as observed in \cite{KK}) the variable $\eta$ does have a physical interpretation as (up to the factor $-kT$) the value of the gradient of the Hamiltonian at points of equilibrium.
On substituting $W=W(\eta)$ the equation~(\ref{e:newprob}) becomes a {\em fixed point problem} for $\eta$: it does not involve inverting the function $W$ explicitly, and solutions $\eta$ can then be substituted into $W(\eta)$ to give the required solutions in terms of the order parameter $W$.  In the particular choice of quadratic $H(W)$ as in~(\ref{e:defH}) the equation~(\ref{e:newprob}) simplifies to
\[
BW=kT\eta
\]
which defines a linear subspace in $L(E)\times L(E)$ of dimension $n=\dim L(E)$: its intersections with the graph of $W$ as in~(\ref{e:logZ}) (typically isolated points) correspond to the critical points of the free energy. 

\subsection{Reduced models}  \label{s:reduced}
If a linear subspace of $D\subset E$ is invariant under the action $\alpha$ then the above formalism applies equally well to the action $\alpha$ restricted to $D$.  However, even if $D$ is not $\alpha$-invariant the key constructions can still be carried out using orthogonal projection $\pi:E\to D$ as follows, albeit with likely loss of symmetry and possible physical significance. We associate to every linear transformation $M\in L(E)$ a linear transformation $M_D\in L(D)$ defined by
\[
M_D\,u = \pi Mu\in D
\]
for $u\in D\subset E$.  The earlier general constructions can now be repeated,
using $\alpha(g)_D\in L(D)$ and arbitrary Lagrange multiplier $\zeta\in L(D)$ 
in place of $\alpha(g)\in L(E)$ and $\eta\in L(E)$.
In particular, we construct just as before:
\begin{align*}
\text{partition function}\quad Z_D&\,:\,L(D)\to\R  \\
\text{ensemble average}  \quad W_D&= \nabla_\zeta \log Z_D : L(D)\to L(D)  \\
\text{entropy}           \quad S_D&= k\bigl(\log Z_D - \zeta\cdot W_D\bigr).
\end{align*} 
If $D$ is {\em not} $\alpha$-invariant then the association of $\alpha(g)_D\in L(D)$ to $g\in G$ is {\em not} necessarily a group action because $\pi\alpha(g)\,\pi\alpha(h)$ need not coincide with $\pi\alpha(gh)$ for $g,h\in G$. However, by restricting to a smaller group the situation may be rescued.
\msk 

Let $G_D$ denote the subgroup of $G$ that (under $\alpha$) preserves $D$, that is
\[
G_D = \{g\in G:\alpha(g)u\in D \text{\ for all\ } u\in D\}.
\]
The arguments of Section~\ref{s:sym} show that $Z_D$ (and consequently $S_D$)
is invariant under the left action of $G_D$ on $L(D)$ and also (given that the projection $\pi$ is orthogonal) under transposition $\tau$ applied to $L(D)$.
From Proposition~\ref{p:wreath} we conclude that the natural symmetry group $\wt G_D$ for the restricted partition and entropy functions on $L(D)$ is a wreath product 
\[
\wt G_D=G_D\wr Z_2\cong(G_D\times G_D)\rtimes Z_2
\]
while the symmetry of the Hamiltonian term in the free energy is merely~$G_D$.
\section{Classification programme}  \label{s:classprog}
Armed with these symmetry results we are now in a position to formulate a general programme for classifying bifurcations of critical points of the free energy function close to isotropy for a general field theory of KKLS type. 
This proceeds as follows:
\begin{enumerate}
\item[(i)]  Find a Hilbert basis for the action of the group $\wt G$ on $L(E)$ or, for a reduced model, the action of $\wt G_D$ on $L(D)$.
\item[(ii)] Expand the Taylor series at $0$ for an arbitrary 
$\wt G$-invariant smooth function on $L(E)$ (or $\wt G_D$-invariant function on $L(D)$) in terms of the appropriate Hilbert basis functions. (In the KKLS model the entropy part $-TS$ of the free energy has this form, with specific coefficients.)
\item[(iii)] Add the Hamiltonian function $H$.  This may have less symmetry, but will typically consist only of low-order terms (for example, quadratic).
\item[(iv)] Carry out a bifurcation analysis of critical points of the free energy $F=H-TS$ as a function of $W$ from $W=0$, using methods of symmetry-invariant singularity theory in order deal rigorously with higher-order terms.
\item[(v)] Substitute physically meaningful values for certain coefficients in order to describe the interactions of equilibrium states as other coefficients (for example, the temperature) are varied.
\end{enumerate}
The feasibility of this programme naturally depends on the choice of the shape space $E$ and/or its distinguished subspace $D$.  In the first instance we naturally take $E=V=Sym_0(\R^3)$ with $G=\wt{SO}(3)$. However, in order to carry out the classification programme we are faced with calculations involving at least $25-6=19$ variables (here $\dim L(V)=25$ and $\dim(\wt G)=6$) as there will be at least that many functions in the Hilbert basis.  In fact it turns out that a reasonable choice of Hilbert basis has 1,453,926,067 elements, underlining the pertinent observation in~\cite{LN} that this particular approach is \lq wholly impractical'. Details of the Molien series and discussion of low-order invariants for this representation can, however, be found in~\cite{CLT}.  
\msk

All is not lost nevertheless, because if we restrict attention to bifurcation from the isotropic state then, as we now show, the lower symmetry of the  Hamiltonian as indicated in Section~\ref{s:Ham} leads to a lower-dimensional reduction for bifurcation analysis at the origin.
\subsection{Bifurcation from the origin} \label{ss:origin}
First we call upon a classical result. (For the case $G=SO(n), E=\R^n$ see~\cite[10.2]{KP}.) 
\begin{prop} \label{p:quadinv} Suppose $G$ acts absolutely irreducibly on $E$. Then the quadratic invariants for the left action of $G$ on $L(E)$ are 
generated by the scalar products $Mx\cdot My $ for $M\in L(E)$ and $x,y\in E$.
\end{prop}
Equivalently, when a basis is chosen for $E$ the quadratic invariants are generated by scalar products of columns of the matrix $M$. 
\proof
The gradient of an invariant quadratic function is an equivariant linear map.
With the linear space $L(E)$ regarded as $E^n, n=\dim E$,
the left action of $G$ on $L(E)$ can be seen as the diagonal action of $G$ on $E^n$ where $n=\dim E$. The matrix for an equivariant linear map $E^n\to E^n$ decomposes into $n^2$ blocks of size $n\times n$. By absolute irreducibility each of these blocks must be a scalar matrix, from which the result follows directly. \endproof
\msk

Now suppose that $H:L(E)\times L(E)\to\R$ is a symmetric and $G$-invariant bilinear form on $L(E)$ with associated quadratic function $h(M)=H(M,M)$ for $M=(m_{ij})\in L(E)$. In view of Proposition~\ref{p:quadinv} there exist scalars $a_{ij}\in\R$ with $a_{ij}=a_{ji}$ for $1\le i,j\le n$ such that
\begin{align}
h(M)&=\sum_{i,j=1}^n a_{ij}\sum_{k=1}^m m_{ki}m_{kj}  \\
    &=\sum_{k=1}^n\sum_{i,j=1}^m m_{ki}a_{ij}m_{kj}  \\
    &=\sum_{k=1}^n m_{k*}Am_{k*}^T \label{e:hker}
\end{align}
where $m_{k*}$ is the $k^{{\mathrm th}}$ row of the matrix $M$ and where $A=(a_{ij})$.
\begin{cor} \label{c:dimn}
Suppose $\dim\ker A=p$. Then $\dim\ker H=np$ with
\begin{equation} \label{e:kereq}
\ker H = E\otimes \ker A.
\end{equation}
\end{cor}
\proof  From the block diagonal form~(\ref{e:hker}) we see that $M\in\ker h$ precisely when every row of $M$ belongs to $\ker A$, which is another way of expressing~(\ref{e:kereq}). \endproof
\msk

Consider now a $G$-invariant function $f:L(E)\to\R$ with a quadratic relative minimum at the origin, so the origin represents a locally stable equilibrium for $-\nabla f$.  As $f$ varies with respect to a single parameter $\mu$ this equilibrium may become unstable, and will do so along an unstable manifold with tangent space at the origin equal to $\ker H$.
\begin{prop}   \label{p:unstabledim}
Generically the origin becomes unstable along a $G$-invariant manifold of dimension equal to the dimension of $E$. 
\end{prop} 
\proof  This is a consequence of Corollary~\ref{c:dimn} and the fact that generically $A$ becomes unstable in dimension~$1$ because the set of symmetric matrices $A$ having rank $m-1$ is a smooth submanifold of codimension~$1$ in the space of all such matrices (see e.g. \cite[\S2.2]{AGV},\cite[Prop. 5.3]{GG}: the adaptation to symmetric matrices is elementary). The $G$-invariance of the unstable manifold is automatic, as its uniqueness implies it is unchanged by averaging over the $G$-action. \endproof 
\msk

In particular if $G=SO(3)$ with the conjugacy action on $V=Sym_0(\R^3)$ 
then a function on $L(V)\cong\R^{25}$ invariant under the left action of $G$ loses stability at the origin generically along a smooth submanifold of dimension~5. Moreover, as we show in Appendix~A (Lemma~\ref{l:split}), critical point analysis close to the origin then reduces to that of a smooth $G$-invariant function on $\R^5$.  
We therefore conclude the following result.
\begin{prop} \label{p:simpler}
Although a complete bifurcation study for the KKLS model based on the shape space $V=Sym_0(\R^3)$ requires dealing with an $SO(3)$-invariant Hamiltonian and an $SO(3)\wr Z_2$-invariant entropy function on $L(V)\cong\R^{25}$, as far as local bifurcation from isotropy is concerned it suffices to analyse the local bifurcation from the origin of critical points of a function on $V$ invariant with respect to the usual conjugacy action of $SO(3)$.
\end{prop}
Thus the unfeasible analysis of a $\wt G$-invariant free energy function of~25 variables reduces to the more feasible analysis of a $G$-invariant function of~5 variables when studying bifurcation from isotropy.
\section{Reduced models}  \label{s:reducedmodels}
As Sections~\ref{s:reduced} and~\ref{ss:origin} have shown, there are two routes from the full and intractable $25$-dimensional model with $SO(3)\wr Z_2$ acting by left/right multiplication and transposition on $L(V)$) to the rather more amenable $2$-dimensional model with $D_3$ acting in the natural way on $D\cong\R^2$, namely: 
\begin{itemize}
\item[(1)] applying the projection $\pi:V\to D$ to reduce to the $4$-dimensional space $L(D)$ and then using Proposition~\ref{p:unstabledim} to reduce to $D$ when studying bifurcation from isotropy, or 
\item[(2)] first applying Proposition~\ref{p:unstabledim} to reduce to $V$ when study bifurcation from isotropy, and then using the observation at the end of Section~\ref{s:molintro} to reduce to $D$.
\end{itemize}

Despite the absence of a firm justification for the first approach (it is not obvious that the ensemble average over the orbit of a diagonal matrix will again be diagonal, while the projection $\pi:V\to D$ does not have a frame-independent interpretation) it is~(1) that we shall follow since it is 
widely employed in the literature on biaxiality~\cite{SVD},\cite{MV},\cite{KK},\cite{LN} and enables us to compare our results with others that use four order parameters. It is worth pointing out that the two methods of reduction are not identical, since in~(1) the plane $\R^2$ corresponds to $D\oplus\ker A_D$ where $A_D$ is the matrix describing the Hessian of the free energy function on $L(D)$ in terms of basic quadratic invariants (see Section~\ref{ss:origin}), while in~(2) it corresponds to $D\oplus\ker A\subset V\oplus\ker A$ with $A=A_V$.  There is no \emph{a priori} reason why $\ker A_D$ and $\ker A$ should be conveniently related. 
\msk

From the standpoint of Landau - de Gennes theory, which interprets the free energy as an expansion about the origin of an arbitrary real-valued function with appropriate symmetry, the method of reduction is immaterial, but in order to relate the present methods to the KKLS formalism it is important that a choice be made. The systematic study of the resulting bifurcation scenario, emphasising the key role of symmetry, occupies the remainder of this paper.
\subsection{Four order parameters: geometry and algebra of the group action} 
Every orbit of $G=SO(3)$ in $V$ intersects $D$ since every symmetric matrix can be diagonalised by an orthogonal matrix.  This intersection consists typically of six points, corresponding to
diagonal matrices whose entries are the eigenvalues of the matrices on
that orbit; if two nonzero eigenvalues coincide (uniaxiality) the intersection
consists of only three points.  The subgroup $D_3$ of $SO(3)$ that
corresponds to {\em permuting the three axes} preserves $D$ under
conjugation and permutes the six (or three) points of
intersection; since the conjugacy action $\alpha$ is orthogonal on $V$ the
configuration in $D$ has equilateral triangular symmetry.  Such a triangle exhibiting eigenvalue relationships is a familiar object in the liquid crystal literature (e.g.~\cite{RA},\cite{ZP}), although where coordinates are chosen so that the $D_3$-action is {\em not} orthogonal (as in~\cite{LN},\cite{MR}, for example) then the triangle is not equilateral and the description of the group action becomes unnecessarily cumbersome.
\msk

The symmetries of the partition function $Z_D$ (and hence the entropy $S_D$) that remain in the reduced model are the shadows of the
symmetries (i),(ii) in Proposition \ref{p:symZ}, namely:
\begin{enumerate}
\item[(i)$_D$] The left action of $D_3$ on $L(D)\,$;
\item[(ii)$_D$] transposition in $L(D)$.
\end{enumerate}
We choose an explicit orthonormal basis ${\mathcal E}=\{Q,B\}$ for the
$2$-dimensional space $D$ where
\begin{equation}
Q=\frac1{\sqrt6} \begin{pmatrix}
                                 -1 & 0 & 0 \\ 
                                 0 & -1 & 0 \\ 
                                 0 & 0 & 2
                 \end{pmatrix}\,,\quad
B=\frac1{\sqrt2} \begin{pmatrix}
                                  1 & 0 & 0 \\ 
                                  0 & -1 & 0 \\ 
                                  0 & 0 & 0
                 \end{pmatrix}.
\end{equation}
With respect to this basis the elements of $L(D)$ are represented by
$2\times2$ matrices which we denote by
\begin{equation}  \label{e:matrixdef} 
\begin{pmatrix}
      s & d \\ 
      p & c  
\end{pmatrix}.
\end{equation}
In terms of other notation commonly used in the literature we
find for example
\begin{align*}
(s,\sqrt3p,\sqrt3d,3c)&=(S,P,D,C)\qquad\qquad\enspace\text{(cf. \cite{DT},\cite{LN})} \\
     &=(S_{00}^2,S_{20}^2,S_{02}^2,S_{22}^2) \qquad\text{(cf. \cite{BL},\cite{DG})} \\
\text{or}\qquad (s,p,\sqrt3d,c)&=(S,\sqrt3T,S',T')\quad\qquad\text{(cf. \cite{SVD})} 
\end{align*}
(see also~\cite{RO} for a table of terminology used elsewhere), while in terms of 
Wigner matrices $R_{ij}$ as in~\cite{LN} we have
\[
(s,p,d,c)=(\langle R_{00}\rangle,\sqrt{2}\langle R_{20}\rangle,\sqrt{2}\langle R_{02}\rangle,2\langle R_{22}\rangle).
\]
These four coefficients are in our setting the {\em order parameters} for the reduced problem.
\medskip
  
The symmetries of the free energy function $F=F(s,p,d,c)$ are most conveniently expressed using complex notation.  We write 
\begin{equation}
z=s+ip\,,\ w=d+ic
\end{equation}
and take $D_3$ to be generated by a rotation $\rho$ through $\frac{2\pi}3$ and a reflection
$\kappa$ in the real axis. Then $D_3$ acts on
$L(D)\cong\R^4\cong\C\times\C$ by the usual action as symmetries of an
equilateral triangle in each factor $\C\cong\R^2$, that is $D_3$ is
generated by $\{\rho,\kappa\}$ where
\begin{align} 
  \rho &: (z,w)\mapsto e^{\frac{2\pi i}3}(z,w) \label{e:rot} \\
\kappa &: (z,w)\mapsto (\bar z,\bar w)         \label{e:bar}
\end{align}
where the bar denotes complex conjugate.  Also, the transposition operator
acting on the matrix~(\ref{e:matrixdef}) clearly acts in this setting by
\begin{equation}  \label{e:tau}
\tau: p \leftrightarrow  d\,.
\end{equation}
For future reference we note here the following easily-verified relations:
\begin{align}
\tau\circ\kappa\circ\tau&:(z,w)\mapsto(z,-w)  \label{e:rel1}  \\
\tau\circ i \circ\tau&:(z,w)\mapsto(-w,z)      \label{e:rel2}
\end{align}
where here $i$ denotes multiplication by the complex number $i$.
\msk

The subgroup\footnote{The group $\wt D_3$ is denoted by [70,40] in the GAP SmallGroups library~\cite{GAP}.}
$\wt D_3=D_3\wr Z_2$ of $SO(4)$ generated by $\{\rho,\kappa,\tau\}$ is of order $72$ and contains a subgroup of index $2$ isomorphic to $D_3\times D_3$ whose two factors act on rows and columns respectively of matrices in $L(D)$.  
In~[Section VII]\cite{LN} this subgroup is stated to be the full symmetry group, although in~[Section VI]\cite{LN} the presence of the further $\tau$-symmetry in the entropy terms is also noted.
\msk  

The first step in the bifurcation analysis of $F$
is to find a Hilbert basis for the $\wt D_3$-invariant functions on $L(D)$.   
\subsection{Invariant functions for the group $\wt D_3$}  \label{s:invs}
With the help of Maple we find that an explicit expression 
for the Molien series for the given action of
$\wt D_3$ on $\R^4$ is
\begin{equation}  \label{e:molien1}
\P_{\wt D_3}(t)=\frac{1+t^5}{(1 - t^2)(1 - t^3)(1 - t^4)(1 - t^6)}.
\end{equation}
This indicates that there is a Hilbert basis with just
one invariant function of each homogeneous degree $2,3,4,6$ and that
there is a secondary invariant of degree $5$ which is not expressible
as a polynomial combination of the others but is nevertheless
algebraically dependent on them.
\medskip

As an instructive comparison, be find also that the Molien function
for the subgroup $D_3\times D_3$ is
\begin{equation}   \label{e:molien2}
\P_{D_3\times D_3}(t)=\frac{(1+t^5)(1+t^6)}{(1 - t^2)(1 - t^3)(1 - t^4)(1 - t^6)}
\end{equation}
which informs us that a Hilbert basis for $D_3\times D_3$ contains a
secondary invariant of homogeneous degree $6$ that is {\em not} an invariant
for $\wt D_3$: thus it is only at degree $6$ that the transposition $\tau$
plays an explicit role in the symmetries of the free energy. 
\msk

The Molien functions~(\ref{e:molien1}),(\ref{e:molien2}) tell us how many invariants to look for at
each homogeneous degree.  To find an explicit basis we use a standard technique
exploiting the complex structure (see~\cite{GS}). Every real
homogeneous polynomial of degree $a$ in $s,p,d,c$ can be expressed as
a linear combination of the real and imaginary parts $\Re h$ and $\Im h$ of
monomials 
\[
h(z,w)=z^k\bar z^lw^m\bar w^n
\]
where $k,l,m,n$ are positive integers with $k+l+m+n=a$. The
$\kappa$-invariance dictates that we take only the real part $\Re h$
and the $\rho$-invariance implies that $k-l+m-n$ is a multiple of $3$.  
Moreover, it follows from the symmetry~(\ref{e:rel1}), noted also in~\cite{MR}, that a $\tau$-invariant polynomial must be of {\em even} degree in the $w,\bar w$
terms. Furthermore, if $f(z,w)=\Re h(z,w)$ is $\tau$-invariant it must
satisfy in particular 
\begin{align} 
f(s,p) &= f(s+ip,0)\quad \text{and} \label{e:tau1} \\
f(id,ic) &=  f(0,d+ic)              \label{e:tau2}
\end{align}
for real $s,p,d,c$.  Using these facts a Hilbert basis for 
the $\wt D_3$-invariant functions on $\R^4$ is straightforward to construct
as follows.
\msk

Clearly there are no nonzero linear invariants, and~(\ref{e:tau1}) shows that
the only quadratic invariants are scalar multiples of $f_2:=|z|^2+|w|^2$.
Likewise the only cubic invariants are scalar multiples of $f_3:= \Re(z^3-3zw^2)$.
\ssk

At degree $4$ some elementary linear algebra shows that the only
linear combinations of $|z|^4,|z|^2|w|^2,|w|^4$ and $z^2\bar w^2+\bar z^2w^2$ that satisfy~(\ref{e:tau1})
are linear combinations of $f_2^2$ and $f_4:=(z\bar w-w\bar z)^2\,$; note that $z\bar w-w\bar z$ is pure 
imaginary so $f_4$ is real.
\ssk

At degree $5$ a similar approach shows that in order to satisfy~(\ref{e:tau1})
an arbitrary linear combination of appropriate monomials
in $z,w$ and their conjugates must take the form of a linear combination of
$f_2f_3$ and a polynomial of the form $f_5:=\Im(z\bar w-w\bar z)\Im(qw^3-rwz^2)$ for real coefficients $q,r$.
The fact that~(\ref{e:tau2}) must also be satisfied then shows that $r=3q$.  Since for {\em any} complex
numbers $z,w$ we have $\Im z\Im w=\Re z\Re w-\Re(zw)$ it follows that we can also express $f_5$ as
$f_5=\Re\bigl((w\bar z-\bar wz)(w^3-3wz^2)\bigr)$.
\ssk  

Finally, at degree 6, a similar method exploiting~(\ref{e:tau1}) and~(\ref{e:tau2}) shows (after some 
linear algebra) that a $\wt D_3$-invariant function  must be a linear combination of
$f_2^3,f_3^2,f_2f_4$ and 
\[
f_6:=\bigl(\Im(w^3-3wz^2)\bigr)^2=\bigl(\Re(w^3-3wz^2)\bigr)^2-\Re\bigl((w^3-3wz^2)^2\bigr).
\]

To summarise, a Hilbert (algebraic) basis for the $\wt D_3$-invariant polynomials consists of the following 
polynomial functions of $(s,p,d,c)\in\R^4$:
\begin{align*}
f_2\qquad & \qquad |z|^2+|w|^2  \\
f_3\qquad & \qquad \Re(z^3-3zw^2)    \\
f_4\qquad & \qquad (z\bar w-\bar zw)^2  \\
f_5\qquad & \qquad \Im(z\bar w-\bar zw)\Im(w^3-3wz^2)  \\
f_6\qquad & \qquad \bigl(\Im(w^3-3wz^2)\bigr)^2
\end{align*}
where $(z,w)=(s+ip,d+ic)$.  Here we observe the syzygy $f_5^2+f_4f_6=0$ 
showing that, as predicted by
the Molien function~(\ref{e:molien1}), the degree-$5$ invariant is linearly but not 
{\em algebraically} 
independent of those of degrees $2,3,4$ and $6$.  A {\em linear} basis for the $\wt D_3$-invariant functions up to degree $6$ is given by $\{f_2,\ldots,f_6\}$ together with $f_2^2$ (degree $4$), $f_2f_3$ (degree $5$) and $\{f_2^3,f_3^2,f_2f_4\}$ (degree $6$).
\msk

The $\tau$-invariance \eqref{e:tau} of the free energy implies that the terms of each homogeneous degree $\ge 3$ in the expansion given in~\cite{LN} must remain unaffected by interchanging the order parameters $P,D$, which is indeed found to be the case up to and including degree~6~\cite{ST}. In contrast, a Landau - de~Gennes expansion based only on the assumptions of invariance under left and right actions of $SO(3)$ would necessarily include a further invariant of degree~$6$.  Indeed, we find that the polynomial
\[
\hat f_6:= \Re\bigl((z^2 + w^2)^3\bigr)
\]
is $D_3\times D_3$-invariant but not $\tau$-invariant.  However, its square is $\tau$-invariant and so can be expressed as a polynomial function of $f_2,\ldots,f_6$ above. 
\section{Expansion of the free energy function}  \label{s:expandf}
We now consider a Taylor expansion about the origin of an arbitrary
function of the order parameters $(s,p,d,c)\in L(D)$ exhibiting the relevant
symmetry, that is with quadratic terms invariant under the left action of $D_3$ and higher order terms invariant under the full $\wt D_3$-action.  Although in the spirit of a general Landau-de~Gennes expansion, familiar in liquid crystal theory, the use to which we put the expansion has the following particular features:
\begin{itemize}
\item inclusion of the transposition symmetry $\tau$ (relevant at degree~$6$);
\item use of singularity theory methods to take rigorous account of high order terms;
\item explicit reduction to a $D_3$-invariant function on $\R^2$ for analysis of bifurcation from isotropy;
\item explicit identification of the resulting coefficients in terms of those of the KKLS formalism;
\item identification of critical point bifurcation geometry in terms of standard models.
\end{itemize}
In the bifurcation analysis we do not, in the present context, keep systematic track of {\em global} minima which represent stable liquid crystal phases.  For this, further tools from singularity theory are required.  However, we draw attention here to the thorough analysis in~\cite{AL} that uses efficient algebraic and numerical techniques to display many important features of the phase diagram.
\msk

With quadratic terms invariant under  $\{\rho,\kappa\}$ and those of degree at least three invariant under $\tau$ also, the free energy function may be written as 
\begin{align}  \label{e:fdef}
f(s,p,d,c)=f(z,w)&= \alpha |z|^2 + \beta |w|^2 + 2\gamma Re(z\bar w) \notag \\
          &\quad + a_3f_3 + a_4f_4 + b_4f_2^2 + a_5f_5 + b_5f_2f_3 \notag \\
          &\qquad + a_6f_6 + b_6f_2^3 + c_6f_3^2 + d_6f_2f_4 + O(7)
\end{align}
where the coefficients $\alpha,a_i$ {\em etc.} are all real and where the
functions $\{f_1,\ldots,f_6\}$ are those of the Hilbert basis specified in Section~\ref{s:invs}. 
\msk

\subsection{Comparison with the KKLS expansion}
The entropy contribution $-kTS$ to the free energy given in~\cite{KK},\cite{LN} takes this form with $\alpha=\beta=\frac52 kT$ and $\gamma=0$,
the nonzero $\gamma$-term in the free energy \cite[eq.(62)]{LN} arising only from the Hamiltonian.  The coefficients of the terms with degree from $3$ to~$6$ (all of which come from the entropy) are of the form $a_i=-kTa_i'$ etc. as given by the following list.  Those of degree $3$ and~$4$ appear in~\cite{LN}, while those of degree $5$ and~$6$ are unpublished~\cite{ST}.
\begin{align}
a_3' &= \tfrac{25}{21} \sim 1{\cdot}19  \label{e:a3prime}\\
(a_4',b_4') &= \tfrac{25}{16.49}(-5,68) \sim(-0{\cdot}16,2{\cdot}17)  \\
(a_5',b_5') &=C_5(125,498)\sim(-1{\cdot}73,-6{\cdot}87)  \\
(a_6',b_6',c_6',d_6') &=C_6(-419600,3099312,716640,612405) \\
  &\qquad \sim(-0{\cdot}53,3{\cdot}92,0{\cdot}91,0{\cdot}77) \label{e:a6prime}
\end{align}
where $C_5=-840\,C_6$ with $C_6= \tfrac{5^3}{2^5.3^2.7^4.11.13}
                             =\tfrac{125}{98882784}$.
\subsection{Stability of the isotropic state}   \label{ss:quadstability}
\begin{figure}[!ht]  
\begin{center}
 \scalebox{.60}{\includegraphics{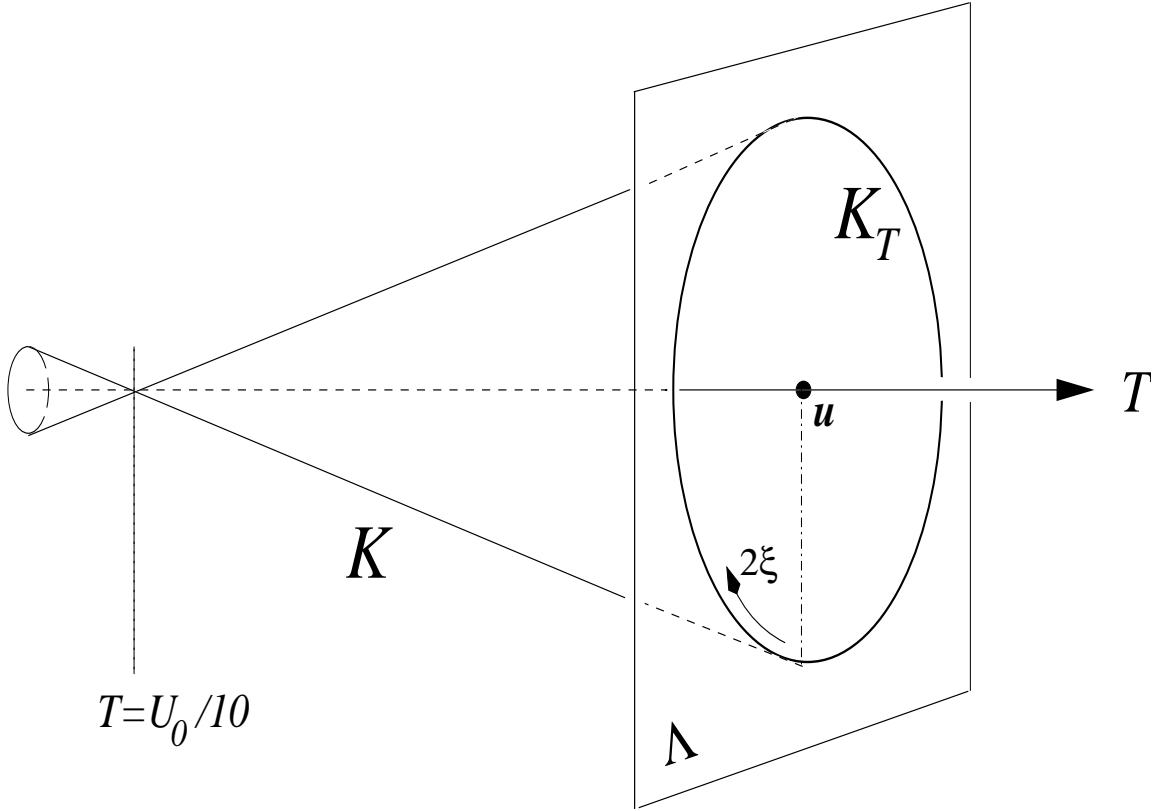}}
\end{center}  
\caption{The stability cone $K$ and its circle $K_T$ of intersection with the $\bl$-plane $\Lambda$ at constant $T$.  
For $T>T_0=U_0/10$ the isotropic state loses stability in two directions as $(T,\bl)$ crosses $K$ from inside to outside.}   
\label{fi:conepicture}
\end{figure}

The equilibrium at the origin (isotropic state) loses local stability when the Hessian matrix $Hf(0)$ of $f$ at the origin drops rank. 
This occurs precisely when
\begin{equation}  \label{e:locus}
\alpha\beta-\gamma^2=0
\end{equation}
which is equivalent to the statement that the matrix identity
\begin{equation}  \label{e:cmatrix}
\begin{pmatrix}
   \alpha  &  \gamma  \\
   \gamma   &  \beta
\end{pmatrix}
=2\mu
\begin{pmatrix}
   \cos^2\xi  &  \cos\xi\sin\xi  \\
   \cos\xi\sin\xi   &  \sin^2\xi
\end{pmatrix}
\end{equation}
holds for some $\mu,\xi\in\R$.  
The symmetries in the problem are now further exploited, following the change of coordinates proposed in~\cite{TS}, by replacing the coefficients $\alpha,\beta,\gamma$ by the parameters $\bl=(\lam_1,\lam_2,\lam_3)$
with $\lambda_1+\lambda_2+\lambda_3=1$ given by 
\begin{align*}
\alpha&=\frac52 T-\frac12
U_0\bigl(\frac14\lam_1+\frac14\lam_2+\lam_3\bigr)\\
\beta &=\frac52 T-\frac38 U_0(\lam_1+\lam_2) \\
\gamma &= -\frac{\sqrt3}8 U_0(\lam_1-\lam_2)
\end{align*}
from which we see in particular
\begin{equation}  \label{e:muterm}
2\mu=\alpha+\beta=\tfrac12(10T-U_0).
\end{equation}
In $(T,\bl)$ coordinates the locus~(\ref{e:locus}) is a circular cone $K$ in $\R^3$ with axis parallel to the $T\,$-axis and vertex $(T_0,\bu)$
with $T_0=U_0/10$ and $\bu=\tfrac13(1,1,1)$.  For fixed $T\ne T_0$ the cone section in the $\bl$-plane $\Lambda$ is a circle $K_T$ with centre $\bu$ and radius $R_T$ that increases linearly with $T$: 
explicitly we may write $\bl$ in terms of $\xi$ as
\begin{equation}  \label{e:conepolar}
(\lam_1,\lam_2,\lam_3)=\tfrac13(1,1,1)       +R_T\sqrt{\tfrac23}\bigl(\cos(2\xi+\tfrac\pi3),\cos(2\xi-\tfrac\pi3),-\cos2\xi\bigr)
\end{equation}
and we find
 $R_T=\sqrt{\frac23}\bigl(\frac{10T}{U_0}-1\bigr)$.
See Figure~\ref{fi:conepicture}.  
\msk

For parameter values $(T,\bl)$ inside the cone with $T>T_0$ the origin (isotropic state) in $\R^4$ is a nondegenerate minimum (stable), while outside the cone the origin is a $2,-2$ saddle point with the two directions of instability depending on the direction of $\bl$.  Inside the cone with $T<T_0$ the isotropic state is unstable in all four directions.  
\msk

\subsection{Reduction of the free energy at bifurcation}
We now study more closely the bifurcations of critical points of the free
energy function $f=f(s,p,d,c)$ as the parameter values $(T,\bl)$ cross the stability cone~$K$. At a point $(T,\bl)$ where $\bl\in K_T$ we find  
\[
\ker Hf(0) = {\mathrm span}\,\left\{
             \begin{pmatrix}
                     -\sin\xi \\ 0 \\ \cos\xi \\ 0
             \end{pmatrix}, 
             \begin{pmatrix}
                     0 \\ -\sin\xi \\ 0 \\ \cos \xi 
             \end{pmatrix}
                              \right\}.
\]
After the linear coordinate change (note the ordering of variables)
\begin{equation}  \label{e:coordmatrix}
\begin{pmatrix}
 s \\ d \\ p  \\ c
\end{pmatrix}
=
\begin{pmatrix}
 \sin\xi & \cos\xi & 0 & 0 \\
-\cos\xi & \sin\xi & 0 & 0 \\
 0 & 0 & \sin\xi & \cos\xi \\
 0 & 0 & -\cos\xi & \sin\xi
\end{pmatrix}
\begin{pmatrix}
 x \\ y \\ u \\ v 
\end{pmatrix}
\end{equation}
the quadratic terms of $f$ reduce to 
\[
2\mu(y^2+v^2). 
\]
Next, from the Splitting Lemma
or Reduction Lemma~\cite{CH},\cite{G},\cite{GM},\cite{PS} we
know that there exists a local diffeomorphism (smooth invertible
coordinate transformation with smooth inverse) in a neighbourhood of the
origin in $\R^4$, and which is the identity to first order, that converts $f$
into a function of the form
\begin{equation}  \label{e:reduce}
\wt f(x,y,u,v)=2\mu(y^2+v^2) + q(x,u)
\end{equation}
where $q$ has no linear or quadratic terms; that is, the coordinate transformation removes all higher order terms that involve $y,v$.  The full local bifurcation behaviour of $f$ (as the parameters $\bl$ and $T$ are varied) is then captured by that of the {\em residual function} $q$ of the two coordinate variables $x,u$ in $\ker Hf(0)$. 
\msk

In our setting it is important for $q$ to inherit as much symmetry as possible from the original function $f$. From Section~\ref{ss:origin} we expect $q$ to be $D_3$-invariant, and to prove this we need the symmetry-preserving version of the Splitting Lemma. In Appendix~A we give a proof of the result in a general setting, while stating below the form that we require here.
\begin{prop}   \label{p:split}
The coordinate transformation in $\R^2$ can be chosen to be $D_3$-equivariant, after which the residual function $q:\R^2\to\R$ is $D_3$-invariant.   \endproof
\end{prop}   
\rem To obtain the residual function $q(x,u)$ it is not sufficient merely to 
set $y=v=0$ in the expression for $f$ after making the coordinate change~(\ref{e:coordmatrix}). For example, this would lead to the erroneous conclusion that a function such as $y^2+v^2+4yx^2+x^4+u^4$ has a minimum at the origin.
\subsubsection{A note on uniqueness}  \label{ss:unique}
It is not {\em a priori} obvious that 
the reduced function $q(x,u)$ in~(\ref{e:reduce}) is itself unique up to
$D_3$-equivariant coordinate change in $\ker Hf(0)$: possibly two different methods of reduction could lead to two inequivalent functions $q_1,q_2$. A proof that in the general setting (no symmetry constraint) this cannot occur is given 
in~\cite[Ch.4]{CH}.  We give a different proof in the symmetry-preserving setting also in Appendix A.
\subsection{The residual free energy function}
To reflect the circular symmetry inherent in the use of \lq polar' coordinates~(\ref{e:conepolar})
in the $\bl$-plane $\Lambda$ it is convenient to introduce new complex coordinates\footnote{Of course these are not the same as the partition function $Z$ and the ensemble average $W$ of Section~\ref{s:entropydefs}. We trust no confusion will arise.}
\begin{align*}
Z&= x + iu\\
W&= y + iv
\end{align*}
related to the original complex order-parameter coordinates $(z,w)$ by 
\begin{equation} \label{e:newzw1}
\tau(z,w) = -ie^{i\xi}\tau(Z,W)
\end{equation}
or equivalently
\begin{equation}  \label{e:newzw2}
\begin{pmatrix}
   z  \\ w
\end{pmatrix}
=
\begin{pmatrix}
  \sin\xi &  \cos\xi  \\
  -\cos\xi & \sin\xi
\end{pmatrix}
\begin{pmatrix}
   Z  \\  W
\end{pmatrix}.
\end{equation}
Our procedure will be to examine in turn the terms of the Taylor
series of $f$ of increasing degree, using suitable coordinate changes to
systematically eliminate all terms involving $y,v$ apart from the
initial quadratic term 
\[
2\mu(y^2+v^2)=2\mu|W|^2
\] 
in order to obtain to any required order an expression for the residual function $q(x,u)$ 
in terms of $Z$ only.
\msk

An obvious question arises: how many terms are enough?  In
other words, at what degree will the truncated Taylor series capture
all the local bifurcation behaviour?  There are mathematical tools to
answer this question, which we now briefly review.
\subsection{Determinacy: dealing with higher order terms}   \label{ss:deter}     
Let $h$ be a an arbitrary smooth  ($C^\infty$) function of $n$ real variables $x_1,\ldots,x_n$, defined on some \nhd of the origin in $\R^n$.  The
Taylor series of $h$ at the origin truncated
at degree $k$ is a polynomial in $x_1,\ldots,x_n$ called the $k${\it-jet}
of $h$ at $0\in\R^n$ denoted by $j_kh$.  We are concerned with 
how accurately $j_kh$ captures the qualitative behaviour of $h$ close to the
origin.  Specifically, if there is a local smoothly invertible change
of coordinates ({\em local diffeomorphism}) $\phi$ around the origin in
$\R^n$ that converts $h$ into $j_kh$, that is such that
\begin{equation}  \label{e:kdet}
h(\phi(x))= j_kh(x)
\end{equation}
for all $x=(x_1,\ldots,x_n)$ in some \nhd of the origin,
then $h$ and $j_kh$ are $\RR${\it-equivalent} at the origin, and the function $h$ is said to be $k$-$\RR${\it-determined} at
the origin: $\RR$ here refers to the fact that the symbol $\phi$
appears to the {\em right} of $h$ in the expression~(\ref{e:kdet}). If
we also allow a similar coordinate change around $h(0)\in\R$, which would appear on the {\em left} of $h$ in~(\ref{e:kdet})
and does not affect its critical point behaviour, then $h$ is 
$k$-$\RR\LL${\it-determined} (also called $k$-$\A${\it-determined}) at the origin. 

Finally, to complicate the notation a little further, in a context where $h$ and therefore $j_kh$ are invariant with respect to a (linear) action of a group $G$ on $\R^n$ (and also on $\R$, although this is often taken to be the {\em trivial} action that has no effect), and where the local diffeomorphism $\phi$ 
on $\R^n$ can be chosen to be $G$-equivariant, then $h$ and $j_kh$ are
$G$-$\RR$-(or $\A$-){\it equivalent} and $h$ is $G$-$k$-$\RR$-(or $\A$-){\it determined}.
\msk

There are algebraic criteria for deciding whether a
given $G$-invariant function $h$ is $G$-$k$-$\RR$- or
$G$-$k$-$\A${\it-determined}. As illustration 
we apply these now to the key example, central to the study of phase transitions in liquid crystals, of the natural action $\del$ of $D_3$ on $\R^2$.
Here coordinate changes in $\R$ play no significant role, so we restrict attention to  $G$-$k$-$\RR$-determinacy which for ease of notation we denote simply by $k$-{\em determinacy}.  The details of the algebra are left to Appendix B.
\paragraph{example} 
Let $n=2$ with coordinates $(x,u)\in\R^2$, and with the action $\del$ of $D_3$ 
on $\R^2$ generated by rotation through $\tfrac{2\pi}3$ about the origin and
reflection in the $x\,$-axis.  A Hilbert basis for the $D_3$-invariant functions on $\R^2$ is  given by $\{X,Y\}$ ({\em cf.} Section~\ref{s:molintro}) where
\begin{align}
X(x,u)&=x^2+u^2   \label{e:h2}\\
Y(x,u)&=x^3-3xu^2.  \label{e:h3}
\end{align}
Thus every smooth $D_3$-invariant function $h(x,u)$ vanishing at the origin has a Taylor expansion at the origin of the form 
\begin{align} \label{e:norform}
h&=e_2X + e_3Y + e_4X^2 + e_5XY + e_6X^3 + d_6Y^2 \qquad\notag \\
 &\qquad\quad + e_7X^2Y + e_8X^4 + d_8XY^2 + O(9).
\end{align}

We now consider systematically at what point it is appropriate to truncate the expansion while preserving the behaviour of $h$ up to a $D_3$-invariant local diffeomorphism close to the origin in $\R^2$.
\begin{prop}  \label{p:determinacy}
\hfill
\begin{enumerate}
  \item  If $e_2\ne0$ then $h$ is $2$-determined. 
  \item  If $e_2=0$ and $e_3\ne0$ then $h$ is $3$-determined. 
  \item  If $e_2=e_3=0$ then $h$ is in general {\em not} $4$-determined.
  \item  If $e_2=e_3=e_4=0$ then $h$ is in general {\em not} $5$-determined. 
  \item  If $e_2=\cdots=e_5=0$ then $h$ is in general {\em not} $6$-determined.
However, if $e_6d_6(e_6+d_6)\ne0$ then $h$ is $G$-$\RR$-equivalent to $j_6h+e_8X^4$.
 \item  If $e_2=\cdots=e_6=0$ then $h$ is in general {\em not} $7$-determined.
 \item  If $e_2=\cdots=e_7=0$ and $e_8d_8(e_8+d_8)\ne0$ then $h$ is $8$-determined.
\end{enumerate}
\end{prop}
\proof  See Appendix B.  \endproof  
\msk

We now apply these ideas and results to the free energy function $f$ in~(\ref{e:fdef}) or, more specifically, to the Taylor series of the residual free energy function $q(x,u)$ in~(\ref{e:reduce}). 
It suffices to restrict attention to $q$ since it
follows from the Splitting Lemma (and the fact that a local diffeomorphism 
cannot reduce the order of the lowest order terms in a function) that the
function $f_S$ in~(\ref{e:reduce}) is $k$-determined if and only if the same holds for $q$. Cases~1 and~2 of Proposition~\ref{p:determinacy} show that truncation of the Taylor series at degree~$2$ or~$3$ is justified when 
the respective coefficients are nonzero, the former being merely an instance of the Equivariant Morse Lemma~\cite{AR}, while Cases~3 and~4 show this is not the case at degree~$4$ or at degree~$5$ when lower-degree coefficients vanish. In Case~5 it turns out that the (unique) degree~7 term $X^2Y$ is dispensable and that all terms of degree $\ge 9$ can be removed, but the term $X^4$ of degree $8$ cannot. Case~6 shows that higher order terms may affect the behaviour of $X^2Y$, and, finally, Case~7 shows that apart from in degenerate cases the degree~$8$ terms determine the behaviour of the function. 
\msk

The answer to the question on where to truncate the Taylor expansion is therefore: at degree $8$. However, in order to limit the complications and make the bifurcation analysis reasonably tractable we shall focus attention on the Taylor expansion up to degree~$6$, assuming the conditions of Case~5 apply, and use separate arguments later to describe the effect of the degree~$8$ term $X^4$.  In the next Section we carry out an explicit determination of the residual function $q$, expressing its coefficients $e_2,\ldots,e_5,e_6.d_6$
in terms of those of $f$, which in turn can be used to compare directly with the results of~\cite{LN}.
%
\subsection{The $6$-jet of the residual free energy function}
In complex $(Z,W)$ coordinates the expression~(\ref{e:fdef}) for the
free energy function $f$ takes the form
\begin{equation}  \label{e:freeZW}
f_\xi(Z,W)=2\mu|W|^2+\til q(Z,W)
\end{equation}
where $j_2\til q=0$.  In this section we implement the Splitting Lemma up to degree~6 
in order to obtain an explicit nonlinear change of variables, coinciding with the identity at first order, that converts $f_\xi(Z,W)$ into a function of the form \eqref{e:reduce} up to degree $6$, that is
\begin{equation}  \label{e:freeZWred}
f(Z,W)=2\mu|W|^2+q(Z) + O(7).
\end{equation}
To carry this out we use the elementary technique of {\it completing the
  square}.
\subsubsection{Completing the square}
First we need the explicit expression for $\til q(Z,W)$ in terms of the original data for the function $f(z,w)$.  The following identities are straightforward to verify:
\begin{align}  \label{e:ZWlist}
f_2(z,w) &= f_2(Z,W)  \\
f_3(z,w) &= Cf_3(W,Z) - Sf_3(Z,W)   \\
f_4(z,w) &= f_4(Z,W)  \\
f_5(z,w) &= \Im(\oz W-Z\ow)\bigl(C\Im(Z^3-3ZW^2)+S\Im(W^3-3WZ^2)\bigr)\\
f_6(z,w) &= \bigl(C\Im(Z^3-3ZW^2)+S\Im(W^3-3WZ^2)\bigr)^2
\end{align}
where $(C,S)=(\cos 3\xi,\sin 3\xi)$ and $S$ is of course not to be confused with the entropy in Section \ref{s:KKLSmodel}.
Next, we define
\begin{equation}  \label{e:wsubs}
W_0:=W+\vf  
\end{equation}
where 
\begin{equation} \label{e:phiexpand}
\vf=\vf(Z)=\vf_2(Z)+\vf_3(Z)
\end{equation}
with $\vf_j(Z)$ a polynomial in (components of) $Z$ of homogeneous degree $j$ for $j=2,3$, and then substitute for $W$ into $f_\xi(Z,W)$ to obtain a function $f_\xi^0$ of $Z$ and $W_0$:
\begin{align}
f_\xi^0(Z,W_0)&=f_\xi(Z,W_0-\vf(Z))\\
              &=2\mu|W_0|^2 + \til q_0(Z,W_0).
\end{align}
The key step is now to choose $\vf(Z)$ in such a way that the terms in $\til q_0(Z,W_0)$ that contain $W_0$ with degree $1$ all vanish up to degree~$5$.
In Table~\ref{t:Wterms} we provide a list of all terms of degree at most~$5$ in $f_\xi^0$ that involve a single $W_0$, as well as a list of terms up to degree~$6$ that do not involve $W_0$.  From this it is evident that the result can be achieved by choosing $\vf(Z)$ to satisfy
\begin{align} \label{e:phieqn}
4\mu\bar\vf &=  3a_3(C\vf^2-CZ^2-2SZ\vf)+4a_4(|Z|^2\bar\vf-\oz^2\vf) \notag \\ 
           &\qquad\quad -4b_4\bar\vf|Z|^2 -3b_5C|Z|^2Z^2+2a_5C\Re(iZ^3)i\oz \notag
\end{align}
up to degree~$3$, where we have used the fact that
\[
\Im(W_0\oz)\Im(Z^3)=\Re(iZ^3)\Re(W_0i\oz).
\]
Comparing terms of degree $2$ and of degree $3$ in $Z,\oz$ we find
\begin{align} 
4\mu\bar\vf_2 &=-3a_3CZ^2  \\
4\mu\bar\vf_3 &=-6a_3SZ\vf_2  
\end{align}
and so writing 
\begin{equation} \label{e:sigmadef}
\sigma:=-3a_3/4\mu
\end{equation}
we have
\begin{align}
\vf_2 &=C\sigma\oz^2    \label{e:phi2}  \\
\vf_3 &=2SC\sigma^2|Z|^2Z.  \label{e:phi3}
\end{align}
Observe that the coordinate transformations 
\[
Z\mapsto\oz^2,\qquad Z\mapsto |Z|^2Z
\]
are each equivariant with respect to the action of $D_3$ on $\C\cong\R^2$.

\begin{table}  [!ht] 
   \begin{center}
     \begin{tabular}{ |c|c|c| }
\hline
\rule[-0.8ex]{0ex}{3.0ex} source & terms with $W_0$ once  & terms without $W_0$  \\ \hline
\rule[-0.8ex]{0ex}{3.0ex} $|W|^2$ & $-2\Re(W_0\bar\vf)$ & $|\vf|^2$  \\ \hline
\rule[-0.8ex]{0ex}{3.0ex}  $f_3$  & $3\Re W_0(C\vf^2-CZ^2-2SZ\vf)$ & 
$-C\Re(\vf^3-3\vf Z^2)-S\Re(Z^3-3Z\vf^2)$   \\ \hline
\rule[-0.8ex]{0ex}{3.0ex}$f_4$  & $4\Re W_0(|Z|^2\bar\vf-\oz^2\vf)$ & $(Z\bar\vf-\oz\vf)^2$ \\ \hline
\rule[-0.8ex]{0ex}{3.0ex}$f_2^2$  & $-4\Re(W_0\bar\vf|Z|^2)$ & $|Z|^4+2|Z|^2|\vf|^2$  \\ \hline
\rule[-0.8ex]{0ex}{3.0ex}$f_5$  & $2C\Im(W_0\oz)\Im(Z^3)$  &  $2C\Im(\oz\vf)\Im(Z^3)$   \\ \hline
\rule[-0.8ex]{0ex}{3.0ex}$f_2f_3$ & $-3C\Re(W_0|Z|^2Z^2)$  & $-S\Re(Z^3|Z|^2)+3C\Re(|Z|^2Z^2\vf)$ \\ \hline
\rule[-0.8ex]{0ex}{3.0ex}$f_2^3$  &   *    &     $|Z|^6$  \\ \hline
\rule[-0.8ex]{0ex}{3.0ex}$f_3^2$  &   *    &   $S^2(\Re Z^3)^2$  \\ \hline
\rule[-0.8ex]{0ex}{3.0ex}$f_2f_4$ &   *    &  \\ \hline
\rule[-0.8ex]{0ex}{3.0ex}$f_6$    &   *    &   $C^2(\Im Z^3)^2$ \\ \hline
     \end{tabular}
     \caption{Terms of degree at most~$6$ in $\til q_0$ that involve a single $W_0$, and terms that do not involve $W_0$. To save space $\vf(Z)$ is not expanded, so some redundant terms are implicitly included in the table.  Terms of degree~$6$ that involve $W_0$ are not needed.}  \label{t:Wterms}
    \end{center}
\end{table}
\msk

Having constructed $\vf$ to eliminate the relevant $W_0$-terms in $\til q_0(Z,W_0)$ we finally substitute
$\vf=\vf(Z)$ into the right-hand column of Table~\ref{t:Wterms} and disregard any resulting terms of degree greater than~$6$ in $Z$. In addition to the original terms in $\til q(Z,0)$ we obtain the terms displayed in Table~\ref{t:csqterms}.
Assembling this information and using the facts that 
\begin{align*}
(\Re Z^3)^2+(\Im Z^3)^2 &=|Z|^6=X^3  \\
              \Re(Z^6) &=-X^3+2Y^2
\end{align*} 
we arrive at the following result:
\begin{prop}  \label{p:splitf}
After the change of coordinates 
\begin{equation}  \label{e:newzw3}
(Z,W) \mapsto (Z,W+\vf(Z)),
\end{equation}
which in real form is
\begin{align}
(x,u)&\mapsto(x,u)\,,\\
(y,v)&\mapsto(y,v)+\sigma  C(x^2-u^2,-2xu)+2\sigma^2SC(x^2+u^2)(x,u)
\end{align}
with $\sigma=-3a_3/4\mu$ and $(C,S)=(\cos 3\xi,\sin 3\xi)$,
the Taylor expansion of the free energy function~(\ref{e:fdef}) about the origin with $(T,\bl)\in K$ takes the form~\eqref{e:reduce} where
\begin{equation}  \label{e:nicef}
q(x,u) = e_3Y + e_4X^2 + e_5XY + e_6X^3 + d_6Y^2 + O(7)
\end{equation}
with as before
\[
(X,Y)=(x^2+u^2,x^3-3xu^2)
\]
and where the $(\xi-\tfrac\pi2)$-rotated variables $(x,y,u,v)$ are given in terms of the original order parameters $(s,p,d,c)$ by~(\ref{e:coordmatrix}).
The coefficients $e_3\dots d_6$ are related to the original coefficients $a_3\dots d_6$ in~(\ref{e:fdef}) as follows:
\begin{align}
e_3 &= -a_3S    \label{e:e3term}\\
e_4 &= 2\mu C^2\sigma^2 + 3a_3C^2\sigma + b_4 = \tfrac32a_3C^2\sigma + b_4 \label{e:e4term}  \\
e_5 &= 8\mu SC^2\sigma^3-9a_3SC^2\sigma^2-b_5S  =3a_3SC^2\sigma^2 - b_5S \label{e:e5term}  \\
e_6 &= 8\mu S^2C^2\sigma^4 + a_3(12S^2C^2\sigma^3+C^4\sigma^3) + 2b_4C^2\sigma^2 -4a_4C^2\sigma^2 \notag \\
    & \qquad\quad +2a_5C^2\sigma + 3b_5C^2\sigma + b_6 \\
   &= a_3(6S^2+C^2)C^2\sigma^3 + 2(b_4-a_4)C^2\sigma^2+(2a_5 + 3b_5)C^2\sigma +   b_6 \label{e:pterm} \\
d_6 &= -2a_3C^4\sigma^3 + 4a_4C^2\sigma^2 - 2a_5C^2\sigma + c_6S^2. \label{e:qterm}
\end{align}
\endproof
\end{prop}  
It is no surprise that the $6$-jet of the residual function in~(\ref{e:nicef}) has the form of~(\ref{e:norform}): this must be the case in view of the Example in Section~\ref{ss:deter} and Proposition~\ref{p:split}, together with the uniqueness property (Section~\ref{ss:unique}). The point of Proposition~\ref{p:splitf} is, for arbitrary $(T,\bl)\in K$, to give explicitly up to order~$3$ the equivariant coordinate transformation that reduces that original $D_3\wr Z_2$-invariant Landau-de~Gennes function $f$ on $L(\R^2)\cong\R^4$ to a $D_3$-invariant function on $\R^2$, and to relate
the coefficients of the latter to the numerical coefficients of the KKLS free energy function using the identifications~(\ref{e:a3prime})--(\ref{e:a6prime}).
\msk

In the particular case $\xi=0$ we have $(y,v)=(s,p)$ and $(x,u)=-(d,c)$, while for $\xi=\frac{\pi}2$ we have $(y,v)=(d,c)$ and $(x,u)=(s,p)$.  These cases correspond to the two points of the instability circle $K_T$ that lie on the line of symmetry $\lam_1=\lam_2$ (the {\em Sonnet-Virga-Durand limit} in the terminology of~\cite{LN}) and play a key role in understanding bifurcation structure at nearby points on $K_T\,$: see Section~\ref{s:bifan} below. 
\msk
\begin{table}  [!ht]
   \begin{center}
     \begin{tabular}{ |c|c|c|c| }
\hline
\rule[-0.8ex]{0ex}{3.0ex} source & $\deg 4$    &  $\deg 5$  &  $\deg 6$   \\ \hline
\rule[-0.8ex]{0ex}{3.0ex} $|W|^2$ & $\om^2|Z|^4$ & $2\om\eta|Z|^2\Re Z^3$ &$\eta^2|Z|^6$  \\ \hline
\rule[-0.8ex]{0ex}{3.0ex}  $f_3$  & $3C\om |Z|^4$ & $9S\om^2|Z|^2\Re Z^3$ & $-C\om^3\Re Z^6+6S\eta\om|Z|^6$ \\ \hline
\rule[-0.8ex]{0ex}{3.0ex}$f_4$    &    &    & $ 4\om^2(\Im Z^3)^2$   \\ \hline
\rule[-0.8ex]{0ex}{3.0ex}$f_2^2$  &      &          & $2\om^2|Z|^6$      \\ \hline
\rule[-0.8ex]{0ex}{3.0ex}$f_5$    &     &       &  $2C\om(\Im Z^3)^2$   \\ \hline
\rule[-0.8ex]{0ex}{3.0ex}$f_2f_3$ &    &   & $3C\om|Z|^6$    \\ \hline
     \end{tabular}
     \caption{Higher degree terms of $p(Z)$ of degree up to~6 in $Z$
        created by completing the square. Here $\om=C\sigma$ and $\eta=2SC\sigma^2$.}  \label{t:csqterms}
    \end{center}
\end{table}
%
\subsection{Points of increasing degeneracy}
As $(T,\bl)$ crosses the stability cone $K$ the isotropic state becomes unstable, but the configuration of the local equilibrium solution branches 
(bifurcation geometry) depends on the nature of the degeneracy of $f$ at the crossing point on $K$.  We now inspect this degeneracy structure more closely.
\msk

In terms of coordinates $R_T,\xi$ on $\Lambda$ as in~(\ref{e:conepolar}) the quadratic terms of the function $f$ from~(\ref{e:fdef}) take the form
\begin{equation}  \label{e:quadterms}
\tfrac12(10T-U_0)(|z|^2+|w|^2) 
      +\tfrac{\sqrt{3}}{4\sqrt{2}}U_0R_T\bigl((|z|^2-|w|^2)\cos2\xi + 2\Re(z\bar w)\sin2\xi\bigr).
\end{equation}
On the stability cone $K$ we have 
\begin{equation}  \label{e:Rval}
U_0R_T=\sqrt{\tfrac23}(10T-U_0)
\end{equation}
and so the quadratic terms of $f$ reduce to
\[
\tfrac12(10T-U_0)|z\cos\xi+w\sin\xi|^2=2\mu|W|^2
\]
just as in~(\ref{e:freeZW}).  Perturbing $(T,R_T)$ to
$(T+t,R_T+\rho)$ and exploiting the rotational symmetry of $K$ by here taking
$\xi=0$ without loss of generality, we see that~(\ref{e:quadterms})
becomes
\begin{equation}  \label{e:quadchange}
\tfrac12(10(T+t)-U_0)\,|z|^2 + (\tfrac52t-\tfrac{\sqrt3}{4\sqrt2}U_0\rho)|w|^2,
\end{equation}
confirming that stability is lost (resp. gained) as $\rho$
(resp. $t$) alone increases through zero, with the cone generator (stability boundary) given locally by $10t=\sqrt{\tfrac32}U_0\rho$.
\subsubsection{Cubic terms} \label{s:cubicterms}
We next investigate higher order terms of the residual function $h$ in~(\ref{e:norform}), given that $(T,\bl)\in K$.
In terms of the original coefficients (\ref{e:fdef}) we see
from~(\ref{e:e3term}) that $e_3=-a_3\sin3\xi$, and therefore
item~(2) of Proposition~\ref{p:determinacy} immediately yields the following result up to rescaling.
\begin{prop}
If $(T,\bl)\in K$ and $a_3\sin3\xi\ne0$ then $q(x,u)$ is locally 
$D_3$-$\RR$-equivalent to $\pm Y$.  \endproof
\end{prop}
This means that if we fix $T$ and avoid the three points on $K_T$ where $\sin3\xi=0$ then up to local diffeomorphism (which is the identity to first order) we may take the free energy function to have the form of the cubic polynomial
\[
f_\xi(x,y,u,v)= 2\mu(y^2+v^2)\pm(x^3-3xu^2)
\]
where the sign is that of $a_3\sin3\xi$ and $(x,y,u,v)$ are given in terms of the original order parameters $(s,p,d,c)$ by the linear transformation~(\ref{e:coordmatrix}).
\msk

\rem The apparent anomaly that $\xi$ and $\xi+\pi$ correspond to the same point on $K_T$ while $\sin3\xi$ and $\sin3(\xi+\pi)$ have opposite signs is resolved by remembering that the coordinate transformation~(\ref{e:coordmatrix}) depends on $\xi$.  Replacing $\xi$ by $\xi+\pi$ reverses the sign of all coordinates $(x,y,u,v)$ so that $f_\xi$ is unchanged.
\msk

The branching diagram for $f$ showing critical point behaviour as $(T,\bl)$ passes through $K$ is naturally a very familiar object in the study of phase transitions of nematic liquid crystals as it represents the simplest 1-parameter bifurcation for critical points of functions with $D_3$ symmetry~\cite[Theorem XIV 4.3]{GS},\cite[Sect. 3.5]{SA}. (Those references deal with zeros of general non-gradient vector fields, but the result here coincides with that for the gradient case.) In the liquid crystal setting the parameter is typically taken to be the temperature~$T$, although the behaviour is qualitatively equivalent along any smooth path transverse to~$K$.  Usually the $D_3$ action is factored out so that behaviour is represented on a slice such as (in our coordinates) $u=0$: see e.g.~\cite[Figure 2]{MR},\cite[Figure 1]{LN},\cite[Figure 13]{SL}. For an example of the full picture obtained by $3$-fold rotation about the $T\,$-axis see~\cite[Figure 2]{CW}.  
\subsubsection{Higher order terms}  \label{s:hoterms}
Next suppose that $(T,\bl)\in K$ and $a_3\sin3\xi=0$. 
Assume $a_3\ne0$, otherwise the whole function takes on an extra degenerate character, regardless of the choice of $\bl$, which is not realistic
in the liquid crystal context.  Thus $S=0$ while $C^2=1$, and from~(\ref{e:e4term}) we have
\begin{equation}  \label{e:e4}
e_4=b_4-\frac{9a_3^2}{8\mu}
\end{equation}
which (since $4\mu=10T-U_0$) is nonzero provided $T\ne T_1$ where
\begin{equation}  \label{e:T1}
T_1=T_0 + \frac{9a_3^2}{20b_4}
\end{equation}
with $T_0=U_0/10$. From Proposition~\ref{p:determinacy} and the discussion in Appendix~B we cannot conclude, however, that in this case higher order terms of any order may be disregarded.  Instead, we adopt a standard approach in bifurcation theory by supposing that $e_4=0$, so the function $f_\xi$  has higher degeneracy at the origin, and then regarding the case $e_4\ne0$ as a perturbation of this more degenerate  {\em  organising centre} which we {\em are} able to analyse. The
cost of this method is that conclusions are valid only for
sufficiently small $e_4\ne0$, although experience shows that \lq
sufficiently small' can often be quite large.  
\msk

The condition $e_4=0$ is the condition that, after completing the square, there is no degree~$4$ term in the residual function.  This is described elsewhere~\cite{MS},\cite{MV},\cite{LN} as the condition for {\em tricriticality}.
As we see below, there are several reasons why the bifurcation analysis at tricriticality is particularly sensitive to arbitrarily small perturbation, and so it is no surprise that investigations are complicated~\cite[Appendix C]{MS} and in
some respects inconclusive~\cite[Figure 2]{MV}.
\msk

Let us then take $T=T_1$ as in~(\ref{e:T1}). 
Suppose also $b_4>0$ so that $T_1>T_0$ and the isotropic state is stable for $\bl$ inside the stability circle $K_{T_1}$.
We already have $S=0$ and so $e_5=0$.
From~(\ref{e:pterm}) and~(\ref{e:qterm}) we find 
\begin{align}
e_6 &=a_3\sigma^3+2(b_4-a_4)\sigma^2+(2a_5+3b_5)\sigma+b_6
 \label{e:newpterm}  \\
d_6 &=-2a_3\sigma^3+4a_4\sigma^2-2a_5\sigma
 \label{e:newqterm}
\end{align}
where $\sigma=-3a_3/4\mu$.  From Proposition~\ref{p:determinacy} we therefore conclude the following.
\begin{prop}
If $(T,\bl)\in K$ with $T=T_1$ and $\sin3\xi=0$ 
then the residual function $q(x,u)$ is locally 
$D_3$-$\RR$-equivalent to 
\[
e_6X^3+d_6Y^2+e_8X^4
\]  
provided $e_6d_6(e_6+d_6)\ne0$.
\endproof
\end{prop}
We assume that the coefficients in the free energy function~(\ref{e:fdef}) are such that this last condition is indeed satisfied when $\mu=\tfrac14(10T_1-U_0)=\tfrac1{18}b_4a_3^{-2}$. 
\subsubsection{Normal form and versal deformation}  
Rather than carry out further explicit \lq bottom up' calculations we now
invoke the techniques of singularity theory to provide a \lq top down' {\em normal form} for the bifurcation structure. 
First, we replace the coefficients $e_6,d_6$ by new symbols $m,n$ respectively, to emphasise that they are assumed nonzero unlike (possibly) the coefficients $e_2,\ldots,e_5$. 
(We do not need to suppose as in~\cite{LN} that the coefficients of the highest-order terms considered are independent of the temperature $T$: as we see, the $T$-dependence can be absorbed into lower-order coefficients.)
Then, regarding the free energy function $\wt f$ in~(\ref{e:reduce}) as a perturbation of the function
\begin{equation}  \label{e:tidyf6}
f_0(x,y,u,v):=2\mu(y^2+v^2)+q_0(x,u)
\end{equation}
where $q_0(x,u)$ consists of the degree-$6$ terms
\[
q_0(x,u):=\,mX^3+nY^2\,=\,m(x^2+u^2)^3+n(x^3-3xu^2)^2
\]
and it is assumed $mn\ne0$, we invoke the theory of {\em versal deformation} (or {\em universal unfolding})
to exhibit a $7$-parameter family $f_e$ of perturbations of $q_0$ that captures {\em all possible local bifurcation behaviour} for~(\ref{e:tidyf6}).  More precisely, any other multiparameter perturbation of~(\ref{e:tidyf6}) can be written in terms of $f_e$ by an appropriate local coordinate-mapping (invertible in the $x,u$-variables) that respects the $D_3$-symmetry. 
For further explanation and a sketch of the proof of the Proposition see  Appendix~C. 
Stating the result in more formal terms:
\begin{prop}  \label{p:versaldef}
The $7$-parameter family of polynomial functions 
\begin{equation} \label{e:vdef}
q_e(x,u):=e_0 + e_2X + e_3Y + e_4X^2 + e_5XY + e_6X^3 + e_8X^4 + q_0(x,u)
\end{equation}
is a $D_3$-$\RR$-{\em versal deformation} of $q_0$ at the origin in $\R^2$.
\endproof
\end{prop}
The same arguments extend easily to show also
\begin{prop}  \label{p:vdefplus}
The $7$-parameter family of polynomial functions
\begin{equation}  \label{e:finalnorform}
f_e(x,y,u,v):= 2\mu(y^2+v^2) + q_e(x,u)
\end{equation}
is a $D_3$-$\RR$-versal deformation of the free energy function 
\[ 
f_0(x,y,u,v):= 2\mu(y^2+v^2)+q_0(x,u)
\]
at the origin in $\R^4$.
\endproof
\end{prop} 
Therefore, in view of Proposition~\ref{p:vdefplus} and the
fact that the constant $e_0$ is irrelevant for locating critical
points and for assigning {\em relative } function values to those points, we have the following consequence.
\begin{cor} \label{p:F6}
In order to study bifurcation behaviour of critical points of the free energy
function~(\ref{e:fdef}) close to isotropy and around the points of highest
degeneracy (sixth order) on the singular cone $K$ it suffices to study
bifurcation of critical points of the family of polynomials 
\begin{equation}  \label{e:F6}
P_e(X,Y):=e_2X + e_3Y + e_4X^2 + e_5XY + +e_6X^3 + e_8X^4 + mX^3 + nY^2
\end{equation}
where $(X,Y)=(x^2+u^2,x^3-3xu^2\,)$ and nonzero $m,n$ are fixed with $m+n\ne0$.
\endproof   
\end{cor}
This $6$-parameter family of $D_3$-invariant functions on
$\R^2$ is the {\em normal form} for the bifurcation problem.  In our application the parameters $e_2,\ldots,e_6,e_8$ and the coefficients $m,n$ are expressed in
terms of the original coefficients and the $(T,\bl)$ parameters. Writing 
\[
(T,R_T)=(T_1+t,R_{T_1}+\rho)
\]
the perturbed quadratic terms~\eqref{e:quadchange} show that if we first rescale $z$ so that the coefficient of $|z|^2$ remains equal to $2\mu=\tfrac12(10T-U_0)$ then 
\begin{equation}  \label{e:e2term}
e_2=\tfrac52t-\tfrac{\sqrt3}{4\sqrt2}U_0\rho
\end{equation}
while $e_3,e_4,e_5$ and $m,n$ are given by~(\ref{e:e3term})--(\ref{e:qterm}).
In particular the quadratic term $e_2X$ corresponds to applying the perturbation
\[
(\alpha,\beta,\gamma)\mapsto(\alpha,\beta,\gamma)
                       +\eps(\beta,\alpha,-\gamma)
\]
to the original quadratic coefficients in~(\ref{e:fdef}) where $e_2=2\mu\eps$. 
We discuss later the roles played by $e_6$ and $e_8$.
\msk

Since the $D_3$-symmetry arises automatically from the independence of the physics on the choice of coordinate frame, this normal formal form (up to degree~$6$) and its bifurcations have naturally been much studied by other authors in different contexts.
In particular, a detailed analysis has been given in~\cite{AL}
making clever use of direct algebraic manipulations and (importantly) distinguishing global minima from other critical points.
Moreover, the several cases of solution branching studied in~\cite{LN} in the original coordinates and under various symmetry assumptions can all be viewed in terms of properties of the function~(\ref{e:finalnorform}).
However, we are not aware of other accounts of the bifurcation structure of the free energy function in the wider setting of equivariant versal deformation theory.
%
\section{Bifurcation analysis of the normal form of the residual function}  \label{s:bifnormal}
The $D_3$ symmetry is most efficiently exploited by moving to polar
coordinates in the $(x,u)$-plane and letting
\[
(x,u)=r(\cos\theta,\sin\theta)
\]   
so that
\[
X=r^2\,,\quad Y=r^3\cos\alpha
\]
with $\alpha=3\theta$.  Then writing $f_e=f$ and $P_e=P$ for simplicity we have
\begin{align}
\dbd fr&=\dbd PX\,2r + \dbd PY\,3r^2\cos\alpha  \\
\dbd f\alpha &= \dbd PX\,0 - \dbd PY\,r^3\sin\alpha.
\end{align}
Critical points $(x,u)$ for $f$ with $r\ne0$ thus arise where 
\begin{equation}  \label{e:rdiff}
\sin\alpha=0 \quad \text{and} \quad2\dbd PX \pm 3r\dbd PY = 0
\end{equation}
(the sign choice depending on the choice $\alpha=0,\pi$) or where
\begin{equation}   \label{e:adiff}
\dbd PX = 0 \quad \text{and} \quad \dbd PY = 0.
\end{equation}
\msk

The first alternative~(\ref{e:rdiff}) corresponds to critical points lying on the
$x\,$-axis or rotations of it by $\tfrac{2\pi}3$.  Without loss of generality  we restrict to the $x\,$-axis which from~(\ref{e:coordmatrix}) with $\xi=0$ means $s=p=c=0$ and $x=-d$.  Critical points here correspond to {\it uniaxial} equilibria. They can be found by substituting $X=x^2,Y=x^3$ into
$P_e(X,Y)$ and then using standard local multiparameter bifurcation analysis provided
by elementary catastrophe theory.  We carry out this analysis in Section~\ref{ss:uniax} below.
\msk

\begin{figure}[!ht]  
\begin{center}
 \scalebox{.50}{\includegraphics{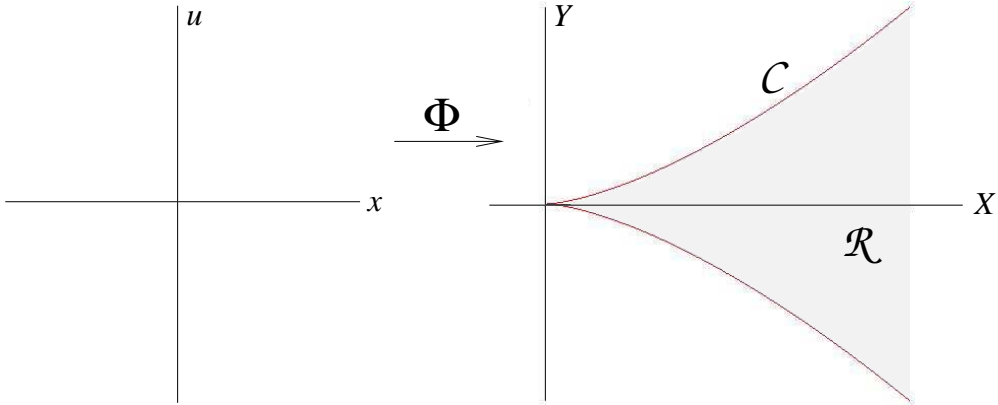}}
\end{center}  
\caption{The region $\mathcal R$ in the $(X,Y)$-plane that is the image of the
$(x,u)$-plane under the map $\Phi$. The bounding curve $\CC$ is the image of the $x\,$-axis.}   
\label{fi:cuspregion}
\end{figure}

The second alternative~(\ref{e:adiff}) identifies critical points of $f$ as 
corresponding to critical points of $P_e$, which (as a function of
$X,Y$) has a much simpler form than $f$ (as a function of $x,u)$.
The subtlety, however, is that not every point $(X,Y)$
corresponds to points $(x,u)$.  To see this explicitly, observe that the
smooth map 
\[
\Phi:\R^2\to\R^2:(x,u)\mapsto(X,Y)=(x^2+u^2,x^3-3xu^2)
\]
is singular along the $x\,$- and $u\,$-axes, and the $x\,$-axis is taken by $\Phi$ to
the cusped curve ${\CC}$ in the $(X,Y)$-plane with equation $X^3=Y^2$. The
image of the $(x,u)$-plane under the map $\Phi$ is the region
${\mathcal R}$ to the right of the curve $\CC$: see Figure~\ref{fi:cuspregion}. 
The problem is therefore to find the critical points of
$P$ that actually lie in ${\mathcal R}$.  Those lying
inside ${\mathcal R}$ and not on $\CC$ correspond to critical points of
$f$ away from the $x\,$-axis, that is with nonzero $u$ component.  Since
$u$ corresponds to $-c$ these represent {\it biaxial} equilibria. 
This construction is well known in liquid crystal theory (see~\cite[Figure 1]{MM} or~\cite[Figure 1]{AL} for example) and in the wider context of bifurcation with symmetry~\cite{MZ}.
\msk   

We now consider in turn the uniaxial and biaxial bifurcation scenarios.
\subsubsection{Uniaxial equilibria}  \label{ss:uniax}
Taking $u=0$ and substituting $X=x^2$ and $Y=x^3$ into~(\ref{e:F6}) gives
\begin{equation}
p_e(x):=P_e(x,0)=e_2x^2+e_3x^3+e_4x^4+e_5x^5+e_6x^6+(m+n)x^6+e_8x^8.
\end{equation}
Critical points of $p_e$ with $x\ne0$ occur where  
\begin{equation}  \label{e:swallow}
0=2e_2 + 3e_3x + 4e_4x^2 + 5e_5x^3 + 6(m+n+e_6)x^4 + 8e_8x^6.
\end{equation}
Making the assumption $m+n>0$ and taking $e_2,\ldots,e_5$ in a \nhd of zero
this is a deformation  of an $A_3$ (that is, $x^4$) singularity.  Standard techniques show that a $3$-parameter $\RR$-versal deformation of $x^4$ is provided
by~(\ref{e:swallow}) with $e_5=0$, and the full bifurcation
geometry is described by the {\em swallowtail} catastrophe~\cite{CH,G,PS}.
The redundancy of the $e_8$ parameter is seen by replacing $x$ by a new variable of the form $\til x=x(1+kx^2)^{1/6}$, while the redundancy of the $e_5$ parameter is seen by applying an explicit translation of coordinates $x\mapsto x-x_0$ (here
$x_0=\frac{5}{24}e_5(m+n)^{-1}$) that removes the $x^3$ term and leaves
$e_2,e_3,e_4$ unaffected to first order.
\msk

\begin{figure}[!ht]  
\begin{center}
 \scalebox{.60}{\includegraphics{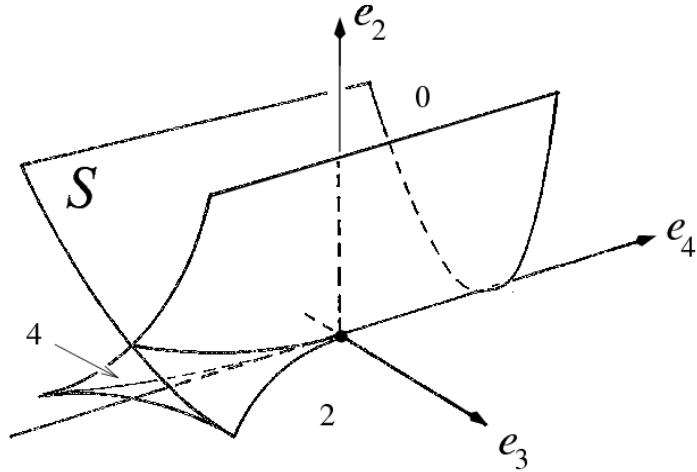}}
\end{center}  
\caption{The swallowtail bifurcation surface $S$. The numbers in each complementary region in $\R^3$ indicate the numbers of critical points of $h_e^0$ (in addition to the one at the origin) for $(e_1,e_2,e_3)$ in that region.}   
\label{fi:swallowtail}
\end{figure}

\begin{figure}[!ht]  
\begin{center}
 \scalebox{.60}{\includegraphics{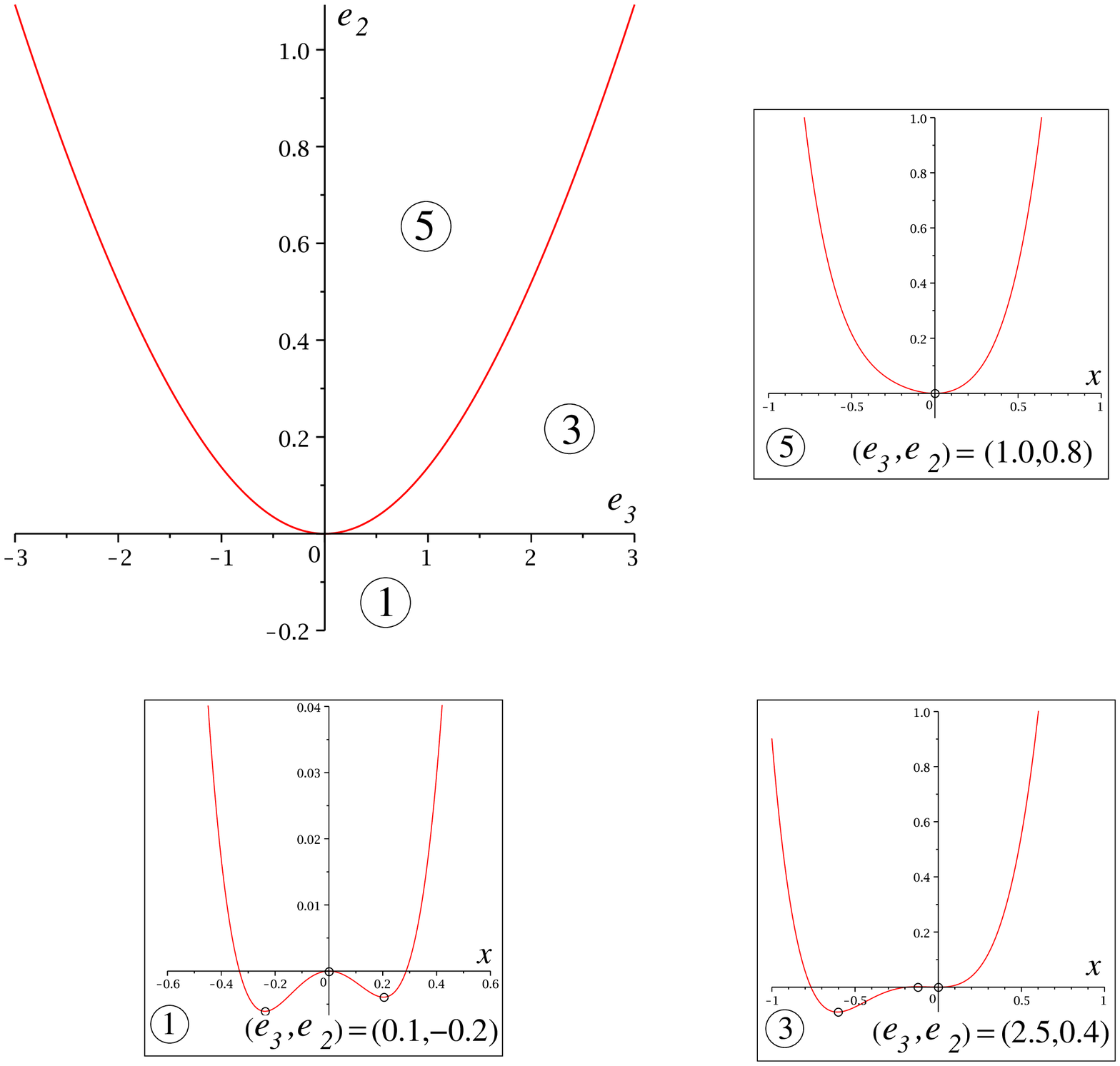}}
\end{center}  
\caption{Section of the swallowtail bifurcation set $S$ for $e_5=0$ and fixed
  $e_4>0$, showing the graph of $h_e^0$ for various choices of $e_2$ and $e_3$.
 The regions are numbered to match with regions in Figure~\ref{fi:fig1neg}.  Critical points represent uniaxial equilibria.}   
\label{fi:fig1pos}
\end{figure}

\begin{figure}[!ht]  
\begin{center}
 \scalebox{.60}{\includegraphics{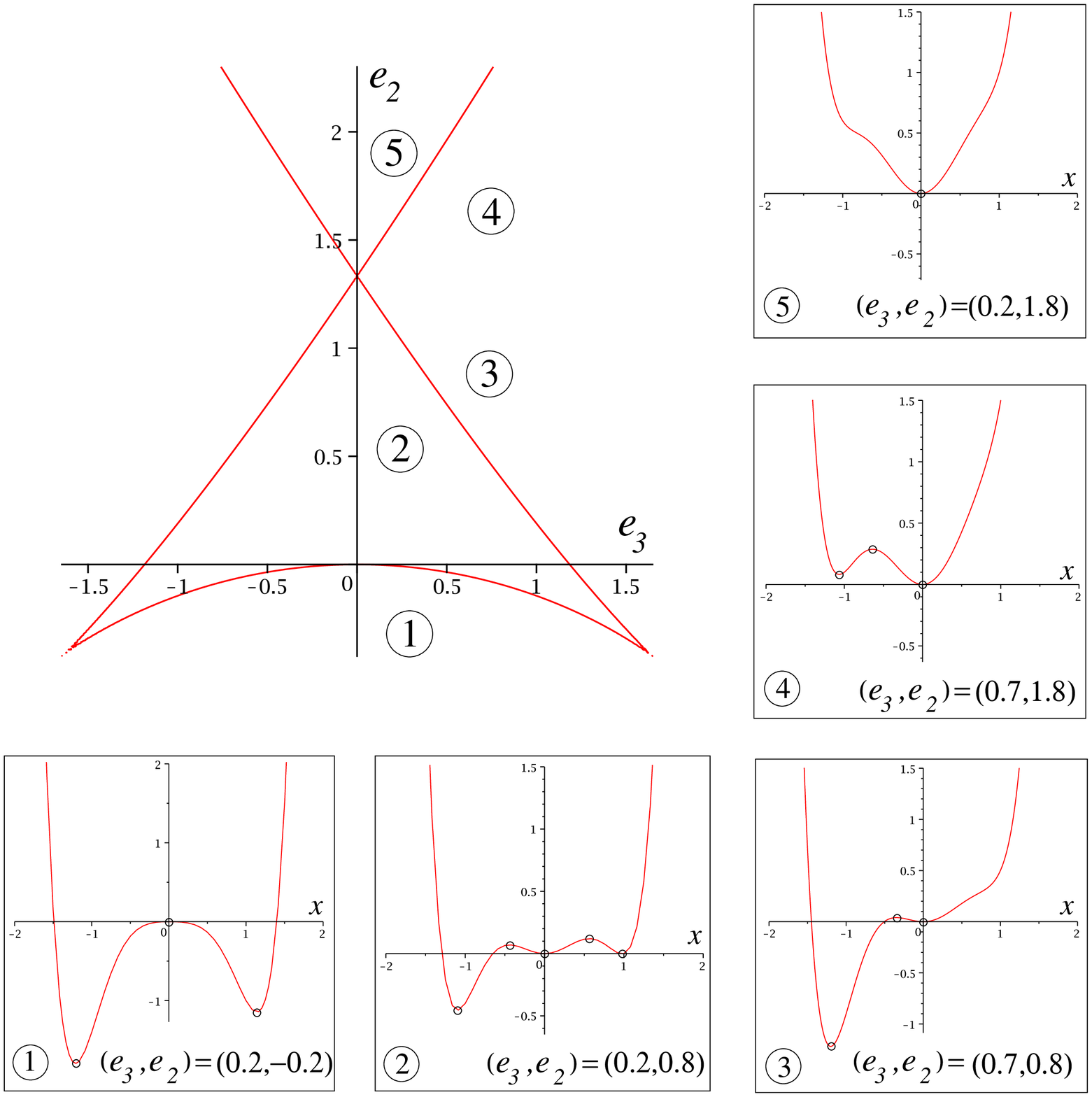}}
\end{center}  
\caption{Section of the swallowtail bifurcation set $S$ for $e_5=0$ and fixed
  $e_4<0$, showing the graph of $h_e^0$ for various choices of $e_2$ and $e_3$.  Critical points represent uniaxial equilibria.}   
\label{fi:fig1neg}
\end{figure}

The $3$-parameter bifurcation set $S$ for the solutions of~\eqref{e:swallow} is sketched in Figure~\ref{fi:swallowtail}.  For fixed
$e_4\ne0$ the form of the bifurcation set $S$ in the
$(e_2,e_3)$-plane depends only on the sign of $e_4$. In
Figure~\ref{fi:fig1pos} and Figure~\ref{fi:fig1neg} we show graphs of $p_e$ for
representative parameter values in different connected components of the complement of the intersection of $S$ with the $(e_2,e_3)$-plane for $e_4>0$, and 
for $e_4<0$ respectively.  Note that, in view of the particular form of
the function $p_e$ for which $x=0$ is always a critical point,
in the each diagram the $e_3\,$-axis also forms part of the
bifurcation set as it controls the transition from local minimum to
local maximum at the origin. In Figure~\ref{fi:fig1neg} the cusp points represent 
values of $(e_2,e_3)$ at which $p_e$ exhibits a $4^{\rm th}$-order critical point, deforming (or {\em unfolding}) inside the cusp to two local minima and a maximum.
\msk

The effect of nonzero $e_5$ is to cause a second-order 
distortion in the geometry of the swallowtail bifurcation set, as we observe below in Section~\ref{s:fsymbr} when considering the full bifurcation set for $f$ including biaxial equilibria.
\subsubsection{Biaxial equilibria}
We have
\begin{align}
0=\dbd PX&=e_2 + 2e_4X + e_5Y + 3(m+e_6)X^2 +4e_8X^3 \label{e:Xderiv}\\
0=\dbd PY&=e_3 + e_5X + 2nY. \label{e:Yderiv} 
\end{align}
The second equation represents a family of straight lines in the $(X,Y)$-plane
that for small $(e_3,e_5)$ may or may not intersect the region
$\RR$ close to the origin.  The corresponding loci 
$\Gamma=\Gamma_{e_3,e_5}$
in the $(x,u)$-plane are cubic curves with $D_3$~symmetry.
Eliminating $Y$ between~(\ref{e:Xderiv}) and~(\ref{e:Yderiv}) gives 
\begin{equation}  \label{e:elimY}
8ne_8X^3 + 6n(m+e_6)X^2 + (4ne_4-e_5^2)X + 2ne_2 - e_3e_5 = 0.
\end{equation}
which for each $e=(e_2,\ldots,e_6,e_8)$ 
represents a set of at most three circles with
centre the origin in the $(x,u)$-plane.
However, with $mn\ne0$ and small $e_6,e_8$ there can be at most two of these circles close to the origin. In order to capture the geometry of bifurcation from isotropy we may therefore suppose $e_8=0$, and at the cost of rescaling parameters $(e_2,\ldots,e_5)$ (changes effective only at second order) we may also take $e_6=0$.
The biaxial equilibria then correspond to the points of intersection of the 
set $C=C_e$ of at most two circle(s) with the cubic curve~$\Gamma$ where 
$\sin 3\theta\ne0$.
\msk

To analyse the intersections $C\cap\Gamma$ in detail it is convenient to factor out the $D_3$~symmetry through replacing the coordinates $(x,u)=(r\cos\theta,r\sin\theta)$
by 
\[
(x',u')=(r\cos\alpha,r\sin\alpha)
\] 
with $\alpha=3\theta$: then $\Gamma$ becomes a curve $\Gamma'$ symmetric abut the $x'\,$-axis and with only one branch instead of the three branches of~$\Gamma$.  Biaxial equilibria correspond to points of $C\cap\Gamma'$ that do not lie on the $x'\,$-axis. 
\msk

\begin{figure}[!ht]  
\begin{center}
 \scalebox{.60}{\includegraphics{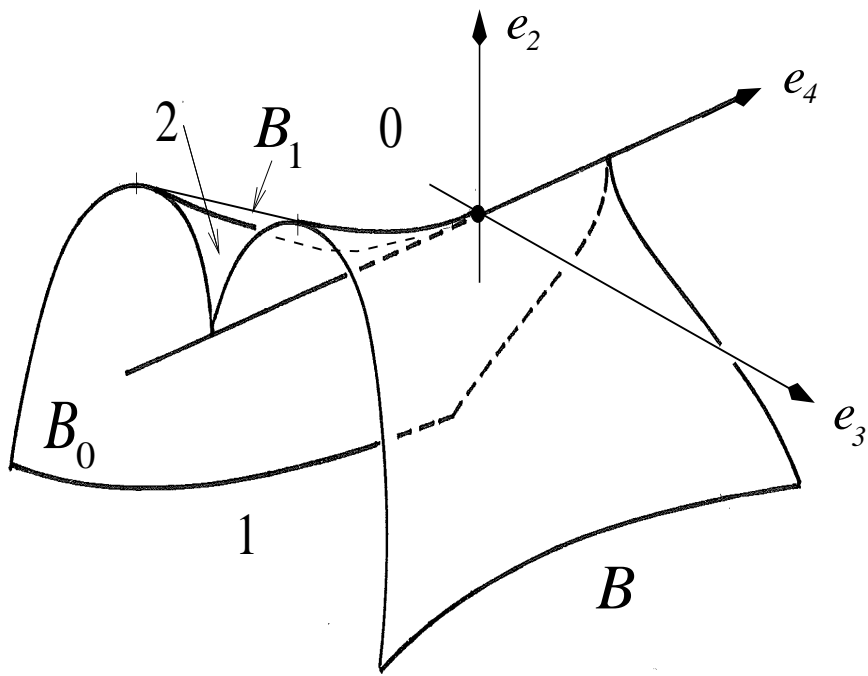}}
\end{center}  
\caption{Bluebird bifurcation surface $B$ for $e_5=0$.
The numbers in each complementary region
in $\R^3$ indicate the numbers of $D_3$-orbits of biaxial critical points of $P_e$.}   
\label{fi:figblu}
\end{figure}

For $e_5=0$ the bifurcation set $B$ in $(e_2,e_3,e_4)$-space
has a geometric form as indicated in Figure~\ref{fi:figblu}.  
It consists of two parts: a surface $B_0$ that is smooth apart from a cusp ridge along the $e_4\,$-axis (from the side $e_2>0$ or $e_2<0$ according as $e_4<0$ or $e_4>0$), and a surface $B_1$ that for fixed $e_4<0$ consists of a straight line segment connecting the two \lq hilltops' of $B_0$. The surface $B_0$ corresponds to bifurcations from the $x'\,$-axis (i.e. from uniaxial to biaxial equilibria) while $B_1$ corresponds to mutual annihilation and creation of critical points with $u'\ne0$ (biaxial equilibria).  In the absence
of a standard name for this bifurcation surface we call it the {\em bluebird}. It is the bluebird that governs the phase transitions into biaxiality, and between biaxial equilibria themselves. 
\msk

\begin{figure}[!ht]  
\begin{center}
 \scalebox{.50}{\includegraphics{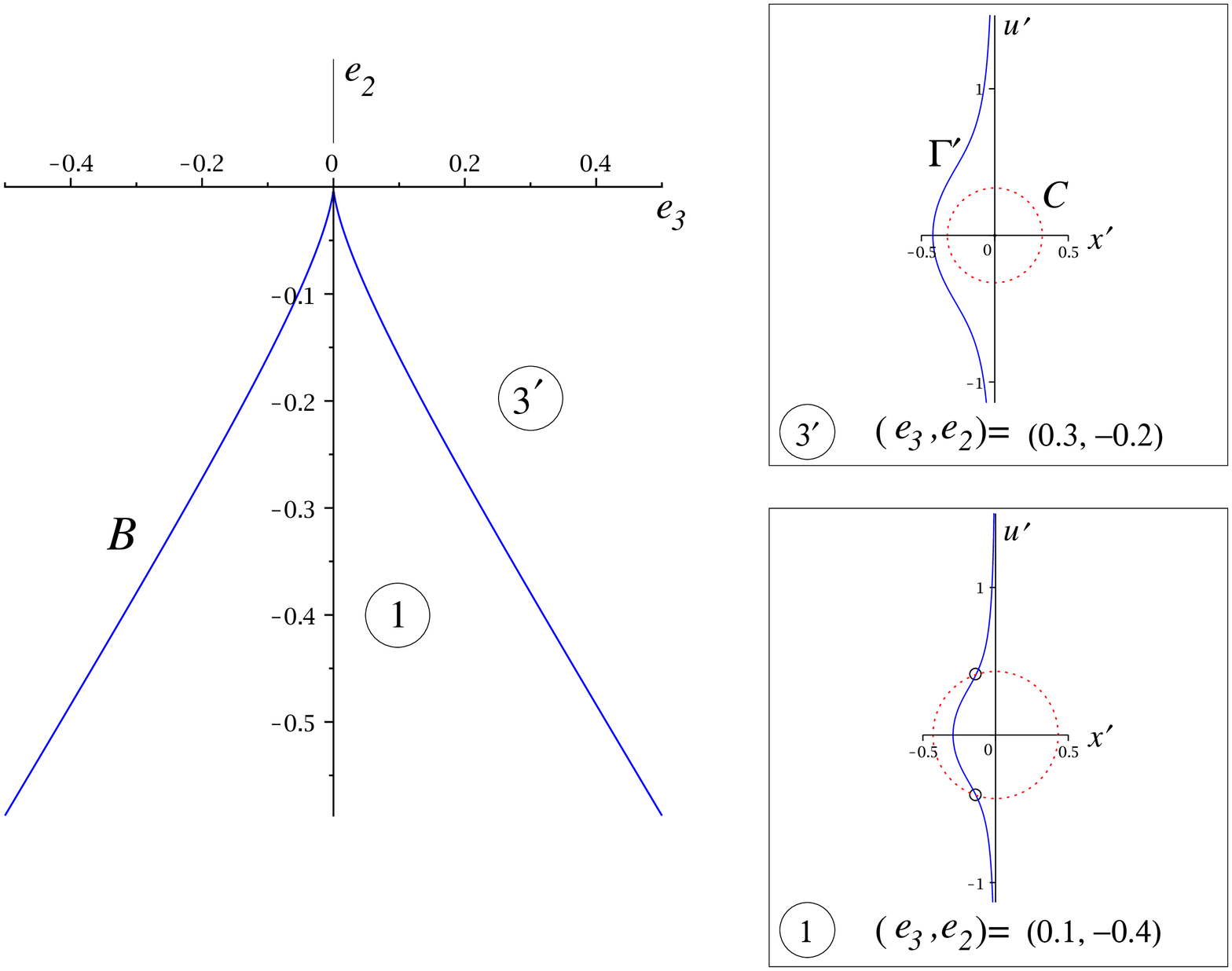}}
\end{center}  
\caption{Bluebird bifurcation set for $e_5=0$ and fixed
  $e_4>0$, showing biaxial equilibria as intersection points of the
  curve $\Gamma'$ and the circles $C$ for two representative choices
  of $e_2$ and $e_3$. }   
\label{fi:figBpos}
\end{figure}

\begin{figure}[!ht]  
\begin{center}
 \scalebox{.60}{\includegraphics{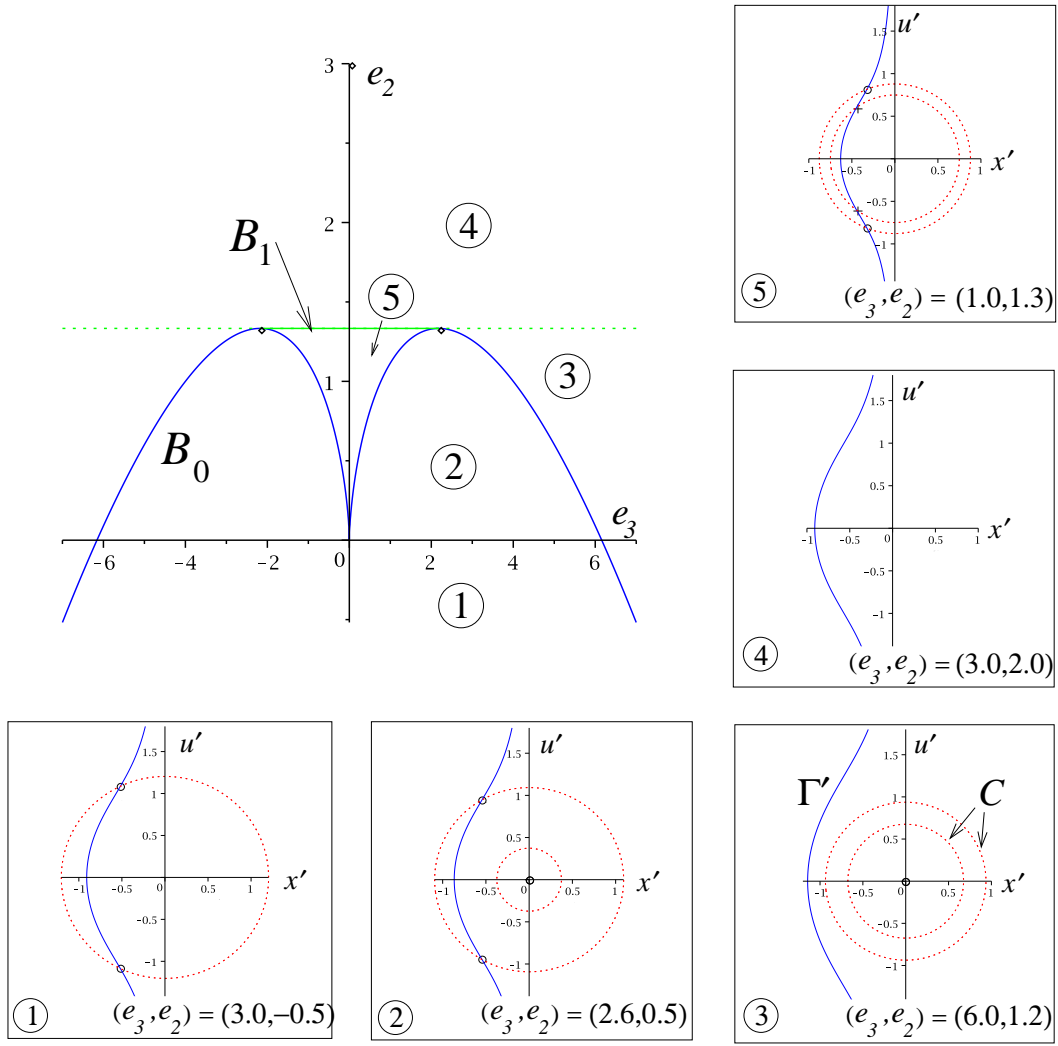}}
\end{center}  
\caption{Bluebird bifurcation set for $e_5=0$ and fixed
  $e_4<0$, showing biaxial equilibria as intersection points of the
  curve $\Gamma'$ and the circles $C$ for various choices of $e_2$ and $e_3$. }   
\label{fi:figBneg}
\end{figure}

Again, for fixed
$e_4\ne0$ the form of the bifurcation set $B$ in the
$(e_2,e_3)$-plane depends only on the sign of $e_4$. In
Figure~\ref{fi:figBpos} and Figure~\ref{fi:figBneg} we show configurations of $C$ and $\Gamma'$ for representative parameter values in different connected components of the complement of $B$ in the $(e_2,e_3)$-plane for a fixed $e_4>0$ and $e_4<0$ respectively. In Figure~\ref{fi:figBneg} transition across the curve $B_0$ from above to below corresponds to creation of a pair of biaxial equilibria bifurcating from a uniaxial equilibrium on the $x'\,$-axis, while transition across the line segment $B_1$ from above corresponds to creation of two pairs of biaxial equilibria 
at simultaneous saddle-node (fold) bifurcations away from uniaxial states.  The two points of $B_0\cap B_1$ at which $B=0$ is tangent to $B_1$ correspond to the more degenerate scenario in which the two biaxial pairs bifurcate simultaneously from a uniaxial equilibrium. Finally, it is important not to forget that these descriptions in $(x',u')$ coordinates reflect behaviour occurring simultaneously at three locations in the original $s,p,d,c$ -space of four order parameter.
\msk

\begin{figure}[!ht]  
\begin{center}
 \scalebox{.50}{\includegraphics{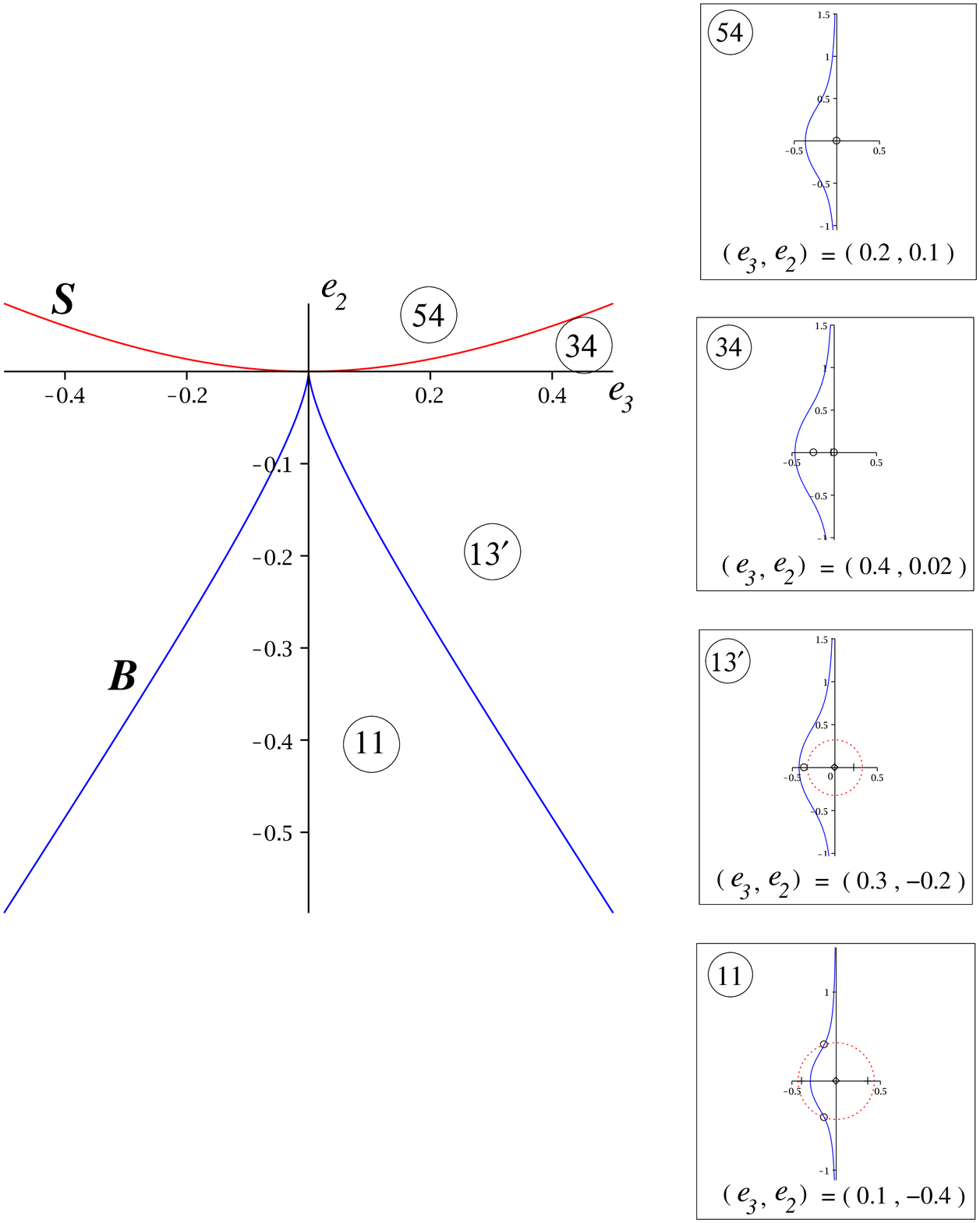}}
\end{center}  
\caption{Combined bifurcation set for $e_5=0$ and fixed
  $e_4>0$, showing uniaxial equilibria on the $x\,$-axis and biaxial
  equilibria off the $x\,$-axis lying on $\Gamma'\cap C$ as in Figure~\ref{fi:figBpos} for representative choices
  of $e_2$ and $e_3$. }   
\label{fi:fig4pos}
\end{figure}
\begin{figure}[!ht]  
\begin{center}
 \scalebox{.50}{\includegraphics{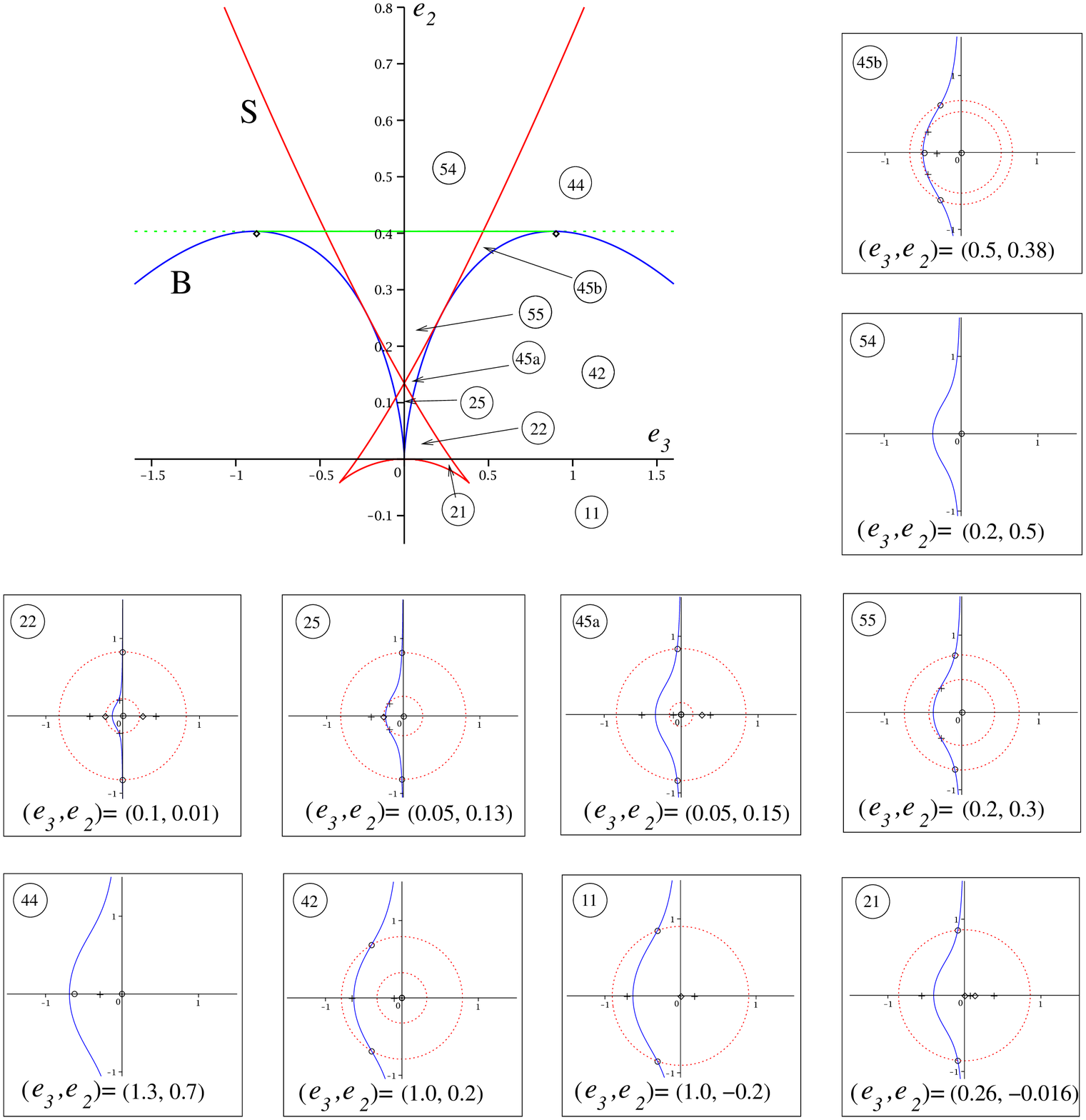}}
\end{center}  
\caption{Combined bifurcation set for $e_5=0$ and fixed
  $e_4<0$, showing uniaxial equilibria on the $x\,$-axis and biaxial
  equilibria off the $x\,$-axis lying on $\Gamma'\cap C$ as in Figure~\ref{fi:figBneg} for representative choices
  of $e_2$ and $e_3$. }   
\label{fi:fig4neg}
\end{figure}
The full bifurcation behaviour for critical points of the family of functions~(\ref{e:vdef}) close to the origin  is therefore given by the superposition of the swallowtail $S$  and the bluebird $B$.
In Figure~\ref{fi:fig4pos} we show (after suitable rescaling) the superposition of Figure~\ref{fi:fig1pos} and Figure~\ref{fi:figBpos}, and likewise for Figure~\ref{fi:fig1neg} and Figure~\ref{fi:figBneg} in 
Figure~\ref{fi:fig4neg}. 
\msk

Before looking more closely at the critical point behaviour that these diagrams represent, we make three observations.
\paragraph{Independence}

Swallowtail and bluebird bifurcations are not independent, since if~(\ref{e:adiff}) is satisfied then so also is the second equation of~(\ref{e:rdiff}). This
reflects the fact that if biaxial equilibria bifurcate from the
$x\,$-axis then they must do so at critical points of $p_e$ (uniaxial equilibria).
\endproof
\paragraph{Tangency}

This concerns simultaneous bifurcation of
uniaxial and biaxial equilibria. Fix $e_5=0$ and $e_4<0$
to simplify the description, and denote by $M_S$ the $2$-dimensional surface (manifold) in $(e_2,e_3,x)$~-space $\R^3$ given by the solution set
to~(\ref{e:swallow}).  Let
$\Sigma_S\subset M_S$ be the set of points (a smooth curve) where the projection
$\pi:M_S\to\R^2$ into the $(e_2,e_3)$-plane fails to be a local
diffeomorphism (smoothly invertible).  These are precisely the points
where bifurcation occurs, and it is the projected image $\Delta_S=\pi(\Sigma_S)\subset\R^2$ that is the swallowtail bifurcation set $S$ as in Figure~\ref{fi:fig1neg}.
Likewise let $M_B\subset\R^4$ be the $2$-dimensional solution set to~(\ref{e:Xderiv}) and (\ref{e:Yderiv}) with, with $\Sigma_B$ and $\Delta_B$ defined analogously: thus $\Delta_B$ is the bluebird bifurcation set $B$ as in Figure~\ref{fi:figblu}. As just noted, the intersection $M_B^0$ of $M_B$ with the hyperplane $u=0$ lies in $M_S$ and is part of $\Sigma_B$.
Now consider points where $M_B^0$ intersects $\Sigma_S$: these are points of bifurcation on the $x\,$-axis of biaxial and uniaxial equilibria simultaneously.  At most points of $\Sigma_S$ the projection $\pi:M_S\to\R^4$
exhibits a fold singularity~\cite{CH,G,PS}, which has the local
geometric property that almost any smooth curve in $M_S$ crossing
$\Sigma_S$ projects by $\pi$ to a curve {\it tangent} to $\Delta_S$.
Consequently at the points of $B\cap S$ that correspond to points of
$M_B^0\cap\Sigma_S$ the bifurcation sets $B$ and $S$ will be
{\em mutually tangent}. 
This property is readily visible in Figure~\ref{fi:fig4neg}.  Note that there
are typically also points of $B\cap S$ that correspond to
distinct (although simultaneous) biaxial and uniaxial bifurcations, and there is no reason why those intersections of $B$ and $S$ should be mutually tangent. \endproof  
\paragraph{Tricriticality}
As indicated in Section~\ref{s:hoterms} the condition called tricriticality elsewhere corresponds here to the condition $e_4=0$.  Figures~\ref{fi:swallowtail} and~\ref{fi:figblu} show that, not only is it necessary to take account of terms of degree at least~$6$ in order to grasp the 
full local $3$-dimensional bifurcation structure where this condition holds, but that complicated bifurcation behaviour takes place in arbitrarily small neighbourhoods of the tricritical point, depending crucially on the sign of $e_4$ and mapped out in Figures~\ref{fi:fig4pos} and~\ref{fi:fig4neg}.
\endproof
\subsubsection{The full bifurcation analysis} \label{s:bifan}
Turning now to the bifurcation geometry for the family of functions $P_e$
as organised by the bifurcation sets $S$ and $B$ in
$(e_2,e_3,e_4)$-space for fixed $e_5$, we take $e_5=0$ as before
and comment later on nonlinear distortions to the picture that
arise with $e_5$ nonzero.
\msk

There are two complementary approaches to studying the bifurcation behaviour.
The first is to fix $T$ and consider the bifurcation geometry in the
$\bl$-plane $\Lambda$.  This gives good geometric
insight into the relationships between various critical point
branches.  However, in applications it is likely to be more relevant
instead to fix $\bl$ (physical constants) and vary $T$ (temperature).
We look at these two approaches in turn.
\subsubsection*{Fixed $T$}
The preceding bifurcation analysis applies to values of $T$ close to $T_1$ given by~(\ref{e:T1}) and $\bl$ close to any of the three points
on the stability circle $K_{T_1}$ where $\sin3\xi=0$.
Again for simplicity we take $\xi=0$ and thus choose the point $\bl=\bl^1$ given by~(\ref{e:conepolar}) with $T=T_1$.
From~(\ref{e:quadchange}) we see that for fixed $t$ the coefficient of $|w|^2$
increases with negative $r$, that is the radial vector pointing {\em
  into} the critical circle $K_T$ corresponds to the $e_2\,$-axis,
while~(\ref{e:e3term}) shows that $e_3$ depends only on $\xi$ and so
$K_T$ itself corresponds locally to the
$e_3\,$-axis. Finally, the expression~(\ref{e:e4}) for $e_4$ together
with~(\ref{e:muterm}) show that $e_4$ increases with $T$. 
\msk

The local bifurcation of critical points close to the degree-6 degeneracy of the free energy at $(T,\bl)=(T_1,\bl^1)$ is described using $(x',u')$ coordinates in Figures~\ref{fi:fig4pos} and~\ref{fi:fig4neg}. Consider first the case $e_4>0$ in Figure~\ref{fi:fig4pos}. For a sample point $e=(e_3,e_2)$ in the region
{\large\textcircled{$\scriptstyle{54}$}} the only equilibrium state is the isotropic state $(x',u')=(0,0)$. 
As $e$ crosses $S$ into the region {\large\textcircled{$\scriptstyle{34}$}} a pair of equilibria is created on the negative $x'\,$-axis, one of these passing through the origin as $e$ crosses the $e_3\,$-axis causing the isotropic state to become locally unstable through a transcritical bifurcation.  Finally as $e$ crosses $B$ into the region {\large\textcircled{$\scriptstyle{11}$}} 
the stable (uniaxial) equilibrium on the $x'\,$-axis bifurcates to give a pair of (biaxial) equilibria with $u'\ne0$ (second order phase transition).  Since this is the generic codimension-2 bifurcation at points on the stability cone with $T>T_1$ it is not surprising that this diagram is ubiquitous in one form or another in the liquid crystal literature: see for example~\cite[Figure 2.2]{TS},\cite[Figure 1]{LP},\cite[Figure 6]{LN},\cite[Figure 6]{AL},\cite[Figure 3]{ZP}.
\msk 

If instead we take $e_4<0$ (that is $T<T_1$) then the situation is much more complicated, as indicated in Figures~\ref{fi:fig1neg},\ref{fi:figBneg} and~\ref{fi:fig4neg}.  As $T$ decreases through $T_1$ a whole bouquet of uniaxial and biaxial critical points is generated.  In particular we note that 
in the small region~{\large\textcircled{$\scriptstyle{35}$}} in the triangular region created by the swallowtail~$S$ but lying above the bluebird curve~$B$ there are two pairs of biaxial critical points as well as five uniaxial critical points including the isotropic state at the origin, giving (before factoring out the $3$-fold symmetry) a total of 25 critical points of the free energy.
The points of tangency of the swallowtail and bluebird correspond to a pair of uniaxial equilibria and a pair of biaxial equilibria bifurcating from the {\em same} point on the $x'\,$-axis. Observe also that the only parameter values giving simultaneous uniaxial and biaxial bifurcation {\em from the isotropic state}, corresponding to intersection of $S\cap B$ with the $e_3\,$-axis, are $e_2=e_3=0$.
Thus Alben's \lq accidental' isotropic--uniaxial--biaxial collision as described in~\cite{LN} takes place only at the overall organising centre where the residual free energy function has degree six.  Of course this does not exclude discontinuous biaxial to isotropic phase transition (competition of global minima) close to points where $B$ intersects the Maxwell set~\cite{AGLV} of~$S$.
\msk

Naturally, in view of the (uni)versality of the bifurcation geometry,  related diagrams corresponding to different slices of the $3$-dimensional bifurcation surface $S\cup B$  have been explored by other authors considering closely related models for the free energy: see~\cite[Figures 1,3]{MS} or~\cite[Figure 3]{LP} and~\cite{AL} for example. As emphasised earlier, however, we do not discuss here which critical points are global minima.
\msk

\begin{figure}[!ht]  
\begin{center}
 \scalebox{.50}{\includegraphics{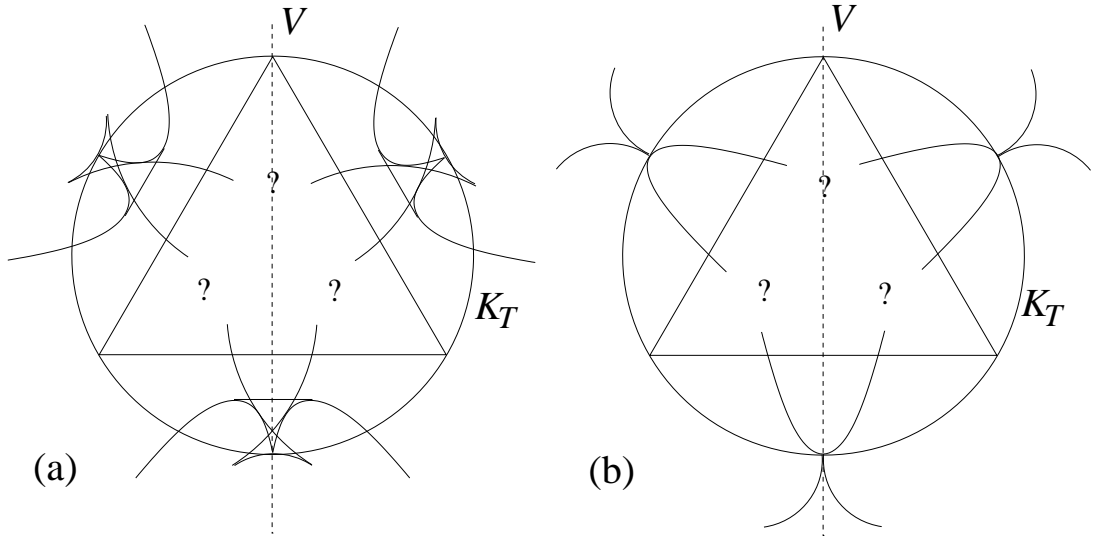}}
\end{center}  
\caption{The bifurcation set close to the three maximally degenerate
  (degree~6) points on the critical circle $K_{T_1}$ in the $\bl$-plane
  $\Lambda$ in the cases (a)~$T<T_1$ and (b)~$T>T_1$.  The question
  marks indicate lack of information on the geometry of the
  bifurcation sets away from their organising centres.}   
\label{fi:3swallow}
\end{figure}

Combining Figure~\ref{fi:fig4neg} with the $3$-fold symmetry in the $\bl$-plane $\Lambda$ we see that the bifurcation set in $\Lambda$ takes the form indicated in Figure~\ref{fi:3swallow} (a) or (b) according as $T<T_1$ or $T>T_1$.
The question marks represent uncertainty as $(T,\bl)$ moves away from
$(T_1,\bl^1)$ or its counterparts: the local bifurcation analysis given earlier
does not determine the intervening geometry.  See, however, \cite[Section  2.3]{TS} (in particular Figure~2.4) for a pertinent discussion.    
\subsubsection*{Fixed $\bl$ with symmetry $\lam_1=\lam_2$}
First choose $\bl$ to lie on the line $V$ (the Sonnet-Virga-Durand limit) given by $\lam_1=\lam_2$, that is $\gamma=0$ in the notation of~\eqref{e:fdef}.  In 
this case it follows from~\eqref{e:rel1} that the whole free energy function (not just the entropy part) is an {\em even} function of $w$, that is of $(d,c)$.
\msk
 
At the point of most interest to us where $T=T_1$ and $\xi=0$ we have from~\eqref{e:newzw2} that $(z,w)=(W,-Z)$, that is
\[
(s,p,d,c)=(y,v,-x,-u)
\]
so that at bifurcation from isotropy $s,p$ are both zero to first order while the \lq active' variables $(x,u)$ represent $(-d,-c)$. By the Splitting Lemma (Appendix~A) the residual function $q$ is an even function of $(d,c)$, justifying the conjecture in~\cite[Section VII]{LN} that this should hold.

The $x'\,$-axis corresponds to the $d\,$-axis and its rotations in the $d,c$-plane by $\pm 2\pi/3$, while the $u'\,$-axis corresponds to the line $d=\sqrt3c$ and its $\pm 2\pi/3$ rotations: it is on these two sets of lines that the $D_3$-orbits of (respectively) uniaxial and biaxial equilibria lie.  The plane containing $V$ and the $T\,$-axis
corresponds to the $e_2,e_4$-plane although (\ref{e:e2term}) shows that the $e_4\,$-axis corresponds to the line $10t=\sqrt{\tfrac32} U_0r$, while it can be deduced from~\eqref{e:e4} and~\eqref{e:T1} that the original $T\,$-axis becomes a curve through $(e_2,e_4)=(0,0)$ with tangent direction $(9a_3^2,8b_4^2)$.  Note also from~\eqref{e:phi3} that with $\xi=0$ the third-order term $\vf_3$ in the square-completing coordinate transformation~\eqref{e:newzw3} vanishes.
\msk

\begin{figure}[!ht]  
\begin{center}
 \scalebox{.60}{\includegraphics{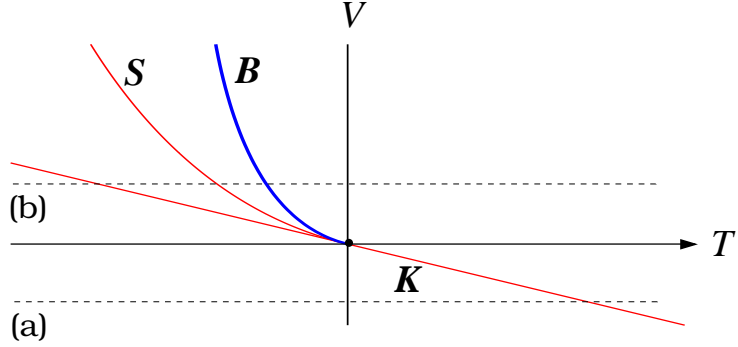}}
\end{center}  
\caption{The bifurcation set in the $T,\bl$-plane with $\bl$
  restricted to the line $V$ of symmetry $\lam_1=\lam_2$.  The dashed
  lines~(a),(b), each representing variation of $T$ for fixed $\bl\in V$, give rise to equilibrium branching diagrams as indicated in Figure~\ref{fi:bifdiag0}.}
\label{fi:symslice}
\end{figure}

\begin{figure}[!ht]  
\begin{center}
 \scalebox{.55}{\includegraphics{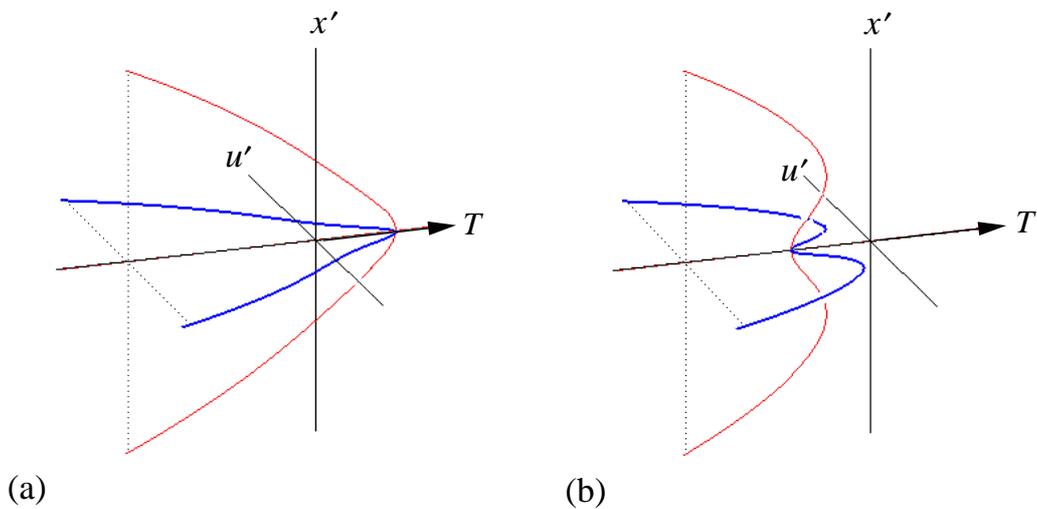}}
\end{center}  
\caption{Bifurcation diagrams corresponding the paths (a) and (b) in 
Figure~\ref{fi:symslice}. The coordinate $x'$ corresponds to
$(d,c)=(-x',0)$ (or $\tfrac12 x'(-1,\pm\sqrt{3})$) while $u'$ corresponds to
$(d,c)=(0,u')$ (or $\tfrac12 u'(\pm\sqrt{3},-1)$).
The light (red) curve in the $(T,x')$-plane
represents uniaxial equilibria while the heavier (blue) curve in the
$(T,u')$-plane represents biaxial equilibria.}
\label{fi:bifdiag0}
\end{figure}

In Figure~\ref{fi:symslice} we show
the intersections of this plane with the swallowtail and bluebird
bifurcation sets $S$ and $B$. The configurations of
critical points of $f_e$ that correspond to the various complementary
regions in the plane are easily deduced from Figures~\ref{fi:fig4pos} and~\ref{fi:fig4neg}: by symmetry the graph of $h_e$ is symmetric about $x'=0$ and the curve $\Gamma'$ is the $u'\,$-axis on which the biaxial equilibria must therefore lie.
For $r>0$ ($\bl$ outside $K_T$) the line~(a) cuts $S$ and $B$ only on the
$e_4\,$-axis, corresponding to simultaneous bifurcation from zero of both uniaxial and biaxial equilibria. For $r<0$ ($\bl$ inside $K_T$) for decreasing $T$ the line~(b) first cuts $B$ at a point of $B_1\subset B$ (see Figure~\ref{fi:figblu}
and  Figure~\ref{fi:figBneg}) corresponding in $(x',u')$ coordinates to simultaneous creation of two pairs of biaxial equilibria, then cuts $S$ at a point corresponding to simultaneous creation of two pairs of uniaxial equilibria, and finally cuts $S$ and $B$ at a common point at which two uniaxial equilibria (one from each pair) and two biaxial equilibria (one from each pair) are simultaneously annihilated at zero.  The corresponding bifurcation (branching) diagrams are shown in Figure~\ref{fi:bifdiag0}.
\msk   

At the opposite point of $V\cap K_T$, that is where $\xi=\tfrac{\pi}2$, we have
$(z,w)=(Z,W)$ so that
\[
(s,p,d,c)=(x,u,y,v)
\]
and the free energy is {\em even} in $W$.  This means that completing the square is superfluous: it suffices to set $W=0$, that is $d=c=0$. (Note that $\vf_2=\vf_3=0$ in~\eqref{e:phi2} and~\eqref{e:phi3}.)  Here we are in the situation of Section~\ref{s:cubicterms}, with uniaxial equilibria given by critical points on the $s\,$-axis and its rotations in the $s,p$-plane by $\pm 2\pi/3$, while biaxial equilibria lie off these lines.
\msk

Note that neither of these descriptions coincides with the analysis of the KKLS model (up to degree~$4$) given in~\cite{LN}. There (following~\cite{SVD}) the assumption is made that $p=d=0$, and conditions are sought for a uniaxial solution with nonzero $s$ to lose stability in the $c$-direction. Our results above show that in the degree~$6$ case $\xi=0$ (where $\lam_1=\lam_2$ with $\lam_3<\tfrac13$) we have $s=p=0$ to first order, while uniaxial equilibria must satisfy $d\ne0$ although biaxial equilibria exist with $d=0$.  At the less degenerate degree~$3$ case $\xi=\tfrac\pi 2$ (where $\lam_3>\tfrac13$) now $d=c=0$ to first order while uniaxial equilibria must satisfy $s\ne0$ although biaxial equilibria exist with $s=0$.
\subsubsection*{Broken symmetry: $\lam_1\ne\lam_2$}
\begin{figure}[!ht]  
\begin{center}
 \scalebox{.70}{\includegraphics{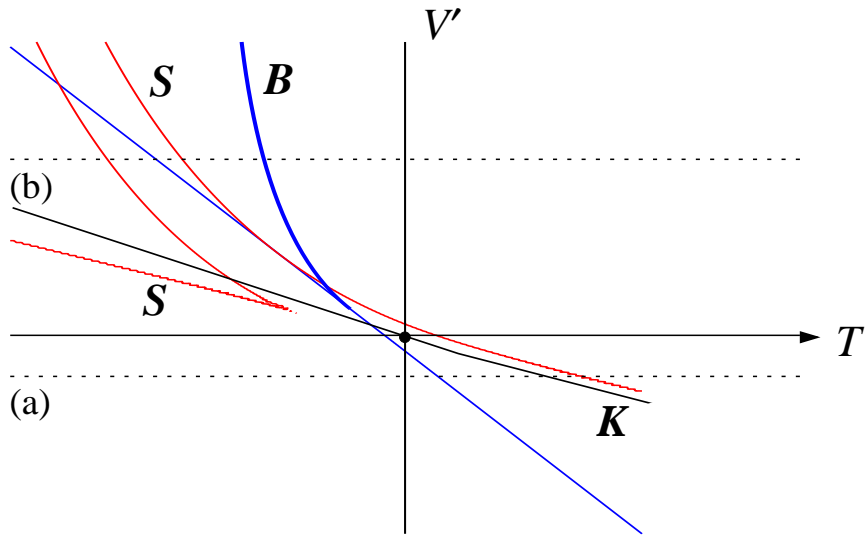}}
\end{center}  
\caption{The bifurcation set in the $T,\bl$-plane with $\bl$ lying on a line $V'$ parallel to but 
  offset from the line $V$ of symmetry $\lam_1=\lam_2$.  The dashed
  lines, each representing variation of $T$ for fixed $\bl$, give rise to
  bifurcation diagrams as indicated in Figure~\ref{fi:bifdiag1}. Note that there are intersections of line~(b) with (as $T$ decreases) first $K$ and then $S$ further to the left of the picture.}
\label{fi:sideslice}
\end{figure}

\begin{figure}[!ht]  
\begin{center}
 \scalebox{.60}{\includegraphics{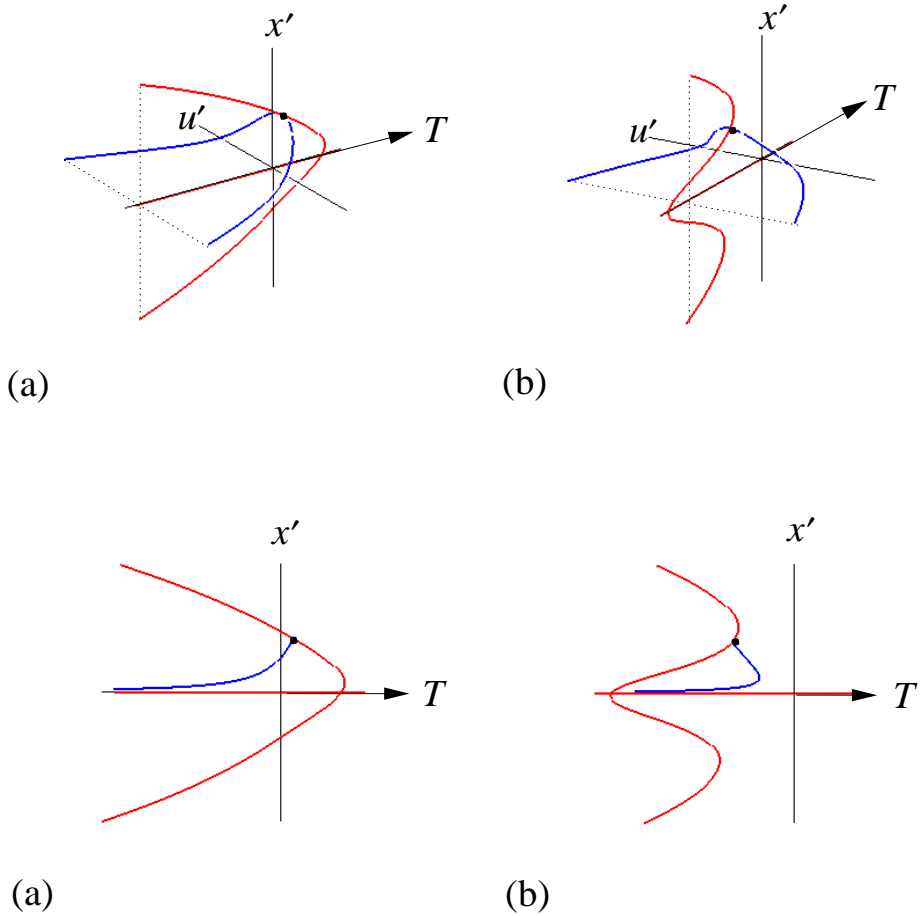}}
\end{center}  
\caption{Bifurcation diagrams corresponding the paths (a) and (b) in 
Figure~\ref{fi:sideslice} together with (lower diagrams) their projections into the $T,x)$-plane.}
\label{fi:bifdiag1}
\end{figure}

If $\bl$ is slightly displaced from $V$ so that 
the $\lam_1,\lam_2$ symmetry is broken the bifurcation set shown in
Figure~\ref{fi:bifdiag0} splits into asymmetric configuration as shown
in~\ref{fi:sideslice}. Some associated branching diagrams are shown in
Figure~\ref{fi:bifdiag1}, together with their profiles in the $T,x'$-plane which show more clearly the order in which branching occurs as $T$ decreases. Note that it remains the case that biaxial equilibria occur in pairs with equal and opposite values of~$u'$. 
\msk

The lower diagram (a) indicates that first a pair of uniaxial equilibria is created at a saddle-node (fold) bifurcation with $x'>0$, and then one these passes through zero (isotropic state) which loses stability.  As $T$ decreases a pair of biaxial equilibria branches from the uniaxial equilibrium with $x'>0$, and for these the value of $x'$ tends to zero as $T$ decreases. Of course each branch corresponds to three symmetrically-placed branches in the original $(x,u)$ coordinates. 
\msk

For the lower diagram (b) the sequence is more complicated.  First, two pairs of biaxial equilibria are created as $T$ decreases at a (double) saddle-node bifurcation.  Next, a pair of uniaxial equilibria is created at a saddle-node bifurcation with $x'>0$.  The \lq inner' pair of biaxial equilibria then coalesces with one of these uniaxial equilibria.  As $T$ decreases further a second saddle-node creation of uniaxial equilibria occurs with $x'<0$, while the origin remains a locally stable equilibrium throughout.  However, a uniaxial equilibrium with $x'>0$ then passes through the origin as the latter equilibrium becomes unstable, and two uniaxial equilibria mutually annihilate at a saddle node with $x'<0$.  Meanwhile the biaxial equilibria have persisted, with $x'$ values tending to zero as $T$ decreases.
\msk

\begin{figure}[!ht]  
\begin{center}
 \scalebox{.60}{\includegraphics{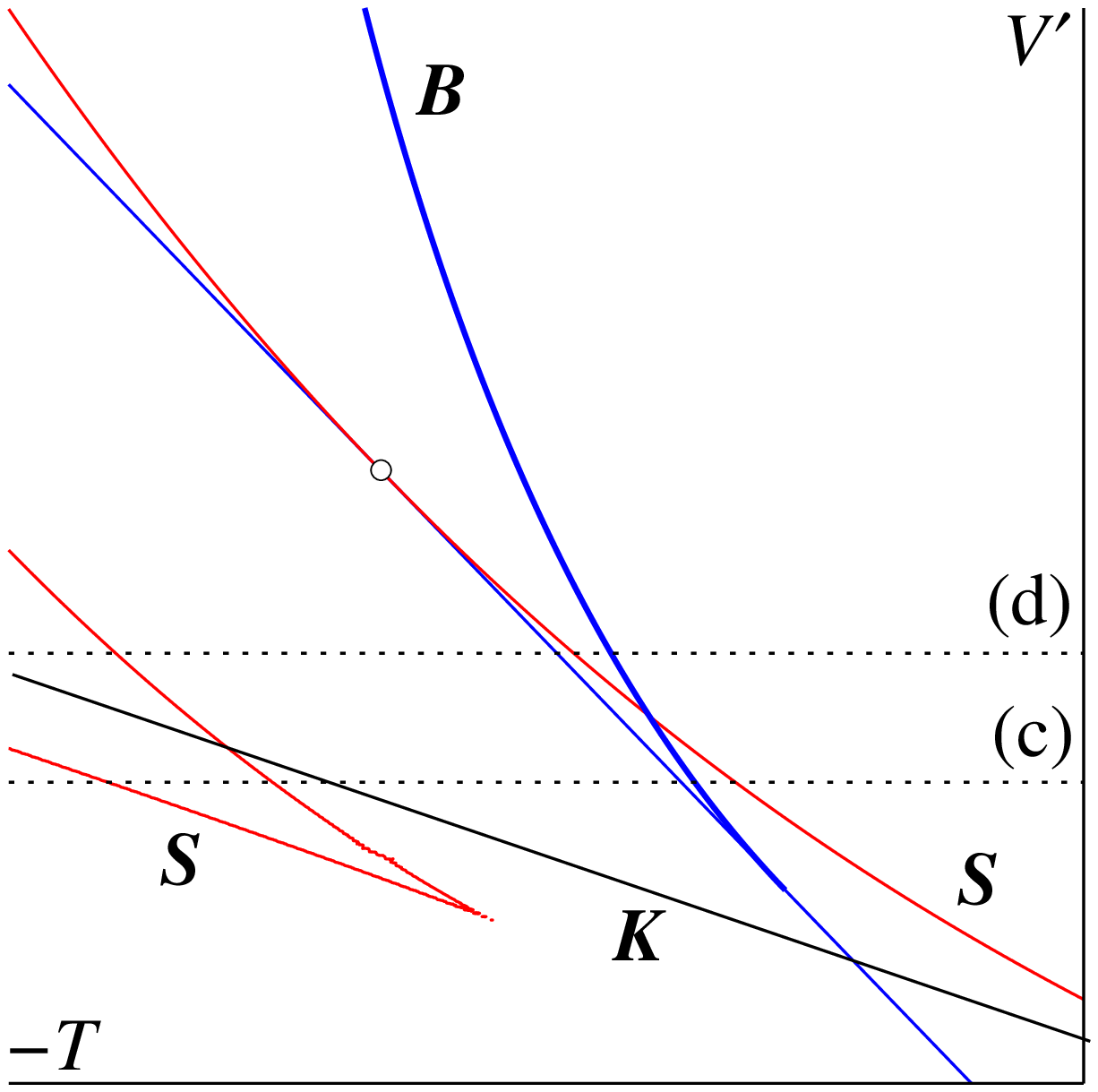}}
\end{center}  
\caption{Zoom to part of Figure~\ref{fi:sideslice}.  The two paths (c) and (d) through the bifurcation set correspond to different sequential order of of uniaxial and biaxial bifurcations as $T$ decreases (further description in text).  The marked point corresponds to simultaneous bifurcation of uniaxial and biaxial equilibria from the same uniaxial state.} 
\label{fi:bifdiagzoom} 
\end{figure}

Figure~\ref{fi:sideslice} being rather congested, we reproduce a zoom image of its central portion in Figure~\ref{fi:bifdiagzoom}.  This indicates the point of tangency of the uniaxial bifurcation set with the biaxial bifurcation set, corresponding (as $T$ decreases) to the simultaneous annihilation of a pair of biaxial equilibria at a point of creation of a pair of uniaxial equilibria. As $T$ decreases the bifurcation sequence along the path (c) is a follows (we omit repetition of \lq equilibrium'): creation of a uniaxial pair, creation of two  biaxial pairs, annihilation of two biaxials at a uniaxial, loss of stability of the origin, creation of a uniaxial pair, annihilation of a uniaxial pair.  Along the path (d) the order of the first two transitions is interchanged, as is the order of the penultimate two.

%
\subsection{Further symmetry-breaking} \label{s:fsymbr}
The preceding study of the swallowtail and bluebird bifurcations assumed $e_5=0$ since, up to certain nonlinear coordinate changes as described in section~\ref{ss:uniax}, there is no loss of generality in doing so as far as overall configurations of critical points of the free energy are concerned.  However, in the present context of phase transitions for liquid crystals the allowed coordinate changes are somewhat too general since the distinguished role of the temperature parameter $T$ may be compromised.  This difficulty may be addressed by building the distinguished character of $T$ into the theory of versal unfolding, using either the techniques of $(r,s)$-unfolding theory of Wasserman~\cite{WG} or those of equivariant bifurcation theory~\cite{GS}.  Unfortunately for our purposes, the former does not explicitly handle  $G$-invariance, while the latter is not attuned to variational problems (critical points of a function).  In both cases the tools are available to fill this gap, but there does not yet appear to be any readily-available account in the literature.
\msk

A further limitation on our analysis is the absence of any significant discussion of absolute (global) minima rather than relative minima. It is the changes of configuration of {\em absolute} minima of the free energy function that correspond to phase transitions in liquid crystals. The mathematical tools to handle this rigorously
involve the use of {\em multijets} (simultaneous Taylor expansions at more than one point), also in the presence of symmetry.  The geometric structures involved are so-called {\em Maxwell sets}~\cite[Ch.2, \S3]{AGLV}.
A general theory of bifurcations of Maxwell sets in the presence of symmetry has yet to be adequately formulated.  Note, however, that the detailed analysis in~\cite{AL} does focus explicitly on Maxwell sets, the bifurcation loci for critical points themselves playing a subsidiary role.
\msk

Without attempting to address these problems here, we limit our discussion to two further observations on symmetry-breaking. In Section~\ref{s:bifnormal} we argued that there was no loss of generality (up to appropriate coordinate changes) in assuming $e_5=0$.  However, if the parameter $T$ is distinguished from the others, then the notion of \lq appropriate' is modified.
In Figure~\ref{fi:swbend}(a) we indicate the bending effect of nonzero $e_5$ on the symmetric swallowtail and bluebird configuration as seen in Figure~\ref{fi:fig4neg}. This has a significant effect on the geometry of the slices of the bifurcation surfaces shown in Figures~\ref{fi:symslice},~\ref{fi:sideslice}, and~\ref{fi:bifdiagzoom}.
Also, in Figure~\ref{fi:swbend}(b) we illustrate the effect on the swallowtail of the inclusion of a linear $x$-term in the function $p_e(x)$ of Section~\ref{ss:uniax} with $e_5=0$ and $e_4<0$, appropriate when considering the effect on phase transitions of the imposition of an electric or magnetic field. Any nonzero linear term makes the full analysis much more difficult because $D_3$-symmetry is broken and it is insufficient to work with the $D_3$-invariant functions $X,Y$ of $x,u$. In particular the inclusion of a nonzero $u$-term in the free energy moves the uniaxial critical points off the $x\,$-axis, which complicates the picture even further.
\msk

A complete singularity-theory analysis of the critical point structure of the free energy and its implications for liquid crystal phase transitions would need to incorporate all these aspects and is pursued no further here. Nevertheless, although a complete account of phase transitions arising through bifurcation from isotropy, under the assumptions of the KKLS model~\cite{KK,LN}, has yet to be written it in principle ought to be obtainable by combining the methods of the present paper with those of~\cite{AL} and~\cite{AGLV}.
\begin{figure}[!ht]  
\begin{center}
 \scalebox{.50}{\includegraphics{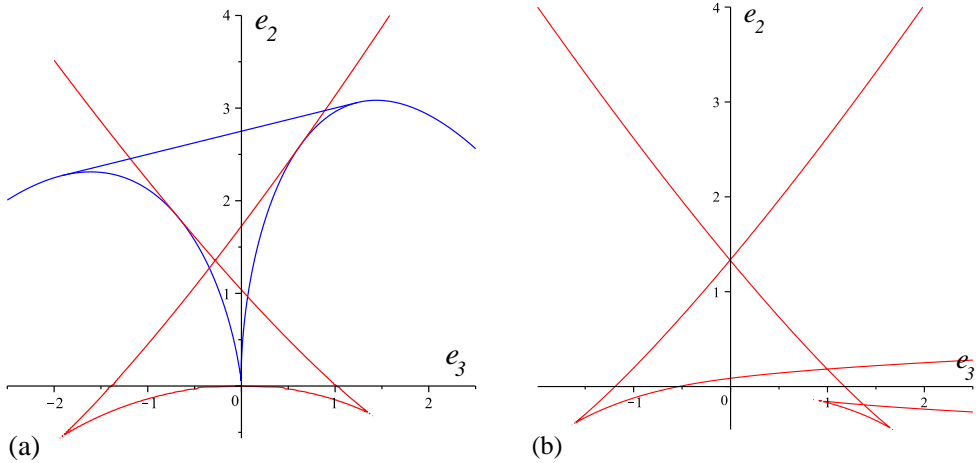}}
\end{center}  
\caption{The effect of (a) nonzero (positive) $e_5$ on the swallowtail and bluebird configurations and (b) nonzero (positive) $x$-term on the swallowtail configuration, both for negative $e_4$.}
\end{figure}  \label{fi:swbend}
\section{Conclusion}
We have attempted to chart a rigorous path from the molecular field theory KKLS model to the bifurcation geometry of a $D_3\,$-invariant function on~$\R^2$, using the language and tools of group actions and equivariant singularity theory, since although many of these ideas and methods are implicit in the literature, sometimes expressed in terms of physics, there has not appeared to be any clear mathematical account of the whole framework.  The results presented here provide a geometric setting in which to study the complete bifurcation behaviour of critical points of the free energy function close to isotropy. The further inclusion into this picture of the Maxwell set geometry, characterising the competition between critical points at the same energy value, will then yield a full map of the associated phase transitions.
\msk

\noindent{\em Acknowledgements}. The author is much indebted to Paolo Biscari, Giovanni De Matteis, Chuck Gartland, Lech Longa, Nigel Mottram, Stefano Turzi, Epifanio Virga and many other liquid crystal experts for numerous valuable and instructive conversations, and especially to Tim Sluckin for highlighting liquid crystal phase transitions as a problem in symmetric bifurcation and for continual inspiration over at least a decade.  Thanks are also due to Reiner Lauterbach, Cormac Long and Gareth Jones for helpful discussions about the $72$-element group $D_3\wr Z_2$ and its representation theory, and to Terry Wall for help with $G$-determinacy.  The author is grateful to the Isaac Newton Institute, Cambridge for supporting the Programme on the Mathematics of Liquid Crystals in 2013, to the Leverhulme Trust for an Emeritus Research Fellowship, and to the Basque Center for Applied Mathematics (BCAM), Bilbao for a hospitable environment enabling completion of the paper.
\bsk

------------------------------------------------------------------------------
\section*{Appendix A: The $G$-invariant Splitting Lemma}
We first prove the existence of a $G$-invariant splitting, and then show the uniqueness of the residual function up to a $G$-equivariant coordinate change. Although the extension of these \lq standard' results to the $G$-invariant case is often assumed to be automatic
(note, for example, the final sentence of~\cite{GM}) there is a subtlety in the completion of the uniqueness proof.
\subsection{Existence of a $G$-invariant splitting} 
Let the compact group $G$ act on the finite-dimensional linear space $E$, and suppose $f:E\to \R$ is a smooth $G$-invariant function defined on a neighbourhood of the origin in $E$.
Assume $f(0)=0$ and $df(0)=0$, and that the Hessian $H(0)$ of $f$ at the origin has kernel~$K$.  Then $K$ is invariant under the action of $G$, and $K$ has a $G$-invariant complement $N$.  We take coordinates $(u,v)\in K\oplus N=E$.
\begin{lemma}   \label{l:splitting} 
\emph{(Splitting Lemma)}
There is a $G$-equivariant local diffeomorphism at the origin in $E$ transforming $(u,v)$ into coordinates $(u,w(u,v))$ such that $f(u,v)=f^*(u,w)$ with 
\begin{equation}  \label{e:splitform}
f^*(u,w)=\tfrac12w^TAw+q(u)
\end{equation}
where $A$ is the Hessian of $f|N$ at the origin and where $q:K\to\R$ is a smooth $G$-invariant function with $q(0), dq(0)$ and $d^2q(0)$ all zero. \label{l:split}
\end{lemma}
\proof  The IFT applied to $\partial_vf:K\oplus N\to N$ implies that the locus of zeros of $\partial_vf$ is locally the graph of a unique smooth map $h:K\to N$ with $h(0)=0$ and $Dh(0)=0$.  Moreover, the uniqueness immediately yields that $h$ is equivariant, since
if $\partial_vf$ vanishes at $(u,h(u))$ it does so at $(gu,gh(u))$ for all $g\in G$ and so $gh(u)=h(gu)$.  Replacing $(u,v)$ by the new variables $(u,v-h(u))$ (a $G$-equivariant coordinate change) we may thus suppose $f$ satisfies $f(0,0)=0$ and $\partial_vf(u,0)=0$ for small $u$.
\ssk

We next apply Taylor's formula (cf.~\cite{CH},\cite{HO}) to write 
\begin{equation} \label{e:Beq}
f(u,v)-f(u,0)=\tfrac12 v^TB(u,v)v
\end{equation}
where $B(u,v)$ is a symmetric matrix varying smoothly with $(u,v)$.  In a star-shaped neighbourhood of the origin, $B(u,v)$ is given explicitly by
\begin{equation}
B(u,v)=2\int_{t=0}^1\int_{s=0}^tH(u,sv)\,dsdt=2\int_{s=0}^1(1-s)H(u,sv)ds
\end{equation}
where $H$ is the Hessian of $f|N$,
as follows easily from the observation that if $k(v)$ is any smooth function that vanishes at $v=0$ then $k(v)=\int_0^1\frac{d}{dt}k(tv)dt$.  The map
$B$ from $E$ to the space $Sym(m)$ of symmetric $m\times m$ matrices ($m=\dim N$)
is equivariant with respect to the given $G$-action on $E=K\oplus N$ and the associated action by conjugacy on $Sym(m)$ because the same is automatically true of $H$.
\ssk

The final step is to construct a smooth map $R:E\to L(N)$, equivariant with respect to
the $G$-action on $E$ and the conjugacy $G$-action on $L(N)$, with the property that $R(0)=I$ and 
\[
B(u,v)=R(u,v)^TAR(u,v)
\]
for all $(u,v)$ close to the origin. Writing $w=R(u,v)v$ then converts $f(u,v)$ into $f^*(u,w)$ where by~(\ref{e:Beq}) we have 
\[
f^*(u,w)=f(u,0)+\tfrac12 w^TAw
\]
which completes the proof of the Splitting Lemma: we note that the coordinate transformation
$(u,v)\mapsto(u,w)$ is $G$-equivariant as 
\[
(gu,gv)\mapsto(gu,R(gu,gv)gv)=(gu,gR(u,v)v)=(gu,gw),
\]
and is locally invertible at the origin by the IFT since its derivative there is 
the identity map.
\ssk
 
The map $R$ is constructed by finding an equivariant local right inverse $\Psi$ to the $G$-equivariant map
\[
\Phi:L(N)\to Sym(N):R\mapsto R^TAR
\]
which we note takes $I$ to $A$, and then defining $R(u,v)=\Psi\,B(u,v)$ so that $\Phi\,R(u,v)=B(u,v)$.  To find $\Psi$ we observe that the derivative
\[
D\Phi(I):L(N)\to Sym(N):S\mapsto SA + AS^T
\]
is surjective (as $\tfrac12 A^{-1}C\mapsto C\in Sym(n)$) and its kernel is $L^{-1}(Skew(m))$ where
$L$ is the linear isomorphism
 \[
L:L(N)\to L(N):S\mapsto SA.
\]
A complement to the kernel is given by $P=L^{-1}(Sym(N))$, that is the set of matrices $S$ for which $SA-AS^T=0$.  Moreover, the linear subspace $P$ of $L(N)$ is invariant under the $G$-action of conjugation.  Therefore by the IFT the map $\Phi|P:P\to Sym(N)$ is locally invertible, and its inverse is a $G$-equivariant right inverse to $\Phi$ as required.  \endproof
\msk

\rem  In the absence of a nontrivial $G$-action, {\em any} right inverse to $\Phi$ may be chosen, as in~\cite{HO}. In~\cite{CH} the map $\Phi$ is restricted to the space $T(m)$ of upper triangular $m\times m$ matrices, easily checked to be injective and hence an isomorphism from $T(m)$ to $Sym(N)$.  Thus neither method immediately provides a $G$-invariant right inverse to $\Phi$ as is required here.  The argument in~\cite{WA} is complete but buried in deeper topological considerations. 
\subsection{Uniqueness of the residual function} \label{ss:Aunique}
It is not immediately apparent that the {\em residual function} $q:K\to\R$ in Lemma~\ref{l:split}  is essentially unique, that is
that if we have two possible expressions for $f$ in the form~(\ref{e:splitform}) then the two residual functions are $G$-right equivalent (i.e. can be converted into each other by $G$-equivariant coordinate changes in $K$).  A proof in the absence of a $G$-action is given in~\cite{CH}.  Here we give a quite different proof, using an argument pointed out in the 1970s by David Kirby.
\begin{lemma}
Suppose there is a $G$-equivariant local diffeomorphism 
\[
\phi:K\oplus N\to K\oplus N: (u,w) \mapsto (\tu,\tw)
\]
such that
\begin{equation} \label{e:bothsides}
\tfrac12\,w^TAw + q(u) = \tfrac12\,\tw^TA\tw + \tq(\tu).
\end{equation}
with $A$ nonsingular. Then there is a $G$-equivariant local diffeomorphism $\chi:K\to K$ such that $q(u)=\tq(\chi(u))$.
\end{lemma} 
\proof  First partition the derivative
$D\phi(0,0)$ into four $2\times2$ blocks according to the coordinate splitting $(u,w)$:
\[
D\phi(0,0)= \begin{pmatrix}  P & Q \\ R & S \end{pmatrix}.
\]
We then see that $S^TAS=A$ and $R^TAS=0$, so $S$ is nonsingular and $R=0$.  
\ssk

Now consider the equation
\[
\psi(u,w):=\tw(u,w)+Sw=0 \in N.
\]
Since $\partial_w\psi(0,0)=2S$ the IFT implies that there is a unique smooth function
$\ell:K\to N$ close to the origin such that $\psi(u,\ell(u))=0$. Thus
\[
\tw(u,\ell(u))^TA\tw(u,\ell(u)) = \ell(u)^TSAS\ell(u) = \ell(u)A\ell(u),
\]
and then substituting $\tw=\tw(u,\ell(u))$ into~(\ref{e:bothsides}) gives
\[
q(u)=\tq(\tu(u,\ell(u))).
\]
The map $\chi:K\to K:u\mapsto\tu(u,\ell(u))$ is a local diffeomorphism since
$D\ell(0)=-\tfrac12  S^{-1}Q=0$. Finally, the $G$-invariance of $\chi$ is immediate from that of $\ell$, which itself follows from
the $G$-equivariance of $\psi$ and the uniqueness clause of the IFT.  \endproof
%
\section*{Appendix B: Determinacy with $D_3$ symmetry}
Using the determinacy criteria set out (in formal language) by Bruce
{\it et al.}~\cite{BPW}, we here give a proof of Proposition~\ref{p:determinacy}.  As in Section~\ref{ss:deter} we simplify terminology by dropping explicit reference to $G$ and $\RR$ in equivalence and $k$-determinacy.
\msk

From the general theory of $G$-invariant functions and $G$-equivariant maps or
vector fields~\cite{SC},\cite{PO},\cite{MA} we know, given an action of $G$ on $\R^n$, not only that there exists a (finite) Hilbert basis for the ring of $G$-invariant polynomials, so that every smooth $G$-invariant function on $\R^n$ can be written as a smooth function of these basis polynomials, but also that there exists a finite basis for the module of $G$-equivariant vector fields on $\R^n$, meaning that there is a finite set of $G$-equivariant vector
fields $\{V_1,\ldots,V_\ell\}$ such that {\it every} $G$-equivariant vector
field $V$ on $\R^n$ can be written as
\[
V(x)=g_1(x)V_1(x)+\cdots+g_\ell(x)V_\ell(x)
\]
for some $G$-invariant functions $g_1,\ldots,g_\ell$.
\msk

Given a $G$-invariant function $f:\R^n\to\R$, let $\WW(f)$ denote the set of all functions $w:\R^n\to\R$ that in  some \nhd of the origin are of the form
$$w(x)=df(x)V(x)$$ 
for some $G$-equivariant vector field $V$, and let $\WW_0(f)\subset\WW(f)$ be the subset for which $V$ has no linear terms (this restriction relates to the 
{\it unipotency} in the title of~\cite{BPW}).  
The function $w$ is automatically $G$-invariant. The methods
of~\cite{BPW} then show
\begin{prop}
The function $f$ is equivalent on some \nhd of the origin to precisely those functions of the form $f+w$ where $w\in\WW_0(f)$.  Therefore
$f$ is $k$-determined at the origin if and only if its $k$-jet $j_kf$ is such that $\WW_0(j_kf)$ contains \emph{all} $G$-invariant functions of degree higher than~$k$.
\end{prop}
We illustrate this algebra in the case of the group $G=D_3$ acting in the usual way on $\R^2$, with Hilbert basis $\{X,Y\}$ where
\begin{align}
X(x,y)&=x^2+y^2 \\
Y(x,y)&=x^3-3xy^2.
\end{align} 
It is not hard to verify using methods similar to those of Section~\ref{s:invs} that a basis for the $G$-equivariant vector fields is given by $\{V_1,V_2\}$ where
\[
V_1(x,y)=\begin{pmatrix} x \\y \end{pmatrix},  \qquad
V_2(x,y)=\begin{pmatrix} x^2-y^2 \\-2xy \end{pmatrix}.
\]
We then find $dX\,V_1=2X,\,dX\,V_2=2Y,\,dY\,V_1=3Y,\,dY\,V_2=3X^2$ and so for example
\begin{align*}
\WW_0(X)&=[X^2,Y]  \\
\WW_0(Y)&=[XY,X^2,Y^2] \\
\WW_0(XY)&=[X^2Y,XY^2,X^3+Y^2]
\end{align*}
where the notation $[H,K,\ldots]$ here means all \lq linear' combinations
of the polynomials $H,K,\ldots$ where the coefficients are 
$G$-invariant polynomials.  
Since every $G$-invariant polynomial in $x,y$ of degree
at least $3$ must have the form $aX^2+bY$ for some $G$-invariant
polynomials $a,b$ it follows that any function with $2$-jet (a nonzero scalar multiple of) $X$ is $2$-determined.  In other words, any $G$-invariant function of the form
\[
f(x,y)=x^2+y^2 \,\,+\,\text{higher order terms}
\]
can be transformed to $x^2+y^2$ (with {\it no} higher order terms) by
a $G$-equivariant change of coordinates in $\R^2$ (the $G$-equivariant Morse Lemma~\cite{AR}).  Likewise, every  
$G$-invariant polynomial in $x,y$ of degree
at least $4$ must have the form $aXY+bX^2+cY^2$ for some $G$-invariant
polynomials $a,b,c$, so it follows that any function with $3$-jet $Y$ is $3$-determined.

On the other hand, a function with $4$-jet $X^2$ may {\it not} be $4$-determined, because every member of $\WW_0(X^2)$ is a multiple of $X$ and so e.g. $Y^2\notin \WW_0(X)$.  
Similarly $\WW_0(XY)$ contains neither $X^3$ nor $Y^2$ and so a function with $5$-jet $XY$ may {\it not} be $5$-determined.

We consider the case of polynomials of homogeneous degree 6 a little more carefully.  Let $f_6=mX^3+nY^2$ where $m,n\in\R$.
Since
\[
df_6=3mX^2\,dX + 2nY\,dY
\]
we have
\begin{align*}
df_6.XV_1&=Xdf_6.V_1=6mX^4+6nXY^2  \\
df_6.YV_1&=Ydf_6.V_1=6mX^3Y+6nY^3  \\
df_6.V_2&=6(m+n)X^2Y
\end{align*}
so that if $mn(m+n)\ne0$ we have
\[
\WW_0(f_6)=[X^2Y,mX^4+nXY^2,Y^3].
\]
From this it follows that $\WW_0(f_6)$ contains all
$G$-invariant polynomials of homogeneous degree at least $9$ and so $f$ is 
$8$-determined if its $8$-jet is $f_6$.  However, if $j_8f=f_6+X^4$
then $f$ is not $6$-determined. Nevertheless, $\WW_0(f_6)$ contains the unique (up to scalar multiple) $G$-invariant polynomial $X^2Y$ of degree~$7$, and so $f_6+cX^2Y$ is equivalent to $f_6$. Moreover, a slight refinement of the argument using Nakayama's Lemma from commutative algebra (cf.~\cite{CH},~\cite{PS} for example) shows that 
\[
\WW_0(f_6+X^4)=\WW_0(f_6)
\]
so that if $j_8f=f_6+X^4$ then $f$ is $8$-determined.

Since every member of $\WW_0(XY^2)$ contains $Y$ it follows that if $j_7f=XY^2$ then $f$ may not be $7$-determined.  However, we find that if $f_8=X^4+XY^2$ then
\[
\WW_0(f_8)=[X^5,X^3Y,X^2Y^2,Y^3]
\]
which contains all $G$-invariant polynomials of degree~$\ge 8$ and so if $j_8f=f_8$ then $f$ is $8$-determined. This completes the argument for the proof of Proposition~\ref{p:determinacy}.
\section*{Appendix C: Versal deformation}
One of the principal notions in singularity theory is that if a function $f$ is finitely $\R$-determined (that is, $k$-$\R$-determined for some $k$) at a point, then it can be placed in a family of functions with finitely many parameters which captures {\em all possible} local perturbations of $f$ in some \nhd of that point: this applies in a wide range of contexts for functions and mappings with various notions of equivalence up to coordinate change, and in particular applies to $G$-invariant functions under $G$-$\R$-equivalence~\cite{DA},\cite{PO}.  We now formulate this a little more precisely. All functions and maps are taken to be~$C^\infty$.
\msk

A real-valued function $H$ defined on a neighbourhood of the origin in $\R^n\times\R^s$ is called an $s$-parameter {\em deformation} of the function $h:=H(\cdot,0)$ on a \nhd of the origin in $\R^n$.
Let $K$ be an $\ell$-parameter deformation of the same function $h$, where 
$\del=(\del_1,\ldots,\del_t)\in\R^t$.  Then $K$ is said to be {\em induced} by $H$ if $K$ can be expressed in terms of $H$, that is to say there is a map
\[
(x,\del)\mapsto (\til x(x,\del),\til\eps(\del))
\]
such that
\[
H(\til x,\til\eps)\equiv K(x,\del)
\]
close to the origin, the map $x\mapsto(\til x,\del)$ being a local diffeomorphism for $\del=0$ and hence for all small $\del$.
The deformation $H$ of $h$ is called $\RR$-{\em versal} if {\em every} deformation of $h$ is induced by $H$ in this way. Moreover, if $h$ is invariant with respect to the action of a group $G$ on $\R^n$, and versality holds even when the map $x\to(\til x,\del)$ is required to be equivariant with respect to this $G$-action, then $H$ is $G$-$\RR$-{\em versal}.
\msk

Remarkably, versal deformations typically exist, and there are explicit algebraic criteria for recognising and constructing them.
For $\RR$-versality and without the $G$-action this is the material of
elementary catastrophe theory (where versal deformations are more
often called {\em universal unfoldings}): see \cite{CH},\cite{G},\cite{PS} for
example. Expositions in the $G$-invariant setting are generally highly technical; the result we need can be found in~\cite{DA}~\cite{PO}.
\begin{theo} \label{t:versal}
Let $\{v_1,\ldots,v_r\}$ be a set of $G$-invariant polynomials with the property that {\em every} $G$-invariant polynomial $p$ can be written in the form
\[
p=w \,+\, \eps_1v_1 + \cdots + \eps_rv_r
\]
for some coefficients $\eps_1,\ldots,\eps_r\in\R$, where $w\in\WW(h)$  as in Section~\ref{ss:deter}.  Then
\[
H:=k +  \eps_1v_1 + \cdots + \eps_rv_r
\]
is a $G$-$\RR$-versal deformation of $h$. 
\end{theo} 
First we apply this to the function $h=Y$ as in Appendix~B. We have $\WW(Y)=[Y,X^2]$ and so we may take $\{v_1,v_2\}=\{1,X\}$ and a $G$-$\RR$-versal
deformation of $Y$ is given (renaming the coefficients) by
\[
H(X,Y,e_0,e_2) = e_0 + e_2X + Y.
\]
Next we apply the Theorem to $h=f_6=mX^3+nY^2$. From calculations in Appendix~B we see that 
\[
\WW(f_6)=[X^2Y,mX^3+nY^2]
\]
and so a $D_3$-$\RR$-versal deformation of $f_6$ is given by 
\begin{equation}  \label{e:F60}
H(X,Y,e_0,e_2,\ldots,e_6,e_8)= e_0 + e_2X + e_3Y + e_4X^2+e_5XY+e_6X^3+e_8X^4+ f_6
\end{equation} 
as claimed in Proposition~\ref{p:versaldef}.
\section*{Appendix D: Proof of the spanning property of the conjugacy action of $SO(3)$ on $V=Sym_0(\R^3)$}
For simplicity of notation we denote the operation of conjugation of matrices in $V$ by the matrix $R\in SO(3)$ by $\wt R$: thus $\wt RA=RAR^T$. Let $A_1,\ldots,B_3$ be the following matrices forming a linear basis for $V$:
\begin{gather}
A_1={\mbox{\scriptsize $\begin{pmatrix} 1 & 0 & 0 \\ 0 & -1 & 0 \\ 0 & 0 & 0  \end{pmatrix}$}},\quad
A_2={\mbox{\scriptsize $\begin{pmatrix} 0 & 0 & 0 \\ 0 & 1 & 0 \\ 0 & 0 & -1  \end{pmatrix}$}},\notag \\
B_1={\mbox{\scriptsize $\begin{pmatrix} 0 & 0 & 0 \\ 0 & 0 & 1 \\ 0 & 1 & 0  \end{pmatrix}$}},\quad
B_2={\mbox{\scriptsize $\begin{pmatrix} 0 & 0 & 1 \\ 0 & 0 & 0 \\ 1 & 0 & 0  \end{pmatrix}$}},\quad
B_3={\mbox{\scriptsize $\begin{pmatrix} 0 & 1 & 0 \\ 1 & 0 & 0 \\ 0 & 0 & 0   \end{pmatrix}$}}. \notag
\end{gather}
Observe that these matrices are mutually orthogonal, with the exception that
$A_1{\cdot}A_2\ne 0$.
Also let $K_0$ be the identity matrix and $K_1,K_2,K_3$ denote the following matrices in $SO(3)$:
\[
K_1={\mbox{\scriptsize $\begin{pmatrix} 1 & 0 & 0 \\ 0 & -1 & 0 \\ 0 & 0 & -1  \end{pmatrix}$}},\quad
K_2={\mbox{\scriptsize $\begin{pmatrix} -1 & 0 & 0 \\ 0 & 1 & 0 \\ 0 & 0 & -1  \end{pmatrix}$}},\quad
K_3={\mbox{\scriptsize $\begin{pmatrix} -1 & 0 & 0 \\ 0 & -1 & 0 \\ 0 & 0 & 1  \end{pmatrix}$}}.
\]
Now consider the operator $L_j=\tfrac12(\wt K_0-\wt K_j)\in L(V), j=1,2,3$.  We find $L_j(A_1)=L_j(A_2)=L_j(B_j)=0$ since $\wt K_j$ fixes $A_1,A_2,B_j$, while
\[
L_j(B_k)=B_k\,, k\ne j.
\]  
Therefore if
\[
M_1:=\tfrac12(-L_1+L_2+L_3)
\]
we see that $M_1(B_1)=B_1$ while $M_1$ takes $A_1,A_2,B_2,B_3$ to zero, so that
$M_1$ represents orthogonal projection of $L(V)$ onto the $B_1\,$-axis.
Likewise we construct $M_2,M_3$ representing orthogonal projections to the $B_2\,$- and $B_3\,$-axes respectively.
\ssk

Since $B_2$ and $A_1$ have the same eigenvalues there exists $N\in SO(3)$ such that $\wt NB_2=A_1$. As $\wt N$ is an orthogonal transformation of $V$ (Proposition~\ref{p:inner}) it follows that the operator $N_2:=\wt NM_2\wt N^{-1}\in L(V)$ is orthogonal projection onto the $A_1\,$-axis that therefore annihilates $B_1,B_2,B_3$ but does not annihilate $A_2\,$: in fact $N_2A_2=-\tfrac12A_1$. However, we can also construct the analogous operator $N_1$ that exchanges the roles of $A_1,A_2$, so suitable linear combinations of $N_1,N_2$ represent orthogonal projections to the $A_1\,$- and $A_2\,$-axes. Finally, since all of $A_1,\ldots,B_3$ are conjugate to each other, a linear combination of operators $\wt R$ for $R\in SO(3)$ can be found to take any one of these five to any other while annihilating the rest. Thus $\{\wt R:R\in SO(3)\}$ spans $L(V)$.
\ssk

It remains to show that the set $\{\wt R:R\in SO(3)\}$ does not lie in a proper affine subspace of $L(V)$.  For this, we observe that if
\[
I_1=K_0={\mbox{\scriptsize $\begin{pmatrix} 1 & 0 & 0 \\ 0 & 1 & 0 \\ 0 & 0 & 1  \end{pmatrix}$}},\quad
I_2={\mbox{\scriptsize $\begin{pmatrix} 0 & 1 & 0 \\ 0 & 0 & 1 \\ 1 & 0 & 0  \end{pmatrix}$}},\quad
I_3={\mbox{\scriptsize $\begin{pmatrix} 0 & 0 & 1 \\ 1 & 0 & 0 \\ 0 & 1 & 0  \end{pmatrix}$}}
\]
then 
\[
(\wt K_0 + \wt K_1 +\wt K_2 + \wt K_3)(\wt I_1 + \wt I_2 + \wt I_3)=0
\]
and so there exists a linear combination of elements $\wt N\in L(V)$ with {\em positive} coefficients that gives zero.  This completes the proof.




\begin{thebibliography}{99}
%
\bibitem{AL}
Allender, D, and Longa, L.,
Landau-de Gennes theory of biaxial nematics reexamined,
{\em Phys. Rev. E} {\bf 78} (2008), 11704.
%
\bibitem{AR}
Arnold, V.I.,
Wave front evolution and equivariant Morse lemma,
{\em Comm. Pure Appl. Math.} {\bf 29} (1976), 557--582.
%
\bibitem{AGLV}
Arnold, V. I., Goryunov, V. V., Lyashko, O. V. and Vasiliev, V. A., 
{\em Singularity Theory II. Classification and Applications} Dynamical systems VIII, Encyclopædia Math. Sci, 39, Springer 1993.
%
\bibitem{AGV}
Arnold, V.I., Gusein-Zade, S.M. and Varchenko, A.N.,
{\em Singularities of Differentiable Maps, Vol.I},
Birkh\"auser 1985.
%
\bibitem{BL}
Bates, M. and Luckhurst, G.L.,
Biaxial nematic phases and V-shaped molecules: A Monte Carlo simulation study,
{\em Phys. Rev. E} {\bf 72} (2005), 051702--051716.
%
%
\bibitem{BD}
Br\"ocker, T. and tom Dieck, T.,
{\em Representations of Compact Lie Groups},
Springer 1985.
%
\bibitem{BPW}
Bruce, J.W., du Plessis, A.A. and Wall, C.T.C.,
Determinacy and unipotency,
{\em Invent. Math.} {\bf 88} (1987), 521--554.
%
\bibitem{CH}
Castrigiano, D.P.L. and Hayes, S.A.,
{\em Catastrophe Theory, 2nd ed.},
Westview Press 2004.
%
\bibitem{CW}
Chillingworth, D.R.J., Vicente Alonso, E. and Wheeler, A.A.,
Geometry and dynamics of a nematic liquid crystal in a uniform shear flow,
{\em J. Phys.} A {\bf 34} (2001), 1393--1404. 
%
\bibitem{CLT}
Chillingworth, D.R.J., Lauterbach, R. and Turzi, S.S.,
Molien series and low-degree invariants for a natural action
of $SO(3)\wr Z_2$ ({\em in preparation})
%
\bibitem{CL}
Chossat, P. and Lauterbach, R., {\em Methods in Equivariant
Bifurcations and Dynamical Systems}, World Scientific 2000.
%
\bibitem{DA}
Damon, J.,
The unfolding and determinacy theorems for subgroups of $\A$ and $\KK$,
{\em Proc. Symp, Pure Math.} {\bf 40} (1983), pp 233--254.
%
\bibitem{DG}
de Gennes, P. G. and Prost, J.,
{\em The Physics of Liquid Crystals},
2nd ed., Clarendon Press 1993.
%
\bibitem{MR}
De Matteis, G., Romano, S. and Virga, E. G.,
Bifurcation analysis and computer simulation of biaxial liquid crystals,
{\em Phys. Rev. E} {\bf 72} (2005), 041706 (13 pages).
%
\bibitem{MS}
De Matteis, G., Sonnet, A. M. and Virga, E. G.,
Landau theory for biaxial nematic liquid crystals with two order parameter tensors,
{\em Continuum Mech. Thermodyn.} {\bf 20} (2008), 347--374.
%
\bibitem{MV}
De Matteis, G. and Virga, E. G.,
Tricritical points in biaxial liquid crystal phases,
{\em Phys. Rev. E} {\bf 71} (2005), 061703 (8 pages).
%
\bibitem{DS}
Dunmur, D. and Sluckin, T,,
{\em Soap, Science and Flat-Screen TVs: a History of Liquid Crystals},
Oxford University Press 2010.
%
\bibitem{DT}
Dunmur, D. and Toriyama, K.,
in {\em Physical Properties of Liquid Crystals},
Demus, D., Goodby, J., Gray, G.W., Spiess, H.-W., Vill, V (Eds),
Wiley-VHC 1999, 87--101. 
%
\bibitem{AE}
Edmonds, A. R., {\em Angular Momentum in Quantum Mechanics}, 3rd printing, with corrections, 2nd ed., Princeton University Press 1974.
%
\bibitem{FI}
Fisch, M.R.,
{\em Liquid Crystals, Laptops and Life},
World Scientific 2004.
%
\bibitem{GAP}
The GAP SmallGroups library,
{\tt www.gap-system.org/Packages/sgl.html} (accessed 14.05.13).
%
\bibitem{G}
Gilmore, R,
{\em Catastrophe Theory for Scientists and Engineers},
Wiley 1981.
%
\bibitem{GG}
Golubitsky, M. and Guillemin, V.,
{\em Stable Mappings and their Singularities},
Springer 1974.
%
\bibitem{GM}
Gromoll, D. and Meyer, W.,
On differentiable functions with isolated critical points,
{\em Topology} {\bf 8} (1969), 361--369.
%
\bibitem{GS}
Golubitsky, M., Stewart, I. and Schaeffer, D.G.,
{\em Singularities and Groups in Bifurcation Theory},
Springer 1988.
%
\bibitem{HO}
H\"ormander, L.,
{\em The Analysis of Linear Partial Differential Operators III},
Springer 1985.
%
\bibitem{KK}
Katriel, J., Kventsel, G.F., Luckhurst, G.R. and Sluckin, T.J.,
Free energies in the Landau and molecular field approaches,
{\em Liquid Crystals} {\bf 1} (1986), 337--355.
%
\bibitem{KP}
Kraft, H. and Procesi, C.,
{\em Classical Invariant Theory}, unpublished 1996.
%
\bibitem{LS}
Lauterbach, R. and Sanders, J.,
Bifurcation analysis for spherically symmetric systems using invariant theory,
{\em J. Dynam. Differential Equations} {\bf 9} (1997), 535--560.
%
\bibitem{LP}
Longa, L., Paj\c{a}k, G. and Wydro, T.,
Stability of biaxial nematic phase for systems with variable molecular shape anisotropy,
{\em Phys. Rev. E} {\bf 76} (2007) 011703 (6 pages).
%
\bibitem{LU}
Luckhurst, G.R.,
Molecular field theories of nematics, in {\em The Molecular Physics of Liquid Crystals}, Luckhurst, G. R. and Gray, G. W. (Eds), Chapter 4, Academic Press 1979.
%
\bibitem{LN}
Luckhurst, G.R., Naemura, S., Sluckin, T.J., Thomas, K.S. and Turzi, S.S.,
A molecular field theory approach to the Landau theory of liquid
crystals.  Uniaxial and biaxial nematics,
{\em Phys. Rev. E} {\bf 85} (2012) 031705 (21 pages).
%
\bibitem{MM}
Macmillan, E. H.,
On the hydrodynamics of biaxial nematic liquid crystals.  Part I: General theory.
{\em Arch. Rat. Mech. Anal.} {\bf 117} (1992), 193--239.
%
\bibitem{MA}
Mather, J., Differentiable invariants,
{\em Topology} {\bf 16} (1977), 145--155.
%
\bibitem{MZ}
Michel, L. and Zhilinskii, B.,
Symmetry, invariants, topology. Basic tools,
{\em Physics Reports} {\bf 341} (2001), 11--84.
%
\bibitem{SL}
Sluckin, T.,
The liquid crystal phases: physics and technology,
{\em Contemp. Physics} {\bf41} (2000), 37--56.
%
\bibitem{PO}
Poenaru, V., 
{\em Singularit\'es $C^\infty$ en Presence de Sym\'etrie},
Lecture Notes in Mathematics {\bf 510}, Springer 1976.
%
\bibitem{PS}
Poston, T. and Stewart, I.N.,
{\em Catastrophe Theory and its Applications},
Pitman 1978.
%
\bibitem{RA}
Rey, A.,
Bifurcational analysis of the isotropic-nematic phase transition
of rigid rod polymers subjected to biaxial stretching flow,
{\em Macromol. Theory Simul.} {\bf 4} (1995), 857--872.
%
\bibitem{RM}
Roberts, R. M. and Sousa-Dias, M. E.,
Symmetries of Riemann ellipsoids,
{\em Resenhas IME-USP} {\bf 4}(2) (1999), 183--221.
%
\bibitem{RE}
Rose, M. E., 
{\em Elementary Theory of Angular Momentum},
Wiley 1957; Dover Publications 1995.
%
\bibitem{RO}
Rosso, R.,
Orientational order parameters in biaxial nematics: Polymorphic notation,
{\em Liquid Crystals} {\bf34} (2007), 737--748.
%
\bibitem{SAT}
Sattinger, D. H.,
{\em Group Theoretic Methods in Bifurcation Theory}, Lecture Notes in Mathematics {\bf 762}, Springer 1979.
%
\bibitem{SA}
Sattinger, D. H.,
{\em Branching in the Presence of Symmetry},
CBMS-NSF Reg. Conf. Ser. in Appl. Math. {\bf 40}, SIAM 1983. 
%
\bibitem{SC}
Schwarz, G.,
Smooth functions invariant under the action of a compact Lie group,
{\em Topology} {\bf 14} (1975), 63--68.
%
\bibitem{ST}
Sluckin, T. J. and Thomas, K. T.: personal communication.
%
\bibitem{SF}
Sturmfels, B.,
{\em Algorithms in Invariant Theory},
Springer 1993.
%
\bibitem{SVD}
Sonnet, A. M., Virga, E. G. and Durand, G. E.,
Dielectric shape dispersion and biaxial transitions in nematic liquid crystals,
{\em Phys. Rev. E} {\bf 67} (2003), 061701 (7 pages).
%
\bibitem{TS}
Turzi, S.S. and Sluckin, T.J.,
Symmetry adapted molecular-field theory for thermotropic biaxial nematic liquid crystals and its expansion at low temperature,
{\em SIAM J. Appl. Math.} {\bf 73} (2013), 1139--1163.  
%
\bibitem{WA}
Wasserman, A.,
Equivariant differential topology,
{\em Topology} {\bf 8} (1969), 127--150.
%
\bibitem{WG}
Wasserman, G.,
$(r,s)$-unfolding theory,
Stability of unfoldings in space and time,
{\em Acta Mathematica} {\bf 135} (1975), 57--128.
%
\bibitem{ZP}
Zheng, X. and Palffy-Muhoray, P.,
One order parameter tensor mean field theory for biaxial liquid crystals,
{\em Discrete Contin. Dyn. Syst. Ser. B} {\bf 15} (2011), 475--490.
%

\end{thebibliography}
\end{document}